\RequirePackage{ifpdf}
\documentclass[hyper,11pt,a4paper]{JHEP3}
\usepackage{cite}
\usepackage{graphicx}
\usepackage{amsmath,amssymb,array}
\usepackage{amsmath,amssymb,multirow,array}
\usepackage{bm}
\usepackage{fancyhdr}
\usepackage{url}
\usepackage{epstopdf}
\usepackage{verbatim}

\newcommand{\Tr}{\text{Tr}}

\newcommand{\seff}[1]{S_{\text{eff}}({#1})}
\newcommand{\udag}[1]{U^{\dagger {#1}}}
\newcommand{\tu}[1]{\Tr U^{{#1}}}
\newcommand{\tud}[1]{\Tr U^{\dagger{#1}}}

\newcommand{\nn}{\nonumber}

\newcommand{\ben}{\begin{eqnarray}\displaystyle}
\newcommand{\een}{\end{eqnarray}}

\newcommand{\be}{\begin{equation}}
\newcommand{\ee}{\end{equation}}

\newcommand{\bc}{\begin{center}}
\newcommand{\ec}{\end{center}}

\newcommand{\eesp}{\end{split}}
\newcommand{\bsp}{\begin{split}}

\newcommand{\Rmnum}[1]{\expandafter\@slowromancap\romannumeral #1@}

\renewcommand{\a}{\alpha}	
\renewcommand{\b}{\beta}		
\renewcommand{\d}{\delta}	
\newcommand{\dow}{\partial}

\newcommand{\g}{\gamma}

\renewcommand{\l}{\lambda}	
\newcommand{\m}{\mu}								
\renewcommand{\o}{\omega}	
\newcommand{\q}{\theta}	
	
\renewcommand{\r}{\rho}		
\newcommand{\s}{\sigma}

\newcommand{\D}{\Delta}

\newcommand{\G}{\Gamma}
\newcommand{\I}{\infty}

\newcommand{\Z}{\mathcal{Z}}

\newcommand{\cA}{\mathcal{A}}

\newcommand{\cD}{\mathcal{D}}

\newcommand{\cN}{\mathcal{N}}
\newcommand{\cO}{\mathcal{O}}
\newcommand{\cP}{\mathcal{P}}

\newcommand{\cT}{\mathcal{T}}

\newcommand{\cZ}{\mathcal{Z}}

\newcommand{\expb}[1]{\exp\left[ #1 \right]}
\newcommand{\zcs}{\cZ_{\text{CS}}}
\newcommand{\lna}[1]{\ln\left( #1 \right)}

\newcommand{\cosa}[1]{\cos\lb #1 \rb}

\newcommand{\cota}[1]{\cot\lb #1 \rb}

\newcommand{\ra}{\rightarrow}

\newcommand{\lB}{\left [}
\newcommand{\rB}{\right ]}
\newcommand{\lb}{\left (}
\newcommand{\rb}{\right )}

\newcommand{\<}{\left\langle}
\renewcommand{\>}{\right\rangle}		

\newcommand{\for}{\text{for}}
\newcommand{\where}{\text{where}}
\newcommand{\with}{\text{with}}
\newcommand{\tand}{\text{and}}

\newcommand{\hmin}{h_{\text{min}}}
\newcommand{\hmax}{h_{\text{max}}}

\newcommand{\bensp}{\begin{eqnarray}\begin{split}}
\newcommand{\eensp}{\end{eqnarray}\end{split}}

\def\Xint#1{\mathchoice
{\XXint\displaystyle\textstyle{#1}}%
{\XXint\textstyle\scriptstyle{#1}}%
{\XXint\scriptstyle\scriptscriptstyle{#1}}%
{\XXint\scriptscriptstyle\scriptscriptstyle{#1}}%
\!\int}
\def\XXint#1#2#3{{\setbox0=\hbox{$#1{#2#3}{\int}$ }
\vcenter{\hbox{$#2#3$ }}\kern-.6\wd0}}

\numberwithin{equation}{section}

\usepackage{wrapfig}

\graphicspath{{Chobi/}}

\usepackage{float}
\usepackage[font=small,labelfont=bf]{caption}
\usepackage{mathrsfs}
\usepackage[parfill]{parskip}
\usepackage{subcaption}

\title{Emergent Phase Space Description of Unitary Matrix Model}
\author{Arghya Chattopadhyay\footnote{arghya@iiserb.ac.in} $^{1}$,
Parikshit Dutta\footnote{parikshitdutta@yahoo.co.in} $^{2,3}$\ and Suvankar
  Dutta\footnote{suvankar@iiserb.ac.in} $^{1}$\\
  $^{1}$Department of Physics \\
   Indian Institute of Science Education and Research Bhopal \\
   Bhopal 462 066, 
   India\\

$^{2}$Department of Physics, Asutosh College\\
  92, Shyamaprasad Mukherjee Road\\
Kolkata 700 026, India\\

$^{3}$School of Physical Sciences\\
  Jawaharlal Nehru University\\
  New Delhi 110067, India}

\abstract{We show that large $N$ phases of a $0$ dimensional 
generic unitary matrix model (UMM) can be described in terms 
of topologies of two dimensional droplets on a plane spanned 
by eigenvalue and number of boxes in Young diagram. 
Information about different phases of UMM is encoded in the 
geometry of droplets. These droplets are similar to phase 
space distributions of a unitary matrix quantum mechanics 
(UMQM) ($(0 + 1)$ dimensional) on constant time slices. We 
find that for a given UMM, it is possible to construct an 
effective UMQM such that its phase space distributions match 
with droplets of UMM on different time slices at large $N$. 
Therefore, large $N$ phase transitions in UMM can be 
understood in terms of dynamics of an effective UMQM. 
From the geometry of droplets it is also possible to 
construct Young diagrams corresponding to $U(N)$ 
representations and hence different large $N$ states of the 
theory in momentum space. We explicitly consider two 
examples : single plaquette model with $\Tr U^2$ terms 
and Chern-Simons theory on $S^3$. We describe phases of 
CS theory in terms of eigenvalue distributions of unitary 
matrices and find dominant Young distributions for them.}

\begin{document}

\maketitle

\section{Introduction and summary}

 Unitary matrix model (UMM) has a wide range of applicability both in
mathematics and physics. Partition functions of different
super-symmetric gauge theories, in particular Chern-Simons theory on
certain manifolds boil down to UMMs. These models also have applications in a
broad class of condensed matter systems.

 A UMM is a statistical ensemble of unitary matrices
(denoted by $U$), defined by the partition function
\ben\label{eq:action-UMM}
 {\cal Z} = \int {[\cal{D}}U]\, \exp\left[S(U)\right],\,\,\,\,
 \text{where, $S(U)$ is action and 
 ${\cal {D}} U$ is Haar measure.}
\een
Both ${\cal D}U$ and $S(U)$ are invariant under unitary
transformations.  Since $U$ does not depend on any parameter, this
partition function defines a $0$-dimensional field theory over unitary
matrices. Solution of UMMs in large $N$ limit ($N$ being dimension of
matrix) renders a distribution of eigenvalues (EVs) of unitary
matrices.  EV distribution captures information of different phases of
the system and transition between them.

On the other hand, it is well know that eigenvalues of hermitian (or
unitary) matrices behave like positions of free fermions
\cite{BIPZ,douglas2}. In hermitian matrix quantum mechanics, dynamics of
eigenvalues can be thought of $N$ non-interacting fermions moving in a
potential $V(\q)$. At large $N$ all the eigenvalues (fermions) occupy the
lowest possible energy states (ground state configuration) of the
potential upto a fermi level. Different large $N$ phases of the
system, therefore, can be explained in terms of shapes of fermi
surfaces in phase space. All the states below that surface are occupied,
like Thomas-Fermi distribution at zero temperature \cite{polchinski}.

Phase space description of matrix quantum mechanics seems to be quite
natural because explicit time dependence of eigenvalues allows one to
define conjugate momenta associated with eigenvalues. Interacting
gauge theories (supersymmetric as well as non-supersymmetric) on
compact manifolds can be written as a 0-dimensional UMM
\cite{Sundborg:1999ue, AMMKR, Basu:2005pj, Yamada:2006rx,
  AlvarezGaume:2005fv, Harmark:2006di}. Eigenvalues of these unitary
matrices do not dependent on time (or any other parameter). Therefore, it
would be interesting to ask if one can provide a similar phase space
description for different large $N$ phases of these theories in terms
of droplets of constant densities in two dimensions. In
\cite{duttagopakumar}, with a bit of surprise, it was observed that
for a very restricted class of matrix models, such description is
indeed possible. Writing the partition function as a sum over all
possible representations of unitary group one can study large $N$
behaviour of the modes in terms of most dominant representation which
is characterised by a distribution function $u(h)$. $u(h)$ measures
how boxes $h$ are distributed in a Young diagram. Solving the model at
large $N$, \cite{duttagopakumar} found that information of different
phases of the theory can be captured in terms of topologies (shapes)
of different droplets in two dimensions. This two dimensional plane is
spanned by $\q$ and $h$. Eigenvalue and Young tableaux distributions
for different phases of the theory can be obtained from a
distributions function $\o(h,\q)$ in $(h,\q)$ plane with property
\ben
\begin{split}
\o(h,\q) &= 1 \quad \for  \quad (h, \q) \in R\\
&=0   \quad \text{otherwise.}
\end{split}
\een
Eigenvalue distribution $\r(\q)$ for a given $\q$ can be obtained from
the distribution by integrating over $h$
\ben \label{eq:rhodef}
\r(\q) = \frac1{2\pi}\int_0^\infty \o(h,\q) dh.
\een 
Similarly, for a given $h$, integrating over $\q$ we obtain Young
tableaux distribution,
\ben \label{eq:udef}
u(h) = \frac1{2\pi}\int_{-\pi}^\pi \o(h,\q) d\q \ .
\een 

The droplet picture of large $N$ matrix model is very interesting, as
these droplets are similar to those of Thomas-Fermi model
where Wigner distribution is assumed to take constant value inside
some region in phase space and zero elsewhere. If eigenvalues of
unitary matrices are like positions of fermions with distribution
function $\r(\q)$, then from droplet picture it seems that $h$ plays
the role of momentum with distribution $u(h)$. Therefore, we see that 
droplet picture raises a very intriguing question if there is any
underlying quantum mechanics (quantum mechanics of $N$ free fermions)
associated with different large $N$ phases of UMMs (or interacting
gauge theories).

Before answering this question, we first need to understand if phase
space description is a universal characteristic of UMM. In other
words, if all large $N$ phases of any UMM can be described in terms of
topologies of droplets. An important relation between character of
conjugacy class and number of boxes in a Young diagram is required to
understand this question. For a class of model considered in
\cite{duttagopakumar, duttadutta}, the relation is known and simple
(character is equal to dimension of representation). However, in
general this relation is nontrivial and there exists no exact formula
for character\footnote{We shall discuss about Frobenius formula in
  section \ref{sec:Young} and see that it offers an expression for
  character in terms of auxiliary variables.}. Therefore, it is
difficult to construct $u(h)$ directly from partition function. However, in
\cite{riemannzero}, a more general class of matrix model was
considered, namely {\it plaquette model} and shown that even if it is
difficult to find $u(h)$ directly from partition function, but a
distribution function $\o(h,\q)$ in $(h,\q)$ plane can be obtained in
large $N$ limit. {\it Our first goal, in this paper, is to establish
  this result for a generic UMM and show that different large $N$
  phases of a generic unitary matrix model also have an emergent
 ``phase space" description.} Our generic model includes multi-trace
terms in action. Therefore, this generic matrix model can describe
weakly coupled gauge theories on compact manifold.

Before delving deep, it is worth taking a digression
and mention about some important works and literature 
in mathematics and their relevance with our work. As 
we have already mentioned, different large $N$ phases of
gauge theory correspond to different asymptotic behaviour 
of dominant Young diagrams.
Asymptotic growth of Young diagrams is a well known process 
in mathematics. It originated from the pioneering 
works by A. Vershik and S. Kerov. Details of some of their 
related works can be found in \cite{kerov_book}. Kerov has 
studied a Markovian growth process of Young diagrams also 
known as \emph{Plancherel growth process} (due to involvement 
of Plancherel measure), and shown that almost all of Young 
diagrams become uniformly close to a common universal curve. 
In the current context the work of \cite{Biane} is also 
worth mentioning, who gave the asymptotic behaviour of the 
characters when the corresponding Young diagram converge to 
some prescribed shape. The key difference between the study 
of Kerov and others and our work is that we are interested 
in large $N$ shapes of the Young diagram in presence of
a potential. As we change the parameters of the theory the
potential changes and hence the asymptotic behaviour of 
the corresponding Young diagram. 
This is the reason why we are not getting any universal 
behaviour of the Young diagrams involved in our case and 
precisely this non-universal behaviour is the reason we have 
different shapes of the droplets which in turn helps one to
differentiate between large $N$ phases of the matrix model 
involved.

Coming back to the main context, in order to understand 
the underlying quantum mechanics associated
with large $N$ phases of UMMs, in this paper we observe that droplet
picture in $(h,\q)$ plane is similar to phase space distribution of 
a unitary matrix quantum mechanics (UMQM) on a constant time slice. 
Partition function of UMQM is
equivalent to a theory of free fermions \cite{BIPZ,douglas2}. Time evolution of
this system can be written as time evolution of two different branches
of fermi surface \cite{polchinski}. These different branches of fermi surface define a
closed region in phase space. At large $N$ all the states inside that
region are filled, otherwise empty. There is a relation between
eigenvalue (position of fermions) and corresponding momentum which
defines the boundary of the region \cite{polchinski}. We see that
droplets of UMM have similar feature as phase space distribution of an
effective UMQM on a constant time slice. Hamiltonians of both the
systems match on that constant time slice. Therefore, given a UMM one
can find an underlying matrix quantum mechanics such that its
evolution with time from $t=t_i$ to $t=t_f$ captures two different
phases of UMM on initial and final time slices. This allows us to
understand large $N$ phase transition in UMM in terms of dynamics of
an effective UMQM. Considering parameters of UMQM to be time dependent
one can find a time evolution in effective UMQM such that
distributions on initial and final time slices correspond to two
different phases of UMM \cite{pallab}. The idea here is similar to
the work of Matytsin \cite{Matytsin}, who showed that leading term in a 
non-trivial large-N limit of the Itzykson-Zuber integral can be 
written in terms of an action functional corresponding to the complex 
inviscid Burgers (Hopf) equation with boundary conditions specifying 
the eigenvalue distribution on two different time slices 
\footnote{\cite{Grossmatytsin} extended this study on QCD2 and associated the 
Douglas-Kazakov phase transition on a cylinder with the presence of 
gap in the eigenvalue distributions for Wilson loops.}. In fact, 
the time evolution of the action functional is given by a Hamilton-Jacobi
form with Hamiltonian similar to collective field theory Hamiltonian
of \cite{das-jevicki,sakita}. As we observe in this paper that
for a one dimensional unitary matrix model or 
equivalently an unitary matrix quantum mechanics one ends up with 
\ref{eq:fermisurfeom}, which can be re-written as the Hopf equation.
Solution of this Hopf equation determines the evolution of asymptotic
shape of Young diagram as the system evolves from one phase to other 
phase in the large $N$ limit. Thus evolution of shape of Young diagrams could be related to time evolution of underlying matrix quantum mechanics.

Above result also shows that parameter $h$ (number of boxes in Young
diagram) does indeed play the role of canonical conjugate of eigenvalue
$\q$. Hence, $u(h)$ captures information about momentum distribution
for free fermions in effective UMQM in the form of geometry of Young
diagram. To explore this further we need to construct Young tableaux
distribution $u(h)$ from droplet, since finding $u(h)$ directly from
partition function is difficult for a generic UMM. {\it In this paper
  we show how one can consistently reconstruct different
  representations which dominate partition function at large $N$ from
  phase space distribution.} We explicitly study two examples to
demonstrate the procedure: $\b_1-\b_2$ model and Chern-Simons theory
on $S^3$.

Organisation of our paper is following. 
\begin{itemize}
\item We start with a very brief introduction of UMM in section
  \ref{sec:UMM}. 
\item In section \ref{sec:Young} we discuss emergent phase space
  description of a generic UMM. 
\item Relation between UMM and UMQM has been studied in section
  \ref{sec:qmec}. 
\item We provide an explicit example of reconstruction of Young
  diagrams from phase space for $\b_1$-$\b_2$ model in section
  \ref{sec:example}.
\item In section \ref{sec:CSonS3} we discuss phase space distribution
  for Chern-Simons theory on $S^3$. Here we present a unitary matrix
  model analysis of large $N$ phase of CS theory. Then we obtain Young
  distribution for different values of 't Hooft coupling. 
\item We end our paper with some discussion and out look in section
  \ref{sec:dis}.
\end{itemize}

\section{Unitary Matrix Model}\label{sec:UMM}

\subsection{Unitary matrix model and eigenvalue distributions at large
  $N$}

A UMM is a statistical ensemble of unitary matrices,
defined by the partition function
\ben\label{eq:UMM}
 {\cal Z} & =& \int {[\cal{D}}U]\, \exp\left[S(U)\right]
 \een
 where $S(U)$ is action and ${\cal {D}} U$ is Haar measure. Both
 action and Haar measure are invariant under unitary transformation,
 hence the partition function. Therefore, one can go to a diagonal
 basis where $U$ is given by 
 \be 
U = \{e^{i\q_i}\}, \quad \text{for}\quad i\in [1,N], \quad
 \text{$\q_i$'s are eigenvalues of $U$.}  
\ee
In this basis Haar measure takes the following form
 \begin{eqnarray}
\int[\cD U] = \prod_{i=1}^N\int_{-\pi}^{\pi}d\theta_{i}\prod_{i<j} \sin^{2}
\left(\frac{\theta_{i}-\theta_{j}}{2}\right)
\end{eqnarray}
and action $S(U)$ becomes function of $N$ variables
$\q_1,\cdots, \q_N$ : $V(\{\q_i\})$.  The partition function in this
basis is given by
\ben
\cZ = \int \prod_{i=1}^N d\q_i \exp\lB -\seff{\{\q_i\}}\rB
\een
where,
\be\label{eq:MMoutential}
\seff{\{\q_i\}} = V(\{\q_i\}) - \sum_{{i=1}\atop{j < i}}^{N} \ln\lB \sin^2
\left(\frac{\theta_{i}-\theta_{j}}{2}\right)\rB.
\ee
The partition function can be thought of as describing a system of $N$
particles interacting by a repulsive potential (Coulomb repulsion)
$-\ln |\sin (\q_i-\q_j)/2|$ and moving in a common potential
$V(\{\q_i\})$. Because of the strong repulsion, two particles
(eigenvalues) do not come close to each other. If we neglected the
Coulomb repulsion, all eigenvalues would have sat at the minima of
potential $V(\q)$. But, due to the Coulomb repulsion they spread
around these minima and fill some finite intervals.

In large $N$ limit one defines a set of continuous variables
\be
\q(x) ={\q_i},\quad \text{where} \quad x = {i \over N} \in[0,1]
\ee
and in terms of these continuous variables the effective action becomes,
\be
\seff{\{\q_i\}} = -N^2 \lB S[\q(x)] -\int_0^1 dx 
\Xint -_0^1 dy \ln \sin^2\lb{\q(x)-\q(y)\over 2}\rb\rB.
\ee
At large $N$, dominant contribution comes from the saddle points of
$\seff{[\q(x)]}$. We define eigenvalue distribution function
\ben
\begin{split}
  \r(\q) &= \frac1N \sum_{i=1}^N \delta(\q-\q_i)\\
  & = \left | {\dow x\over \dow \q}\right |
\end{split}
\een
which captures information about distributions of eigenvalues between
$-\pi$ and $\pi$ in large $N$ limit. Different saddle points
correspond to different eigenvalue distributions and hence different
phases of the theory.

\subsection{Gauge theory at zero coupling}

Thermal partition function of gauge theory on compact manifold ($S^3$)
can be written as a UMM. In zero 't Hooft coupling limit
($\l = g_{YM}^2 N \ra 0$), all the modes of the theory are massive
with mass terms depending on the size of $S^3$.  Only the zero mode of
temporal component of gauge field is massless and strongly
interacting.  Therefore, in zero coupling limit one can integrate out
all the massive modes and obtains an exact partition function in terms
of the zero mode. For example, thermal partition function of a $SU(N)$
gauge theory on $S^3\times S^1$ at zero 't Hooft coupling can be
written as,
\ben\label{eq:pf-zero}
\cZ (\b) = \int [\cD U] \exp\lB \sum_{n=1}^\infty {a_n(\b)\over n}
\Tr U^n \Tr U^{\dagger n}\rB
\een
where $U= e^{i\b \a}$, $\a$ being zero mode of temporal component of
the gauge field and $\b$ (radius of $S^1$) is inverse
temperature. $a_n(\b)$'s are related to single particle partition
function of bosonic and fermionic modes. See \cite{AMMKR, Basu:2005pj,
  Yamada:2006rx, Harmark:2006di} for details.

\subsection{Gauge theory at weak coupling}

The above equation is derived at zero 't Hooft coupling. One considers
only one loop contribution while integrating out massive modes. In
weak 't Hooft coupling limit one can integrate out massive modes order
by order in loop expansion and obtains the following partition
function in terms of the holonomy $U$,
\ben
\begin{split}\label{eq:pf-weak}
\cZ(\b, \l) &= \int [\cD U] \exp[\seff{U}]\\
\seff{U} &= \sum_{\vec n} {a_{\vec n}(\b)\over N^k} \prod_{i=1}^k \Tr
U^{n_i}, \quad 
\text{with} \quad \sum_i n_i =0.
\end{split}
\een
Terms, in $\seff U$, with $k$ traces appears in $(k-1)$ loops in
perturbation theory and have a planner contribution starting with
$\l^{k-2}$. For example, two trace terms (\ref{eq:pf-zero}) comes in
one loop order in perturbation theory.  In this paper, we show that
different large $N$ phases of this model has an emergent phase space
picture.

\subsection{Single plaquette model}

A single plaquette model is given by
\be\label{eq:singleplaq}
\cZ(\b)=\int {\cal{D}}U\, \exp\left[N\sum_{n=1}^{\infty}
\frac{\b_{n}(\b)}{n}(\Tr[U^{n}]+\Tr[U^{\dagger n}])\right]
\ee
where $\b_n(\b)$'s are parameters. This partition function is equivalent to \ref{eq:pf-zero} with $\beta_n(\beta)$ identified as $a_n(\beta)\<\Tr U^n\>$ in the large $N$ limit. They have similar phase structure.

The single plaquette model has lot of applications in lattice gauge
theory. Also, as we shall see, Cherns-Simons theory on $S^3$ can also be written as single plaquette model with
$\b_n$'s, determined in terms of 't Hooft coupling. Single plaquette model has a rich phase structure. We refer to \cite{zalewski, Mandal:1989ry} for a detailed study.

\section{Emergent phase space distribution} \label{sec:Young}

It was first observed in \cite{duttagopakumar} that there exists a
relation between eigenvalue density and Young tableaux distribution
for a restricted class of matrix models.  It was also shown that using
these relations, different large $N$ phases of the theory can be
described in terms of shapes of droplets in two dimensional plane
spanned by eigenvalue and number of boxes in Young diagram.  However,
observation in that paper was accidental and valid for a very limited
class of UMM. Also, it was not clear if such identification follows
from any extremization condition at large $N$. In \cite{riemannzero},
work of \cite{duttagopakumar} was generalized to single plaquette
model.  Most interestingly, \cite{riemannzero} showed that the
relation between Young distribution and eigenvalue distribution
follows from an extremization condition. In this paper we further
extend this work to a most generic matrix model of the form
(\ref{eq:pf-weak}). Although, we work here only upto three trace
terms, our construction can be generalised to any arbitrary number of
traces.

\subsection{Two dimensional droplets for plaquette model}

We start our discussion with a brief review of
\cite{riemannzero}. This discussion will help us to follow the
derivation for the most generic case.

Partition function of single plaquette model (\ref{eq:singleplaq}) is
given by,
\be\label{eq:partfuncplaq}
\cZ = \int [\cD U] \exp\lB N \sum_{n=1}^{\infty}{\beta_{n}\over n}
(\Tr U^{n}+\Tr U^{\dagger n})\rB,\quad \text{$\b_n$'s are some
  arbitrary coefficients.} 
\ee

To obtain phase space distribution, we expand the exponential and write
partition function as a sum over representations $R$ of unitary group
$U(N)$,
\ben \label{eq:ZplaqsumoverR}
\cZ = \sum_R \sum_{\vec k} \frac{\varepsilon(\vec \b,\vec k)}
{z_{\vec k}} \sum_{\vec l} \frac{\varepsilon(\vec \b,\vec l)} {z_{\vec
    l}} \chi_{R}(C(\vec{k}))\chi_{R}(C(\vec{l})).  
\een
Here $\chi_{R}(C(\vec{k}))$ is the character of conjugacy class
$C(\vec{k})$ of permutation group $S_{K}$, $K=\sum_n n k_n$ and
\ben
\varepsilon(\vec \b, \vec k) =
\prod_{n=1}^{\infty}N^{k_n}\b_{n}^{k_n}, \quad
z_{\vec k} = \prod_{n=1}^{\infty} k_{n}! n^{k_n}.
\een
To derive equation (\ref{eq:ZplaqsumoverR}) we have used the
following identity
\be
\prod_n (\Tr U^n)^{k_n} = \sum_{R}\chi_{R}(C(\vec{k}))\Tr_{R}[U]
\ee
and the normalisation condition\footnote{See
  \cite{lasalle,hamermesh,fulton-harris} for details.}
\be \label{eq:normalization_TrU}
\int {\cal{D}}U\, \Tr_{R}[U] \Tr_{R'}[U^{\dagger}]=\delta_{RR'}.
\ee
Sum of representation of $U(N)$ can be written as a sum over different
Young diagrams. If $K$ is the total number of boxes in a Young diagram
with $\lambda_i$ being the number of boxes in $i$-th row then
$\sum_{i=1}^N \lambda_i=K$. Hence, sum over representations can be
decomposed as
\begin{equation}
  \sum_R\longrightarrow\sum_{K=1}^\infty\,\sum_{\{\lambda_i\}}\ 
  \delta\lb \sum_{i=1}^N\lambda_i-K\rb \quad  \text{with}\quad 
  \lambda_1\geq \lambda_2\geq\cdots\geq \lambda_N\geq 0.
\end{equation}
The partition function, therefore, can be written as,
\ben\label{eq:pffinal} 
\cZ=\sum_{\vec \l} \sum_{\vec k, \vec l}
\frac{\varepsilon(\vec \b, \vec k) \varepsilon(\vec \b, \vec
  l)}{z_{\vec k} z_{\vec l}} \chi_{\vec\l}(C(\vec k)) \chi_{\vec
  \l}(C(\vec l)) \ \delta \lb\sum_n n k_n-\sum_i \l_i\rb \delta \lb
\sum_n n l_n-\sum_i \l_i\rb .  
\een

We introduce $N$ variables $h_1, \cdots , h_N$, related to number of
boxes $\l_i$'s as
\begin{eqnarray}\label{eq:h-nrelation}
h_i = \l_i + N -i \qquad \forall \quad i =1, \cdots, N.
\end{eqnarray}
$h_i$'s are shifted number of boxes\footnote{Our terminology is little sloppy. We also call $h_i$ as the number of boxes in the $i$-th row.}. From monotonicity of
$\l_i$s it follows that $h_i$s satisfy the following constraint
\be
\label{hmonotonicity} h_1> h_2> \cdots > h_N \geq 0.  
\ee
From now on we shall use variables $h_i$s in stead of $\l_i$s.

In large $N$ limit we define continuous variables
\begin{eqnarray}\label{eq:contvardef}
  {h_{i}\over N}=h(x), 
  \quad\quad k_{n}=N^{2}k'_n,
  \quad\quad
  x={i\over N} \quad \with \quad x\in [0,1].
\end{eqnarray}
In this limit summation over $i$ is replaced by an integral over $x$,
\be
\sum_{\a=1}^{\infty} \ra N\int_0^1 dx
\ee
and sum over representations ($\vec \l$) and sum over cycles ($k_n$)
are given by path integral over $h(x)$ and integral over
$k_n'$. Writing characters $\chi_{\vec h}$ in terms of $h(x)$ and
$k_n'$, partition function (\ref{eq:pffinal}), in large $N$ limit, can
be written as
\be\label{eq:pfyt1}
\cZ = \int [\cD h(x)] \prod_n\int dk_n' dl_n' \exp\lB -N^2 
S_{\text{eff}} [h(x),\vec {k_n'},\vec{l_n'}]\rB,
\ee
where $S_{\text{eff}}$ is the effective action. Dominant contribution
to partition function comes from those representations which maximise
effective action $S_{\text{eff}}$. To find the most dominant
representations at large $N$, we introduce a function called Young
tableaux density defined as,
\be
u(h) = - {\dow x\over \dow h}.
\ee
Different Young diagrams correspond to different $u(h)$.  Since $h(x)$
is a monotonically decreasing function of $x$, Young density has an
upper and lower cap $1\geq u(h) \ge 0 \ \forall \ x\in[0,1]$.

For the simplest case, $\b_1\neq 0$ and other $\b_n=0$ (for $n\geq 2$)
one can find $u(h)$ for different phases of the model
\cite{duttagopakumar, duttadutta}. In this simple case we need to
calculate character of permutation group in presence of one cycles
only ($k_n=0$ for $n\ge 2$) which is given by dimension of the
corresponding representation. As a result is was possible for
\cite{duttagopakumar, duttadutta} to extremize the effective action to
obtain $u(h)$ for different phases\footnote{A large $N$ phase
  transition corresponds to a qualitative change in nature of this
  dominant representation as one varies the parameters. The
  Douglas-Kazakov phase transition \cite{Douglas:1993iia,
    Kazakov:1995ae} and its generalisations in 2d Yang-Mills theory
  can be understood this way.}.  \cite{duttagopakumar, duttadutta}
also observed that there exists a beautiful identification between
Young tableaux distribution and eigenvalue distribution for all the
phases of the theory,
\ben\label{eq:beta1identification}
 h_\pm =\frac12 +\b_1 \cos\q \pm \pi \r(\q),\quad \text{and} \quad
u(h) = {\q\over \pi}
\een
where $h_\pm$ are solutions of the following equation
\be \label{eq:beta1boundrel}
h^2 - (1+2\b_1 \cos\q)h +\frac14(1+2\b_1\cos\q)^2 = \pi^2 \r^2(\q).
\ee
This identification allows one to write eigenvalue distribution and
Young tableaux distribution in terms of a single constant distribution
function $\omega(h,\q)$ given by equations (\ref{eq:rhodef}) and
(\ref{eq:udef}).  $h_\pm(\q)$ defines the boundary of the distribution. It
is actually the shape (or boundary) of the region $R$ which contains
all the information about two distributions for a given phase of the
model.  Thus, a relation between $h$ and $\q$ defines disjoint
islands (phase space distribution/droplet) in two dimensions for
different phases of the UMM under consideration. Different phases are
distinguished by different topologies of droplets. Constant phase
space distribution is very similar to Thomas-Fermi (TF) model
\cite{thomasfermi1, thomasfermi2} at zero temperature.

The identification, observed in \cite{duttagopakumar, duttadutta},
between two distribution was accidental. It was not known if such
identification holds for a generic model like (\ref{eq:partfuncplaq}).
The main obstacle to apply the idea of \cite{duttagopakumar} to a
generic UMM was finding character in terms of $h_i$s in presence of
arbitrary cycles. Presence of $\Tr U^n$ term in partition function
requires to calculate character in presence of $n$-cycles. Hence,
finding $u(h)$ by extremizing the effective action for a generic
unitary matrix model was a challenging problem.

In \cite{riemannzero} it was shown that, although it is difficult to
find $u(h)$ for a generic plaquette model (\ref{eq:partfuncplaq}), but
one can still find droplets for different phases of the theory and
from the shape of these droplets one can reconstruct dominant Young
diagram.

\subsubsection{Frobenius formula for character}

Let $C({\vec{k}})$ denotes the conjugacy class of
$S_K$ which is determined by a collection of sequence of
numbers
\begin{eqnarray}
\vec{k}=(k_{1},k_{2},\cdots), \qquad \text{with}, \ \
\sum_{n}n k_{n}=K .
\end{eqnarray}
$C({\vec{k}})$ consists of permutations having $k_{1}$ number of
1-cycles, $k_{2}$ number of 2-cycles and so on.  We introduce a set of
independent variables, $x_{1},\cdots,x_N$.  The character
corresponding to a conjugacy class $C(\vec k)$ of $S_K$ for a
representation characterized by $\vec \l$ is given by famous Frobenius
formula
\begin{eqnarray}\label{eq:Frobenius}
  \chi_{\vec h }(C_{\vec{k}})=\bigg[\Delta(x).\prod_{n} \lb P_{n}(x)
  \rb^{k_n}\bigg]_{(h_{1},\cdots, h_{N})} 
\end{eqnarray}
where 
\begin{eqnarray}
  P_{n}(x)=\sum_{i=1}^{N}x_{i}^{n},\quad
  \quad\Delta(x)=\prod_{i<j}(x_{i}-x_{j})
\end{eqnarray}
and $h_i$'s are give by equation (\ref{eq:h-nrelation}). The notation
$[\cdots]_{(h_{1},\cdots, h_{N})}$ implies
\begin{equation}
  \lB f(x)\rB_{(h_{1},\cdots, h_{N})}=\textrm{coefficient of }
  x_{1}^{h_{1}}\cdots x_{N}^{h_{N}}\textrm{ in }f(x) .
\end{equation}

\subsubsection{Extremization of partition function}

Promoting $N$ real variables $x_1,\cdots,x_N$ in (\ref{eq:Frobenius})
to $N$ complex variable $(z_1, z_2, \cdots z_N)$ character can be
written as,
\begin{eqnarray}\label{eq:character2}
 \chi_{\vec h }(C_{\vec{k}})=
\prod_{\a=1}^{N}\oint\frac1{2\pi i}\frac{dz_{\a}}{z_{\a}^{h_{\a}+1}}
\lB \lb\sum_{\a=1}^{N}z_{\a}\rb^{k_{1}}
\lb\sum_{\a=1}^{N} z_{\a}^{2}\rb^{k_{2}} \cdots
\prod_{\a<\b}(z_{\a}-z_{\b})
\rB.
\end{eqnarray}
The contour is taken around origin. Note that, in the above
expressions for character (equation \ref{eq:Frobenius} or
\ref{eq:character2}) variables $x_i$'s or $z_i$'s are auxiliary
variables. They do not carry any physical meaning, as of now.

Integrand in the above 
expression can be exponentiated and written as,
\begin{eqnarray}
\chi_{\vec h }(C_{\vec{k}})=\prod_{\a=1}^{N}\oint
\frac1{2\pi i}\frac{dz_{\a}}{z_{\a}}
\exp\lB \sum_{n=1}^{\infty}k_{n}\ln(\sum_{\a=1}^{N}
z_{\a}^{n})+\frac{1}{2}\sum_{\a\ne \b}\ln|z_{\a}-
z_{\b}|-\sum_{\a=1}^{N}h_{\a}\ln z_{a}\rB.
\end{eqnarray}

Therefore, in large $N$ limit character is given by,
\ben\label{eq:character1}
\chi[h(x), C(\vec{k'})]= \lb\frac1{2\pi i}\rb^N 
\oint {[\cD z(x)]\over z(x)}
\exp[-N^2 S_{\chi}(h[x],\vec{k'})]
\een
where {\it action} $S_{\chi}$ is given by
\ben\label{eq:Schi}
-S_{\chi}(h[x],\vec{k'}) &=& \sum_{n=1}^\infty k_n' \lB (n+1)\ln N
+\ln \lb \int_0^1 dx z^n(x)\rb\rB
+\frac12 \int_0^1 dx \Xint -_0^1 dy \ln|z(x)-z(y)|\nonumber\\
&& \quad -K' \ln N
-\int_0^1 dx h(x) \ln z(x)
\een
and $N^2 K'= K = \text{total number of boxes in a representation}$.

Using the expression for character given in equation (\ref{eq:character1}),
partition function (\ref{eq:pffinal}) can be written as,
\ben\label{eq:pftotal}
\cZ = \cN \prod_n \int dk_n' dl_n' \int \cD h(x)\oint \cD z(x)
\oint{\cal{D}}w(x) \int_{-\I}^\I dt \int_{-\I}^\I ds
\exp\lb - N^2 S_{\text{total}}\lB h,z,w,\vec{k'},
\vec{l'},t,s\rB \rb\nonumber\\
\een
where,
\begin{eqnarray}\label{eq:effStotal}
\begin{split}
- S_{\text{total}}\lB h,z,w,\vec{k'},
\vec{l'},t,s\rB &= \sum_n \bigg[ k_n'\lb1+ \lna{{\b_n Z_n\over n k_n'}}\rb 
+l_n'\lb1+ \lna{{\b_n W_n\over n l_n'}}\rb \bigg]\\
&\qquad +\frac12 \int_0^1 dx \Xint-_0^1 dy\lb \ln|z(x)-z(y)|
+\ln|w(x)-w(y)|\rb\\
& \qquad -\int_0^1 dx h(x) \ln z(x) w(x) 
+i t \lb \sum_n n k_n' -K'\rb +i s \lb \sum_n n l_n' -K'\rb.
\end{split}
\end{eqnarray}
Varying effective action with respect to $z(x)$ we
find\footnote{Similar equation can be obtained when we vary action
  with respect to $w(x)$.},
\ben\label{eq:saddlepoint}
\sum_n {n k_n'\over Z_n} z^n(x) + \Xint-_0^1 dy
{z(x)\over z(x)-z(y)}
-h(x) =0
\een
where,
\be
Z_n = \int_0^1 dx z^n(x).
\ee

Varying effective action with respect to $h(x)$ we find,
\be
z(x)w(x) = e^{-i(t+s)} = \text{constant (independent of} \ x).
\ee
Variation with respect to $k_n'$ and $l_n'$ provides,
\ben
{\b_n Z_n\over n k_n'} = e^{-i t n}, \quad \text{and}
\quad {\b_n W_n \over n l_n'} = e^{-i s n}.
\een
Since, in equation (\ref{eq:pftotal}) $z(x)$ and $w(x)$ vary over some
contour around zero, we take these contours to be unit circle about
the origin and hence a consistent solution to above equations are,
\be
z(x) = e^{i \q(x)}, \quad w(x) = e^{-i\q(x)}, \quad \text{and} \quad
t=s=0.
\ee
Therefore,
\be
k_n' = l_n'= {\b_n \r_n\over n}, \quad \text{where} \quad
Z_n = \int d\q \r(\q) \cos n\q \equiv\r_n.
\ee
Thus saddle point equation (\ref{eq:saddlepoint}) becomes,
\be\label{eq:chaeqn} 
\sum_n {n k_n'\over Z_n} e^{i n \theta}
+\Xint-_{-\pi}^{\pi} d\theta' \r(\theta') \frac{e^{i\theta}}
{e^{i\theta}-e^{i\theta'} } -h(\q)=0, \quad \text{where} \quad
\r(\theta) = {\dow x\over \dow \q}.  
\ee
Evaluating the partition function (\ref{eq:pftotal}) on these 
equations we find,
\begin{eqnarray}
\begin{split}
\cZ &= \cN \int [\cD \q] \exp\lB N^{2}\sum_{n=1}^{\infty}
2\frac{\b_n}{n}\r_n
+N^{2}\frac{1}{2}\Xint-d\q\rho(\q)\Xint-d\q'
\rho(\q')\ln\left| 4\sin^{2}\lb\frac{\q-\q'}{2}\rb\right|\rB.
\end{split}
\end{eqnarray}
This partition function is exactly same as the partition function of
one plaquette model written as an integral over eigenvalues of unitary
matrices. Hence, we identify that the auxiliary variables we used to
write the character of symmetric group with eigenvalues of
corresponding unitary matrix model involved.  Imaginary part of saddle
point equation (\ref{eq:chaeqn}) becomes equation for eigenvalue
density,
\ben 
\sum_n {n k_n'\over Z_n}\sin{n\q} -\frac12
\Xint-_{-\pi}^{\pi} d\theta' \r(\theta') \cota{{\q-\q'\over 2}}=0.
\een
On the other hand, the real part of equation (\ref{eq:chaeqn})
provides a relation between $h$ and $\q$,
\be
h(\q)=\frac12 + \sum_n {n k_n'\over Z_n}\cos{n\q} .
\ee
However, this equation is not complete. It is easy to check that total
action (\ref{eq:effStotal}) is invariant under
\be
h(\q)\ra h(\q) + f(\q)
\ee
where, $f(\q)$ is an even function of $\q$. Hence, the most generic form of
$h(\q)$ is given by,
\be
h(\q)=\frac12 + \sum_n {n k_n'\over Z_n}\cos{n\q} + f(\q).
\ee
Form of $f(\q)$ can be fixed from topological structure of droplets
for different phases \cite{riemannzero} and is given by $f(\q)=\pi
\r(\q)$. Finally we have,
\be \label{eq:boundrelplaq}
h_\pm(\q)= S(\q)\pm \pi \r(\q), \quad 
S(\q) = \frac12 + \sum_n {\b_n}\cos{n\q}.
\ee
This equation defines the shape of constant droplets in $(h,\q)$
plane. This equation is generalization of the relation given in equation
(\ref{eq:beta1identification}). Thus we see that for a generalised
plaquette model all the phases of the theory can be characterised in
terms of topologies of different droplets in two dimensional plane
spanned by $h$ and $\q$.

\subsection{Two dimensional droplets for generic unitary matrix model}
\label{sec:fermisurfaceweak}

We start with an action which contains maximum three traces,
\begin{eqnarray}
S_{eff}=\sum_{{n_1,n_2 \atop \text{all values}}}
{a_{n_1n_2}\over N^3}\Tr{U^{n_1}}\Tr{U^{n_2}}\Tr{U^{\dagger
  (n_1+n_2)}}. 
\end{eqnarray} 
When $n_1$ (or $n_2$) is zero we get back one loop (zero coupling)
action.  The above action is invariant under $U\rightarrow U^{\dagger}$ if $a_{n_1n_2}=a_{-n_1 -n_2}$, and
can also be written as,
\begin{eqnarray}\label{eq:seffthreetrace}
S_{eff}=\sum_{n_1,n_2\ge 0}{a_{n_1n_2}\over N^3}
\Bigg(\Tr{U^{n_1}}\Tr{U^{n_2}}\Tr{U^{\dagger (n_1+n_2)}}+\Tr{U^{\dagger n_1}}
\Tr{U^{\dagger n_2}}\Tr{U^{(n_1+n_2)}}\Bigg)
\end{eqnarray}
which is manifestly invariant under $U\ra \udag{}$.  

Following the same analysis as discussed in previous section we find a
saddle point equation for auxiliary variables appearing in the Frobenius
formula as (see appendix \ref{app:details} for details),
\ben\label{eq:sadeqnthreetrace}
 \sum_{n_1n_2=0}^{\infty} \lb{n_1k_{n_1n_2}'
\over Z_{n_1}}e^{i n_1\q}+{n_2k_{n_1n_2}'
\over Z_{n_2}}e^{i n_2\q}
 +{(n_1+n_2) k_{n_1n_2}'
\over Z_{n_1+n_2}}e^{i (n_1+n_2)\q}\rb && \nn\\
 -{i\over 2}\ \Xint- d\theta'\,\rho(\theta')\cot\lb {\theta-\theta'\over 2}\rb
+\frac12 -h(\q) &=& 0
\een
where, $\r(\q)$ is defined in equation (\ref{eq:chaeqn}).  Imaginary
part of this equation gives eigenvalue equation
\begin{eqnarray}
\sum_{n_1n_2=0}^{\infty}\lB{n_1k_{n_1n_2}'\over Z_{n_1}}
\sin(n_1\theta)+{n_2k_{n_1n_2}'\over Z_{n_2}}\sin(n_2\theta)
+{(n_1+n_2)l_{n_1n_2}'\over Z_{n_1+n_2}}\sin\left((n_1+n_2)\theta\right)\rB \ \ 
&&\nonumber\\
-{1\over 2}\ \Xint-_0^1 d\theta'\,\rho(\theta')\cot\lb{\theta-\theta'\over 2}\rb &=&0.
\ \ \ \ \ \ 
\end{eqnarray}
Integrating over $h(x)$, the partition function can be written as
\begin{eqnarray}
\cZ={N^4\over (2\pi i)^{2N}}\int &&[D\theta]
\exp\Bigg[\sum_{n_1,n_2=0}^{\infty}2N^2a_{n_1n_2}'\rho_{n_1}
\rho_{n_2}\rho_{n_1+n_2}\nonumber\\&&+\frac{N^{2}}{2}
\int d\theta\rho(\theta)\Xint-d\theta'\rho(\theta')
\left(\ln\,4\sin^{2}\left(\frac{\theta-\theta'}{2}\right)
+\ln\,4\right)\Bigg].
\end{eqnarray}
Thus we see, even for a general matrix model auxiliary variables in
Frobenius formula are same as eigenvalues of unitary matrices.

The real part of saddle point equation defines the shape of constant
droplets in $(h,\q)$ plane
\begin{eqnarray}\label{eq:boundrelgen}
  h_\pm(\theta)=S(\q) \pm \pi \r(\q).
\end{eqnarray}
where,
\be 
S(\q)={1\over 2}+\sum_{n_1n_2=0}^{\infty} a'_{n_1n_2}
\r_{n_1}\r_{n_2}\r_{n_1+n_2} \Bigg({n_1\cos n_1\q \over
  \r_{n_1}}+{n_2\cos n_2\q \over \r_{n_2}} + {(n_1+n_2)
  \cos(n_1+n_2)\q \over \r_{n_1}\r_{n_2}}  \Bigg) .
\ee
This is a generalization of equation (\ref{eq:boundrelplaq}). It is
easy to check that all the expressions matches with those of plaquette
model for $n_2=0$ with $2n_1a'_{n_1 0}\r_{n_1}=\b_{n_1}$.

Although here we considered three trace terms only, a generalisation
to arbitrary model is straightforward. However, algebra will be
little difficult.

\section{Unitary matrix quantum mechanics and phase space
  distribution}
\label{sec:qmec}

$h_\pm$ in generalized boundary relation (\ref{eq:boundrelplaq} or
\ref{eq:boundrelgen}) is a solution of the following quadratic
equation
\be\label{eq:quadraticboundrel}
h^2 - 2S(\q)h + S^2(\q)
-\pi^2 \r^2(\q)=0.
\ee
We define a distribution function $\o(h,\q)$ in $(h,\q)$ plane
\be \label{eq:phasespacedistri}
\o(h,\q) = \Theta\lb{(h-h_-(\q))(h_+(\q)-h)\over 2}\rb
\ee
such that $\o(h,\q)=1$ for $h_-(\q)<h<h_+(\q)$ and zero otherwise.
Eigenvalue distribution, by construction, can be obtained by integrating
out $h$ for a given $\q$ (equation \ref{eq:rhodef}) and is given by
\be\label{eq:rhofromphasespace}
\r(\q)= {h_+(\q)-h_-(\q)\over 2\pi}.
\ee
Wilson loops are given by
\be 
\< \Tr U^n\> = \frac1{2\pi} \int_0^\infty dh \int_{-\pi}^{\pi} d\q \
\cos n\q \ 
\o(h,\q).  
\ee
The function $S(\q)$ can also be written in terms of phase space
geometry
\be\label{eq:Sfromphasespace}
S(\q) = \frac1{2\pi \r}\int_0^\infty dh \ h \ \o(h,\q) =
{h_+(\q)+h_-(\q) \over 2}.
\ee

The droplet picture of different large $N$ phases are similar to
droplets of Thomas-Fermi (TF) model at zero temperature. Fermi
distribution at zero temperature is given by
\be
\D(p,q) = \Theta(\mu - \mathfrak{h}(p,q))
\ee
where $\mu$ is chemical potential and $\mathfrak{h}(p,q)$ is single
particle Hamiltonian density. Comparing our phase space distribution
(\ref{eq:phasespacedistri}) with TF distribution we find the
Hamiltonian density is given by
\ben
\mathfrak{h}(h,\q)= {h^2\over 2} - S(\q)h + {g(\q)\over 2} + \mu, \quad 
\where \quad g(\q) = h_+(\q) h_-(\q).
\een
Total Hamiltonian can be obtained by integrating $\mathfrak{h}(h,\q)$
over the phase space
\ben
\begin{split}
H_h &=\frac1{2\pi} \int_{-\pi}^{\pi} d\q
\int_{h_-}^{h_+} dh \ \o(h,\q) \ \mathfrak{h}(h,\q).
\end{split}
\een
Integrating over $h$ we find Hamiltonian as a functional of $\r(\q)$
\ben
H_h[\r] = \frac12\int d\q \lb {\pi^2 \r^3 \over 3} -S^2 \r + g \r\rb +\m.
\een
This is TF energy functional. As a consistency, one can check that
extremization of this Hamiltonian with respect to $\r$ gives eigenvalue
distribution (\ref{eq:rhofromphasespace}).

The Hamiltonian can also be written as
\be
\begin{split}\label{eq:Hamh}
H_h & = \int d\q \lb \frac{S^2 \r}2 + \frac{\pi^2 \r^3}6\rb + V_{eff}(\r)
+\m\\
& = \frac1{2\pi}\int dh\ d\q \lb \frac{h^2}2 + V_{eff}(\q)\rb \o(h, \q) \ + \ \m
\end{split}
\ee
for some $V_{eff}(\theta)$. For example, for simple $\b_1$ model
$S(U) = \b_1 (\Tr U +\Tr U^{\dagger})$,
$V_{eff}(\q) = -\b_1\cos\q-\b_1^2/2\cos2\q$.

\subsection{Unitary matrix quantum mechanics}

We start with a unitary matrix quantum mechanics
\ben\label{eq:mmqmech}
\Z_t = \int [DU] \exp\lB \int dt \lb \Tr\dot U^2 
+ W(U)\rb \rB.
\een
This matrix model is equivalent to a theory of free fermions on circle
with Hamiltonian \cite{BIPZ,douglas2}
\be
H_F = \int d\q \lb \dow\Psi^{\dagger}\dow\Psi + W(U)\Psi^{\dagger}\Psi
\rb 
\ee
and with eigenvalue density $\sim \Psi^{\dagger}\Psi$. The matrix model
(\ref{eq:mmqmech}) can also be described as a real bosonic field
theory of eigenvalue density $\rho(\q,t)$ on $S^1$
\cite{sakita,jevicki,das-jevicki,pallab}. Corresponding Hamiltonian is given
by,
\be\label{eq:HamB}
H_B = \int d\q \lb \frac12 {\dow \pi(t,\q) \over \dow\q} \r(t,\q)
{\dow \pi(t,\q) \over \dow\q} + \frac{\pi^2 \r^3(t,\q)}{6} + W(\q)
\r(t,\q) \rb
\ee
where, $\pi(\q,t)$ is momentum conjugate to $\r(\q,t)$. Equations of
motion for collective fields are given by,
\ben
\begin{split}\label{eq:colleom}
\dow_t \r(t,\q)+\dow_\q \lb \r(t,\q) v(t,\q)\rb &=0\\
\dow_t v(t,\q) + \frac12 \dow_\q v(t,\q)^2 + \frac{\pi^2}2 \dow_\q
\r(t,\q)^2 &= -W'(\q)
\end{split}
\een
where 
\be
v(t,\q) = \dow_\q \pi(t,\q).
\ee
These are coupled, non-linear partial differential equation and hence
difficult to find a solution in general. Therefore, one can translate
the problem into free fermi language \cite{polchinski}. At large $N$,
time evolution of the fermionic system can be described in terms of
time evolution of fermi surface in phase space. Denoting phase space
density by $\varpi(p,\q)$ the Hamiltonian is given by,
\be\label{eq:HamF}
H_p = \frac1{2\pi} \int d\q \ dp \lb \frac{p^2}2 + W(\q) \rb \varpi(p,\q)
\ee
where $\int d\q W(\q) \r(\q,t) = W(U(t))$ and $p$ is momentum
(conjugate of $\q$). This Hamiltonian has the same form as
(\ref{eq:Hamh}) except the fact that $(p(t),\q(t))$ in (\ref{eq:HamF})
are functions of ``$t$". There is a one to one correspondence between
phase space variables and collective field theory
variables. Eigenvalue density and corresponding momentum are given by
\cite{polchinski}
\ben 
\r(\q, t) =\frac1{2\pi} \int dp \ \varpi(p, \q), \quad \dow_\q
\pi(\q,t) = \frac1{2\pi\r} \int dp \ p \ \varpi(p,\q). 
\een
Assuming $p(t,\q)$ ranges from $p_-(t,\q)$ to $p_+(t,\q)$ for a fixed
$\q$ we land up with following dictionary between bosonic and
fermionic variables
\be\label{eq:BFdic}
\r(t, \q) ={p_+(t,\q)-p_-(t,\q)\over 2\pi}, \quad  \tand \quad 
v(t,\q) ={p_+(t,\q)+p_-(t,\q)\over 2}.  
\ee
Using this dictionary the Hamiltonian (\ref{eq:HamF}) reduces to Hamiltonian (\ref{eq:HamB}). Also, using this dictionary one can show that
coupled non-linear differential equations for $\r(t,\q)$ and $v(t,\q)$
take simpler form in fermionic language
\ben\label{eq:fermisurfeom}
\dow_t p_{\pm}(t,\q) + p_{\pm}(t,\q)\dow_\q p_{\pm}(t,\q) +W'(\q) =0.
\een
These are decoupled first order quasi-linear partial differential
equations. $p_\pm(t,\q)$ defines fermi surface in $(p,\q)$ plane. The
above equation (\ref{eq:fermisurfeom}) determines evolution of fermi
surface with time. Thus solving the field theory equations of motion
(\ref{eq:colleom}) is equivalent to solve for upper and lower fermi
surfaces in fermionic picture. In either case, one needs to provide an
initial data on a constant time slice in $(t,\q)$ plane. After that
the problem reduces to a {\it Cauchy problem}. Existence of unique
solution depends on the geometry of initial data curve.

Inverting the dictionary (\ref{eq:BFdic}) we find
\be 
p_\pm(\q,t) = v(t,\q) \pm \pi \r(t,\q).  
\ee
These equations for fermi surface have similarity with generalized
boundary relation (\ref{eq:boundrelgen}) in $(h,\q)$ plane.
Therefore, on a constant $t$ slice\footnote{In finite temperature
  version, constant time slices are constant temperature slices.}, one
can match fermi surfaces $p_\pm(t,\q)$ with boundary of large $N$
droplets (\ref{eq:boundrelgen}) of a generic UMM (\ref{eq:UMM}). In
other words, large $N$ phase space distribution can be used as initial
(or terminal) data for the Cauchy problem,
\be\label{eq:datacurve}
p_\pm(t_0,\q) = h_\pm(\q). 
\ee
In the bosonic language this is equivalent to $v(t_0,\q) = S(\q)$ and
$\r(t_0,\q)=\r(\q)$.  A general phase space distribution
(\ref{eq:boundrelgen}) therefore can be thought of as initial (or
terminal) geometry of an underlying quantum mechanics problem. $h$
which was related to number of boxes in dominant Young representation,
is indeed behave like momentum conjugate to $\q$. In other words
momentum $p$, conjugate to $\q$ in UMQM has a meaning in terms of
boxes in Young diagram. The initial data curve (\ref{eq:datacurve}),
we have chosen, is {\it non-characteristic} and hence the solution to
equation (\ref{eq:fermisurfeom}) is unique.

Thus we see that for any large $N$ phase of a generic UMM
(\ref{eq:UMM}) it is possible to construct an effective UMQM such that
phase space distribution of matrix quantum mechanics matches with that
of UMM on a constant time slice.  The effective matrix quantum
mechanics has a potential $W(\q)=V_{eff}(\q)$. The Hamiltonian
(\ref{eq:HamF}) evaluated on $t=t_0$ slice boils down to
(\ref{eq:Hamh})
\ben
H_p \  &\xrightarrow{t\rightarrow
  t_0}\ H_h
\een
up to a constant piece.

Unitary matrix model (\ref{eq:UMM}) shows a rich phase structure at
large $N$. The model has many possible phases. As we vary the
parameters of the theory the system undergoes a phase transition. It
would be interesting to understand the phase transition in UMM in
terms of dynamics of effective matrix quantum mechanics. One can allow
the parameters of $W(\q)$ in UMQM to depend on $t$ such that dynamics
of UMQM captures phase transition of UMM. We choose UMQM such that on
a constant time slices $W(\q)$ matches with $V_{eff}(\q)$ of UMM for
different phases.
\begin{figure}[h]
\centering
\includegraphics[width=8cm,height=7cm]{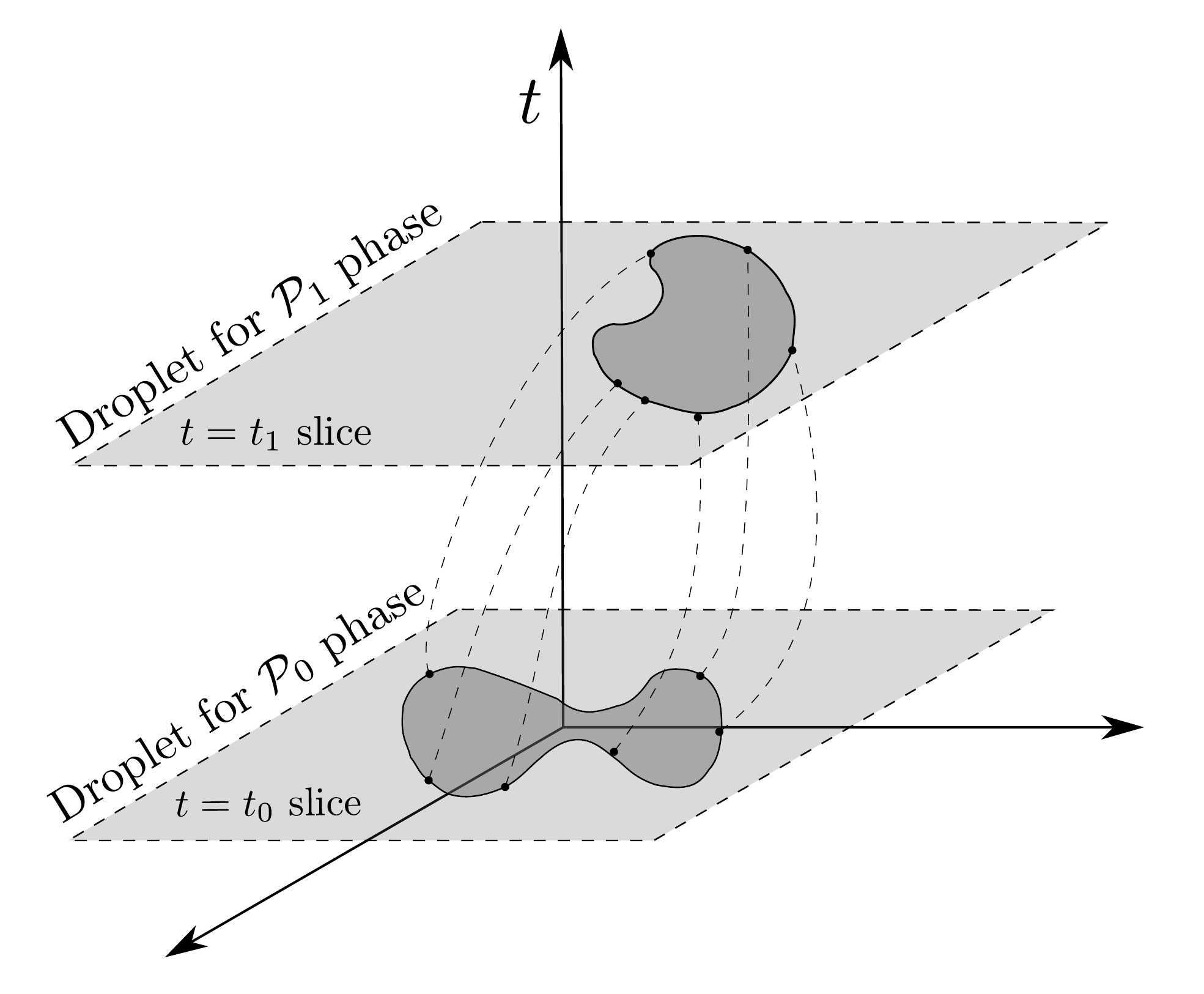}
\caption{Time evolution of phases of UMM}
\label{fig:evolution}
\end{figure}
As depicted in figure \ref{fig:evolution} we can start with a particular phase $\cP_0$ of UMM at $t=t_0$ with
$W(\q(t_0), t_0) = V^{\cP_0}_{eff}(\q)$ and use droplet of $\cP_0$ as
initial data for the Cauchy problem. Then we let the system evolve
with time and some later time $t=t_1>t_0$ we have
$W(\q(t_1), t_1) = V^{\cP_1}_{eff}(\q)$ and shape of fermi surfaces
matches with droplets of $\cP_1$ phase.  This will help us in
understanding the question how dynamics of strongly interacting gauge
theories are encoded in dynamics of free fermions. A work in this
direction was considered in \cite{pallab} for a simpler
model\footnote{Look at \cite{pallabnew} for a recent paper.}, where
they started with a no-gap distribution at $t\ra -\infty$ and showed
that time evolution takes this phase to a one-gap phase at
$t\ra \infty$. However, for a generic model, we need  to thoroughly study the solutions of quasi-linear equations \ref{eq:fermisurfeom}. We leave the issue for future.

For a unitary matrix quantum mechanics, phase space has $\q\ra -\q$
symmetry. Using this fact it is possible to show that phase space area
covered by fermi surface is a constant of motion. Area covered by
fermi surfaces at a time $t$ is given by
\be
\cA(t) = \int_{-\q_0}^{\q_0} \lb p_+(t,\q) -p_-(t,\q)\rb d\q.
\ee
Using equation (\ref{eq:fermisurfeom}) and the fact that
$p_\pm(t,\q)$ and $W(\q)$ are symmetric functions of $\q$ it is easy
to verify that 
\be
\frac d{dt} \cA(t)=0.
\ee
Thus, phase space area is preserved during time evolution. Its only
the shape which changes during evolution.

\subsection{Momentum distribution and Young tableaux}
\label{sec:Youngfromphase}

\begin{figure}[h]
\centering
\includegraphics[width=5cm,height=5cm]{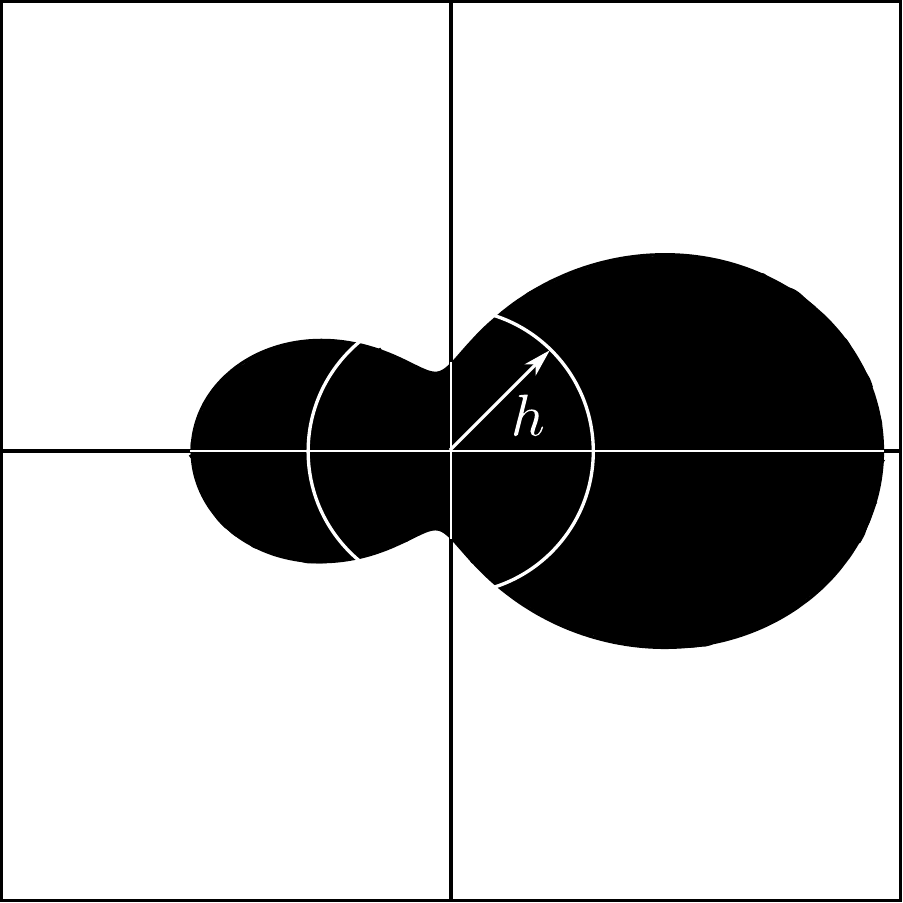}
\caption{$u(h)$ is same the length of white line of radius $h$ for
  no-gap phase.}
\label{fig:thetaintng}
\end{figure}

Since $(h,\q)$ are canonical conjugate of each other, integrating out
$\q$ in phase space gives us momentum distribution for free fermions
under consideration. Thus, momentum distribution is actually encoded in
the arrangements of boxes in the most dominant Young tableaux.  To
obtain Young distribution function $u(h)$ we use the definition
(\ref{eq:udef}). Unlike eigenvalue distribution, finding Young
tableaux distribution directly from partition function is a formidable
task for a generic matrix model. But we assume that the definition
(\ref{eq:udef}) holds in general. This definition boils down to our
earlier identification \cite{duttagopakumar, duttadutta}
$\pi u(h) =\q$ when $\q$ has unique solution for a given $h$ in
equation (\ref{eq:quadraticboundrel}). In general, $\q$ may have
multiple solutions for a given $h$. In that case $u(h)$ is given by
total length of different segments of white arc as shown in figure
\ref{fig:thetaintng}.
This equation is, in general, difficult to solve and find
$u(h)$ for any arbitrary set of parameters $\{\b_n\}$.  In section
\ref{sec:example} and \ref{sec:CSonS3} we use this relation to find
the Young tableaux distribution which dominates the partition function
in large $N$ limit.

Young distribution captures information about momentum distribution of
$N$ fermions. Therefore, large $N$ state of the system can be found
from Young distribution. Since fermions have momentum
$h_1, \cdots , h_N$ a large $N$ state of the system can be recognised
as $|h_1, \cdots, h_N\rangle$ in momentum space\footnote{ We are
  considering a quantum mechanics on circle hence momenta of particles
  are integers.}\cite{douglas2}. Thus a large $N$ state correspond to geometry of
Young diagram. Time evolution of quantum fluctuations about this large
$N$ states can be studied using underlying UMQM.

Finding dominant Young distributions is also important, as local
dynamics of holographic geometry is encoded in Young diagram of dual
theory \cite{koch}.  Supersymmetry plays an important role here, as
weak coupling results are protected. Hence, by determining phase space
distribution and corresponding Young diagrams for a supersymmetry
theory one can reconstruct the dual holographic geometry from that.

\section{Young distribution for $\beta_1 - \b_2$
  model}\label{sec:example}

In this section we discuss an example where we have non-trivial, stable two
gap phases. This example will help us to understand how to
reconstruct Young diagrams from phase space for different large $N$
phases of the theory.

To have a two gap solution we need the potential $V(\theta)$ in equation
(\ref{eq:MMoutential}) to have two minima between $-\pi$ and
$\pi$. This is achieved by considering a plaquette model with
$\beta_1, \beta_2 \neq 0$ and $\b_n=0 \ \text{for} \ n\geq 3$. The
action $S(U)$ is given by
\ben \label{eq:b1b2acn}
{S(U)} & =& N \lB \beta_1 (\Tr U+\Tr
U^{\dagger})+{\beta_2\over 2} (\Tr U^2+\Tr U^{\dagger 2})\rB.  
\een
Phase structure of this model was discussed in \cite{zalewski} and
\cite{Mandal:1989ry}. The model exhibits "five'' different phases
depending on values of parameters $\b_1$ and $\b_2$: one no-gap phase,
two one-gap phases (mirror reflection of each other) and two two-gap
phases. A detailed derivation of eigenvalue distribution for different
phases can be found in \cite{zalewski, Mandal:1989ry}. Here, we 
present a different and independent method (solving Dyson-Schwinger
equation) to reproduce the results (eigenvalue distribution) of \cite{zalewski,Mandal:1989ry}. 
Then we obtain droplets for different phases of the theory. Finally, we
discuss how to reconstruct Young tableaux distributions for five
different phases of this model from the geometry of droplets.

\subsection{The resolvent}\label{sec:resolvent}

Eigenvalue distribution can be obtained from analytic properties of
resolvent which satisfies an algebraic equation in large $N$ limit
\cite{Friedan:1980tu}. This method is quite powerful as one does not
need to solve an integral equation for eigenvalue density. We briefly
review the method here.

``Resolvent" for any generic UMM (\ref{eq:UMM}) is defined as,
\begin{equation}\label{eq:Rzdef}
 R(z)=N^{-1}\langle \Tr[(1-zU)^{-1}]\rangle
\end{equation}
where $z$ is a complex variable.  Expanding the right hand side of
(\ref{eq:Rzdef})
\begin{equation}
 \label{rzexpan}
 R(z) = 1 + \frac{1}{N} \left( z \langle \Tr U\rangle + z^2 \langle
   \Tr U^2\rangle + z^3 \langle \Tr U^3\rangle +\cdots \right). 
 \end{equation}
 One can see that the resolvent is a generating functional of Wilson
 loops in the planar limit.  Since $N^{-1}\langle \Tr U^k\rangle$ lies
 between $-1$ and $+1$ the right hand side of equation (\ref{rzexpan})
 is convergent for $|z|<1$. Therefore, $R(z)$ is an analytic and
 holomorphic function in the interior of the unit disk. In a similar
 way, one can show that the function $R(z)$ is also analytic for
 $|z|>1$ with $R(z)\rightarrow 0$ for $|z|\rightarrow
 \infty$. However, $R(z)$ is discontinuous on $|z|=1$. Inside the unit
 circle one can also write that,
\begin{eqnarray}
R(1/z)=-zN^{-1}\langle Tr[U^{-1}]\rangle-z^{2}N^{-1}\langle
  Tr[U^{-2}]\rangle+\cdots,\quad |z|<1. 
\end{eqnarray}
Hence we have the following property of $R(z)$
\begin{equation}
 R(z)+R(1/z)=1,\quad \text{for}\quad|z|\neq1.
\end{equation}
At large N the resolvent satisfies a quadratic equation
(Dyson-Schwinger equation) which can be solved algebraically
\cite{duttadutta}. The solution, for a particular model
(\ref{eq:b1b2acn}) under consideration, is given by (see
\cite{duttadutta} for details)
\begin{eqnarray}\label{eq:Rzb1b2}
  R(z)={1\over 2}\left[{1\over 2} +\beta_1(z-{1\over z})
  +\beta_2(z^2-{1\over z^2}) +\sqrt{F(z)}\right]\quad\text{$|z|<1$}\\
  R(z)={1\over 2}\left[{1\over 2} +\beta_1(z-{1\over z}) +\beta_2
  (z^2-{1\over z^2})-\sqrt{F(z)}\right]\quad\text{$|z|>1$} 
\end{eqnarray}
where $F(z)$ has the following form,
\be
\begin{split}\label{eq:Fz-a1a2-twocut}
F(z)
&=\bigg[\beta_{1}\lb\frac{1}{z}+z\rb+\beta_{2}\lb\frac{1}{z^{2}}+z^{2}\rb
+1\bigg]^{2} -4\lb\beta_{1}z+\beta_{2}z^{2}\rb \lB\frac{\beta_{1}}{z}
+\frac{\beta_{2}}{z^{2}}\rB \\
& \hspace{3.6cm}+4\beta_{1}R'(0)+4\beta_{2}\lB\frac{R'(0)}{z}
+\frac{R''(0)}{2}+R'(0)z\rB  .
\end{split} 
\ee
Spectral density or eigenvalue density is related to discontinuity of
$R(z)$ on unit circle. Eigenvalue distribution function ($\r(\q)$) is
given by,
\begin{eqnarray}\label{eq:evdef}
2\pi \r(\theta) = \lim_{\epsilon\to 0}
  \left[R((1+\epsilon)e^{i\theta})-R((1-\epsilon)e^{i\theta})\right] =
  2 \Re\lB R(e^{i\q})\rB-1=\sqrt{F(e^{i\theta})}.
\end{eqnarray}
We see that $F(z)$ contains some quantities $R'(0)$ and $R''(0)$ which
can be calculated from (\ref{eq:Rzb1b2}). Therefore one has to
consistently solve these equations to find $R'(0)$ and $R''(0)$ for
different phases of the model and then $\r(\q)$ from
(\ref{eq:Fz-a1a2-twocut}). However, it is difficult to solve equation
(\ref{eq:Rzb1b2}) and (\ref{eq:Fz-a1a2-twocut}) consistently. Rather,
we shall take an ansatz for $F(z)$ for different phases of the theory
and fix the form of $F(z)$ from analytic properties of $R(z)$ :
\begin{enumerate}

\item 

  ${R}(z)$ is analytic inside the unit circle. In equation
  (\ref{eq:Rzb1b2}) we see that $R(z)$ has $1/z$ and $1/z^2$ terms
  which are singular at $z=0$. Therefore, $\sqrt{F(z)}$ should have
  poles of order one and two at $z=0$ to make $R(z)$ analytic
  inside. $F(z)$ can not have zeros inside (or outside) the unit
  circle. In that case $\sqrt{F(z)}$ would have branch points
  inside or outside the circle. However, $F(z)$ can have zeros on
  $|z|=1$ line.

\item 

  For gapped solution $\sqrt{F(z)}$ should have branch points on unit
  circle. Also $F(z)$ is invariant under $z\to 1/z$.

\item 

  Finally, $R(z)$ is normalized as $R_<(0)=1$.

\end{enumerate}

\subsection{Large $N$ phases and eigenvalue distribution}

\subsubsection*{No-gap solution}

This is the most trivial case where there is no branch cuts in
resolvent $R(z)$. Hence, $F(z)$ is a perfect square. Thus we have,
\begin{equation}
F(z)=\lB 1+\beta_1\lb z+{1\over z}\rb+\beta_2 \lb z^2+{1\over z^2}\rb \rB^2.
\end{equation}
The eigenvalue distribution for this phase is given by (from equation
\ref{eq:evdef})
\begin{equation}\label{eq:rhoA0}
\rho_{A_0}(\theta)={1\over 2\pi}(1+2\beta_1\cos\theta+2\beta_2\cos2\theta)
\end{equation}
for
$0\leq \q\leq2\pi$.

Since $\r(\q)\geq0$, a no-gap or zero-gap solution exists for \cite{zalewski}
\ben
\begin{split}
  \b_1^2 -4\b_2(1-2\b_2) &=0, \quad \text{for}\quad \b_2\geq \frac16\\
  |\b_1| -\b_2 -\frac12 &= 0, \quad \text{for}\quad -\frac12 \leq \b_2
  \leq \frac16
\end{split}
\een
We use the notation of \cite{zalewski} to name different phases of
this model. No-gap phase is called $A_0$ phase. Note that there is a
factor of $2$ difference between our set of parameters and
\cite{zalewski}'s set of parameters :
$(2\b_1,\ 2\b_2)_{\text{our}}=(\b_1,\
\b_2)_{\text{\cite{zalewski}}}$.

\subsubsection*{One-gap solution}

For one-gap solution $\sqrt{F(z)}$ has a branch cut on
$|z|=1$. Therefore the most generic ansatz for $F(z)$ for one-gap
phase is given by,
\begin{eqnarray}\label{eq:fzonegapb1b2}
  \sqrt{F(z)}=\alpha  \left(\sqrt{z}+\frac{1}{\sqrt{z}}\right)
  \sqrt{\gamma + \left(\sqrt{z}+\frac{1}{\sqrt{z}}\right)^2}
  \left[\beta+\left(\sqrt{z}+\frac{1}{\sqrt{z}}\right)^2\right],
\end{eqnarray}
where $\a,\ \b,\ \g \ \text{are constants}$.  From analyticity and
normalization condition of $R(z)$ we find
\begin{eqnarray}
  \alpha=\beta_2;\quad\beta = {\beta_1\over\beta_2}-{\gamma\over2} -
  4;\quad\gamma = {2\over
  3\beta}\left(\beta_1-4\beta_2+\sqrt{-6\beta_2+(\beta_1+2\beta_2)^2}\right). 
\end{eqnarray}
Hence, the spectral density for one-gap solution ($A_1$ phase) is
given by
\begin{eqnarray}\label{eq:rhoa1phase}
  \rho_{A_1}(\theta) = {2\over \pi}\sqrt{\sin^2{\q_1\over 2} - \sin^2{\theta\over 2}}
  \cos{\theta\over 2}\left[\beta_1+\beta_2(2\cos\theta+\cos\q_1-1)\right], 
\end{eqnarray}
with
\begin{equation}
\sin^2{\q_1\over 2} = {\beta_1+2\beta_2-\sqrt{-6\beta_2 +
    (\beta_1+2\beta_2)^2}\over 6 \beta_2}. 
\end{equation}
Eigenvalues are symmetrically distributed about $\q=0$ from $-\q_1$ to
$\q_1$. One-gap solution is valid in the following region in
$(\b_1,\b_2)$ plane bounded by the curves \cite{zalewski}
\ben
\begin{split}
 (\b_1+2\b_2)^2-6\b_2 &= 0,\quad \text{for} \quad \b_2 \geq\frac16,\\
 \b_1-\b_2-\frac12 &=0, \quad \text{for} \quad -\frac12 \leq \b_2 \leq\frac16,\\
  (\b_1+2\b_2)^2+2\b_2 &= 0,\quad \text{for} \quad \b_2 \leq -\frac12
\end{split}
\een
and extends in the direction of positive $\b_1$. There exists a common
boundary between this phase and zero-gap phase and is given by
\be\label{eq:A0-A1bound1}
\b_1=\b_2+1, \quad \for \quad -\frac12 \leq \b_2 \leq \frac16.
\ee

$F(z)$ (given by equation \ref{eq:fzonegapb1b2}) for one-gap solution
has a branch cut on unit circle about $z=1$ point ($\q=0$).  If we
replace $z$ by $-z$ then the branch-cut changes its position from
right to left side on the unit circle. However, this is also a valid
ansatz for $F(z)$ for one-gap solution ($B_1$ phase). For this ansatz
the parameters $\a, \b, \g$ take the following values to make $R(z)$
analytic inside.
\begin{eqnarray}
  \alpha=\beta_2;\quad\beta = - {\beta_1\over\beta_2} - {\gamma\over2} 
  - 4; \quad \gamma = {2\over 3\beta} \left(-\beta_1-4\beta_2 +
  \sqrt{-6\beta_2 + (-\beta_1+2\beta_2)^2}\right),
\end{eqnarray}
which in turn gives the spectral density as
\begin{eqnarray}\label{b1phase}
  \rho_{B_1}(\theta) = {2\over \pi} \sqrt{\cos^2{\q_1 \over 2} -
  \cos^2{\theta \over 2}} \, 
  \sin{\theta \over 2}\, \left[\beta_1 - \beta_2 (2\cos
  \theta+\cos\q_1+1)\right], 
\end{eqnarray}
with 
\begin{equation}
\cos^2{\q\over 2} = {-\beta_1+2
  \beta_2-\sqrt{-6\beta_2+(-\beta_1+2\beta_2)^2}\over 6 \beta_2}. 
\end{equation}

Since action (\ref{eq:b1b2acn}) is invariant under $\b_1\ra-\b_1$ and
$\q\ra\q+\pi$ therefore, it is expected that there exists a mirror
image of $A_1$ phase in negative $\b_1$ direction ($B_1$
phase). Eigenvalues, for this branch, are distributed symmetrically
about $\q=\pi$ from $\pi+|\q_1|$ to $\pi-|\q_1|$. Similarly, the
common boundary between this phase and zero-gap phase is given by
\be\label{eq:A0-A1bound2}
\b_1 =- \b_2-1, \quad \text{for} \quad -\frac12 \leq \b_2 \leq \frac16.
\ee
A transition along line (\ref{eq:A0-A1bound1}) or
(\ref{eq:A0-A1bound2}) between zero-gap and one-gap (or its mirror
phase) is a third order phase transition. This is a generalization of
Gross-Witten-Wadia transition \cite{gross-witten, wadia}. In
subsection \ref{sec:b1-b2-fermi} we shall see how phase space topology
changes as we go from one phase to another.

\subsubsection*{Two-gap solution} \label{sec:b1-b2-twogap}

To have a two-gap solution one should take an ansatz for $\sqrt{F(z)}$
such that there is at most a degree four polynomial inside the square
root. In this case there are two possible ways of taking ansatz for
$F(z)$ which give two different two-gap phases namely $A_2$ phase and
$B_2$ phase.

\subsubsection*{$A_2$ Phase}
The simplest ansatz that one can take is
\begin{equation}\label{a2cut}
\sqrt{F(z)} = \sqrt{\alpha^2 \lb z+{1\over z}\rb^2 + \beta \lb z + {1\over z}\rb +
  1} \,\left( \delta + \sigma \lb z + {1\over z}\rb\right).
\end{equation}
$\a,\b,\s,\d$ are constants.  Using the analyticity and normalization
condition of $R(z)$ we find,
\begin{eqnarray}
\sigma = {\beta_2\over \alpha}; \quad \delta = {2\alpha^2 \beta_1 -
  \beta \beta_2\over 2\alpha^{3}}; \quad \alpha^2 =
  \displaystyle{{-\beta\beta_1 - \beta_2 + \sqrt{2\beta\beta_1 \beta_2
  +\beta_2^2+\beta^2\left(\beta_1^2+6\beta_2(-1+2\beta_2)\right)}
  \over-4+8\beta_2}}.\nonumber\\ 
\end{eqnarray}
However, it is not possible to fix all the four parameters from the
properties of $R(z)$. We keep $\b$ undetermined. The eigenvalue
distribution in this case is given by,
 \begin{eqnarray}
   \rho_{A_2}(\theta) = {1\over\pi} \sqrt{(\cos\theta-\cos\theta_1)
   (\cos\theta -\cos\theta_2)}\,\,\,\,\left[\beta_1+\beta_2
   (2\cos\theta +\cos\theta_1+\cos\theta_2)\right]\nonumber\\
 \end{eqnarray}
where
\begin{eqnarray}\label{eq:rangeA2}
\cos\theta_1 = -{\beta\over 4\alpha}-{1\over4\alpha}
  \sqrt{-4\alpha+\beta^2};\quad\cos\theta_2=-{\beta\over
  4\alpha}+{1\over4\alpha}\sqrt{-4\alpha+\beta^2}. 
\end{eqnarray}
$A_2$ phase has eigenvalue distribution symmetric about horizontal
axis. Validity of this phase has been shown in figure
\ref{fig:b1b2droplets}. Eigenvalues are distributed between $-\q_1$
and $\q_1$ and then $\pi-\q_2$ to $\pi+\q_2$.

Eliminating the undetermined parameter $\b$ from equations
(\ref{eq:rangeA2}) one can write a constraint equation relating
$\cos\theta_1$ and $\cos\theta_2$ as
\begin{eqnarray}
-\beta_1(\cos\theta_1+\cos\theta_2) + \beta_2\left[2
  -(\cos\theta_1+\cos\theta_2)^2-{1\over2}
  (\cos\theta_1-\cos\theta_2)^2\right]=1.
\end{eqnarray}
The undetermined parameter $\b$ can be fixed by imposing an extra
condition that total number of eigenvalues between $\q_1$ and $\q_2$
is zero
\be 
\frac1\pi \int_{\q_1}^{\q_2} \r_{A_2}(\q) d\q = 0.  
\ee
A second order phase transition between $A_2$ phase and $A_1$ phase
(or $B_1$ phase) occurs along the lines \cite{zalewski}
\ben
\begin{split}
\b_2 &= \frac1{4\cos^3\q_1}\lB {\cos\frac12\q_1 \over \sin^2\q_1} -\ln
\lb  \frac{1+\cos\frac12 \q_1}{\sin\q_1}\rb \rB, \\
\b_1 &= \frac1{2\sin^2\frac{\q_p}{2}}\lB 1-2 \b_2 \sin^2\frac12 \q_1
(2-3 \sin^2 \frac12 \q_1) \rB.
\end{split}
\een

\subsubsection*{$B_2$ Phase}

The second possible ansatz for $F(z)$ for a two gap solution is given
by,
\begin{equation}\label{b2cut}
\sqrt{F(z)}=\gamma \sqrt{\lb z+\frac1z \rb^2 -4}
\,\,\,\,\sqrt{\alpha^2\lb z+{1\over z}\rb^2+\beta \lb z+{1\over z}\rb
  +1}.  
\end{equation}
with,
\begin{eqnarray}
\alpha^2 = {\beta_2^2\over\beta_1^2 + 2\beta_2};
  \quad\beta={2\beta_1\beta_2\over\beta_1^2+2\beta_2};
  \quad\gamma=-2\,i\,\sqrt{\beta_1^2+2\beta_2}. 
\end{eqnarray}
The spectral density for this case is
\begin{eqnarray}
\rho(\theta) = - {2\beta_2\over\pi}
  |\sin\theta|\sqrt{(\cos\theta_1-\cos\theta)
  (\cos\theta-\cos\theta_2)}
\end{eqnarray}
with
\begin{eqnarray}
\cos\theta_1 = - {\beta_1\over 2\beta_2}+\sqrt{-1\over 2\beta_2};
  \quad\cos\theta_2=-{\beta_1\over 2\beta_2} -\sqrt{-1\over 2\beta_2}.
\end{eqnarray}
Eigenvalues are distributed between $\q_1$ and $\q_2$, and $-\q_{1}$
and $-\q_{2}$.  This phase exists for $\b_2\leq -\frac13$. A parabola
$(-\b_1+2\b_2)^2 +4\b_2=0$ separates this phase from one-gap
phase. The point $(\b_1=0,\b_2=-\frac12)$ is a triple point in the
parameter space. See figure (\ref{fig:b1b2droplets}) for details.

\subsection{Droplets and phase transition in $\b_1-\b_2$
  model}\label{sec:b1-b2-fermi}

We shall discuss about geometries of droplets for different phases
of the model. The boundary relation (\ref{eq:boundrelplaq}) for this
model is given by
\be
h_\pm(\q)=\frac12 + \b_1 \cos{\q} + \b_2 \cos 2\q \pm \pi \r(\q)
\ee
where $h_\pm(\q)$ are solutions of 
\be\label{eq:b1b2boundrel}
h^2 - \lb 1+ \b_1 \cos{\q} + \b_2 \cos 2\q \rb h =\pi^2 \r^2(\q).  
\ee

\begin{figure}[h]
\centering
\includegraphics[width=10cm,height=9cm]{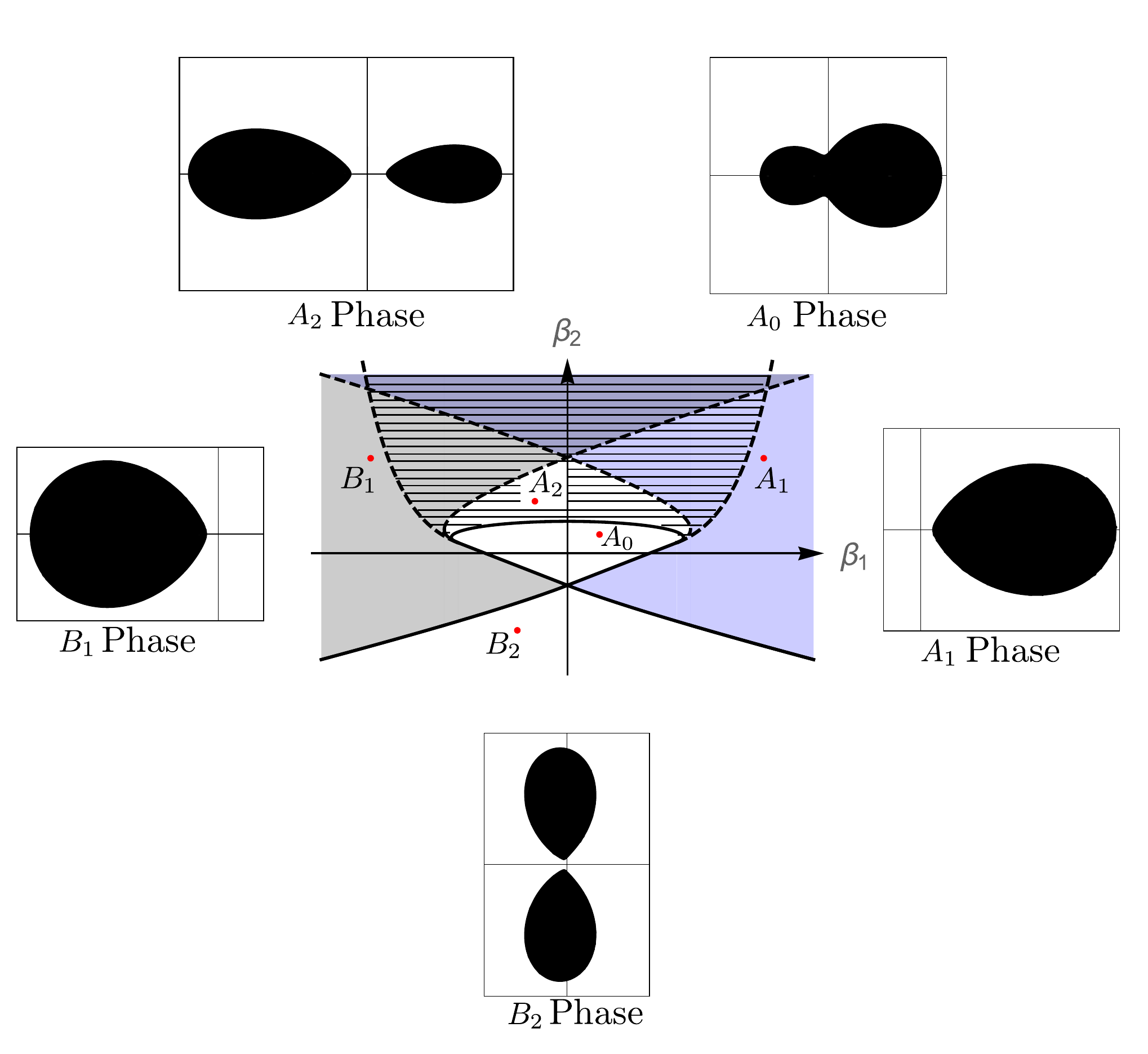}
\caption{Droplets for $\b_1-\b_2$ model. The central figure shows
  parameter space of the model. All the distributions are plotted in
  polar coordinates $(h,\q)$, $h$ radial and $\q$ angular. The black
  regions correspond to density $\o(h,\q)=1$. Figure on top right shows
  droplet for no gap phase. Here we see that distribution is non-zero
  for all values of $\q$. This implies eigenvalue density is non-zero
  for all values of $\q$ and hence the name no-gap (in eigenvalues)
  phase. Distribution on right middle corresponds to one-gap phase
  $A_1$. Unlike no-gap phase, here distribution is non-zero for $\q$
  between $-\q_1$ and $\q_1$. Distribution at the bottom correspond to
  $B_2$ phase. Here, distribution is non-zero for $\q$ between two
  different ranges. Distribution on left middle is a mirror image of
  $A_1$ phase. This is $B_1$ phase. Finally, on top left we draw
  distribution for the second two-gap phase ($A_2$). Here also phase
  space density is non-zero for two different ranges of $\q$. One
  should note that all the distributions are symmetric about $\q=0$
  line. This is because of $U\ra U^{\dagger}$ symmetry of the model.}
\label{fig:b1b2droplets}
\end{figure}
There are five possible droplets with different topologies
corresponding to five different phases. In figure
\ref{fig:b1b2droplets} we draw phase space distributions for different
phases. The diagram in the middle is parameter space ($\b_1-\b_2$
plane) of the theory under consideration. We divide this parameter space
into different parts according to validity of different phases. See
\cite{zalewski} for more detail\footnote{The parameter space in
  \ref{fig:b1b2droplets} is similar to the parameter space given in
  \cite{zalewski}. However, it has been re-drawn according to our
  normalization.}.  Since, $h>0$ (being number of boxes) and
$-\pi <\q \leq \pi$, we plot phase space distribution in polar
coordinates. The black regions correspond to density $\o(h,\q)=1$. We
note that topologies of droplets are different for different
phases. Origin ($h=0$) is inside the distribution for no-gap phase
(since $h_-(\q)$ is zero for all $\q$). For one-gap and two-gap
phases, on the other hand, origin is outside the distribution (both
$h_+$ and $h_-$ are non-zero). We consider this as a topological
property of droplets. One should also note that all the distributions
are symmetric about real axis ($\q=0$ line). This is because of UMM
has $U\ra U^{\dagger}$ symmetry. While deforming the droplets (by
changing values of parameters) this symmetry is always
preserved. Therefore, one can not continuously deform the droplets of
$B_2$ phase to bring them in the shape of droplets of $A_2$
phase. This implies, there is no transition possible between $B_2$ and
$A_2$ phases, which is true for this model. Boundary of a droplet
crossing the origin implies (according to our definition) change of
topology, which can only happen at the time of phase transition.
\begin{figure}[h]
\centering
\includegraphics[width=10cm,height=9cm]{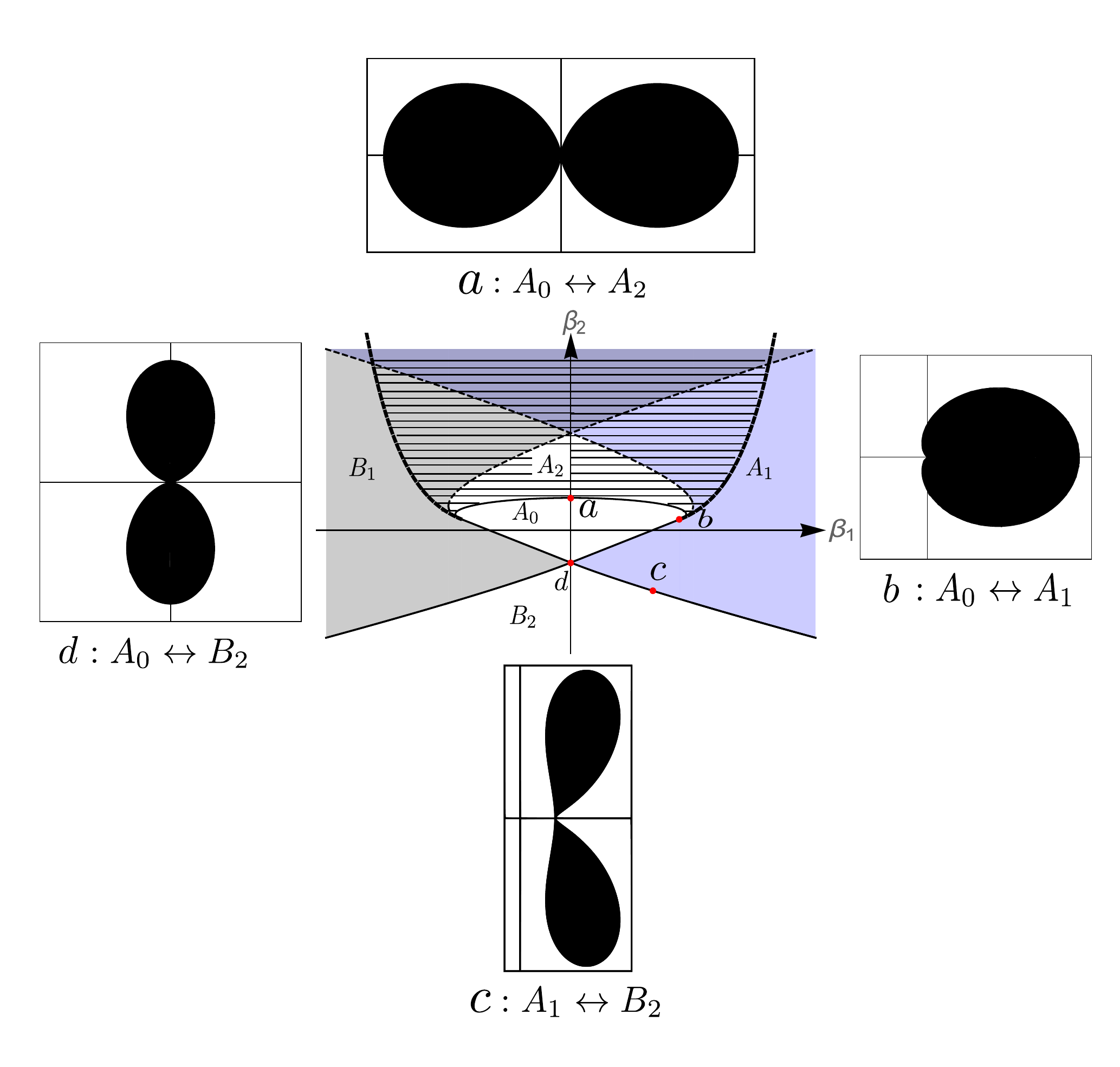}
\caption{Droplets for $\b_1-\b_2$ model at different phase transition
  points.}
\label{fig:b1b2trans}
\end{figure}
We see that at $A_0$-$A_1$ transition point (generalized
Gross-Witten-Wadia transition), left boundary of the droplet touches
the origin (figure \ref{fig:b1b2trans}(b)). From this point as we move
towards $A_0$, the origin goes inside and as we move towards $A_1$ the
origin goes outside. Similarly, cutting a single droplet into two (or
gluing two droplets to make a single one) also corresponds to change
of topology and occurs at the time of phase transitions. For example,
starting from $A_1$ as we move towards $B_2$ phase we see that $A_1$
droplet starts deforming (two mid-points come closer) and finally
becomes two blobs touching each other at a point (as shown in figure
\ref{fig:b1b2trans}(c)) at the transition point. After that as we move
in $B_2$ region two blobs separate out from each other. Similar things
happen for other transitions. Figure \ref{fig:b1b2trans}(d),
\ref{fig:b1b2trans}(a) show distributions at $A_0 - B_2$ and $A_0-A_2$
transition respectively.

Since boundary of droplets are fermi surfaces in underlying MQM, the
above diagrams show how shape of fermi surfaces change during phase
transition.

\subsection{Reconstruction of Young diagrams from droplet
  geometry} \label{sec:reconYoungb1b2}

The phase space description is a powerful approach as the shape of
different droplets not only contains information about eigenvalue
distributions but it also contains information about dominant
representations for different phases of the theory. In this section we
discuss how one can construct dominant Young diagrams for different
phases.

To construct Young tableaux distribution from geometry of droplets we
use equation (\ref{eq:udef}). \cite{duttagopakumar, duttadutta},
considered plaquette model with $\b_n=0$ for $n\geq 2$ and observed
that while solving boundary relation (\ref{eq:beta1boundrel}) for a
given $h$ there exists a unique value of $\q$ and hence Young
distribution was given by $\pi u(h)=\q(h)$ for all phases. However,
solution for $\q$ does not remain unique when we turn on $\b_2$ in
plaquette model (\ref{eq:b1b2acn}). Here we see that boundary relation
(\ref{eq:b1b2boundrel}) for a fixed $h$ has maximum two possible
solution for $\q$. Therefore, the simple identification between
eigenvalue and Young density ($\pi u(h)=\q$) does not hold in this
case. However, we consider equation (\ref{eq:udef}) as a fundamental
definition for $u(h)$ and construct Young distributions from phase
space for different phases.

\subsubsection*{No-gap phase}

For no-gap phase $\r(\q)$ is given by (\ref{eq:rhoA0}). Therefore the
Fermi surface for this phase is given by,
\be
h_+(\q) = 1 + 2\b_1\cos\q+ 2\b_2\cos2\q, \quad h_-(\q) =0.
\ee
This defines a droplet about the origin. For $\b_1>|4\b_2|$, $\q$ has
a unique solution (we denote that by $\bar \q_1(h)$). In that case
$u(h)$ is given by
\be
u(h) = \frac{\bar\q_1(h)}{2\pi}, \quad \for \quad \b_1>|4\b_2|
\ee
as before.

\begin{figure}[h]
\begin{subfigure}{.5\textwidth}
\centering
\includegraphics[width=7cm,height=7cm]{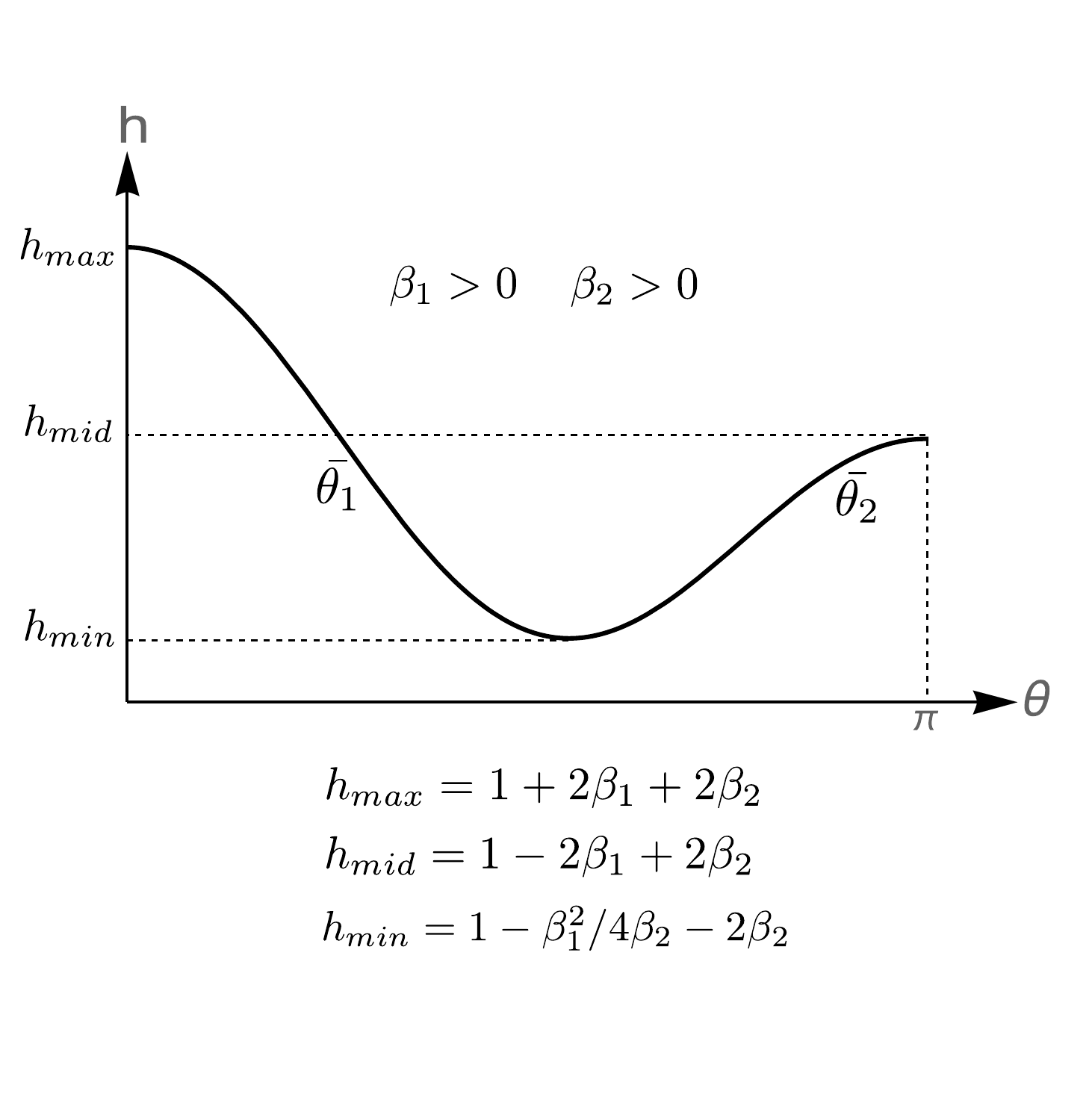}
\caption{$\b_2 >0$ }
\end{subfigure}%
\begin{subfigure}{.5\textwidth}
\centering
\includegraphics[width=7cm,height=7cm]{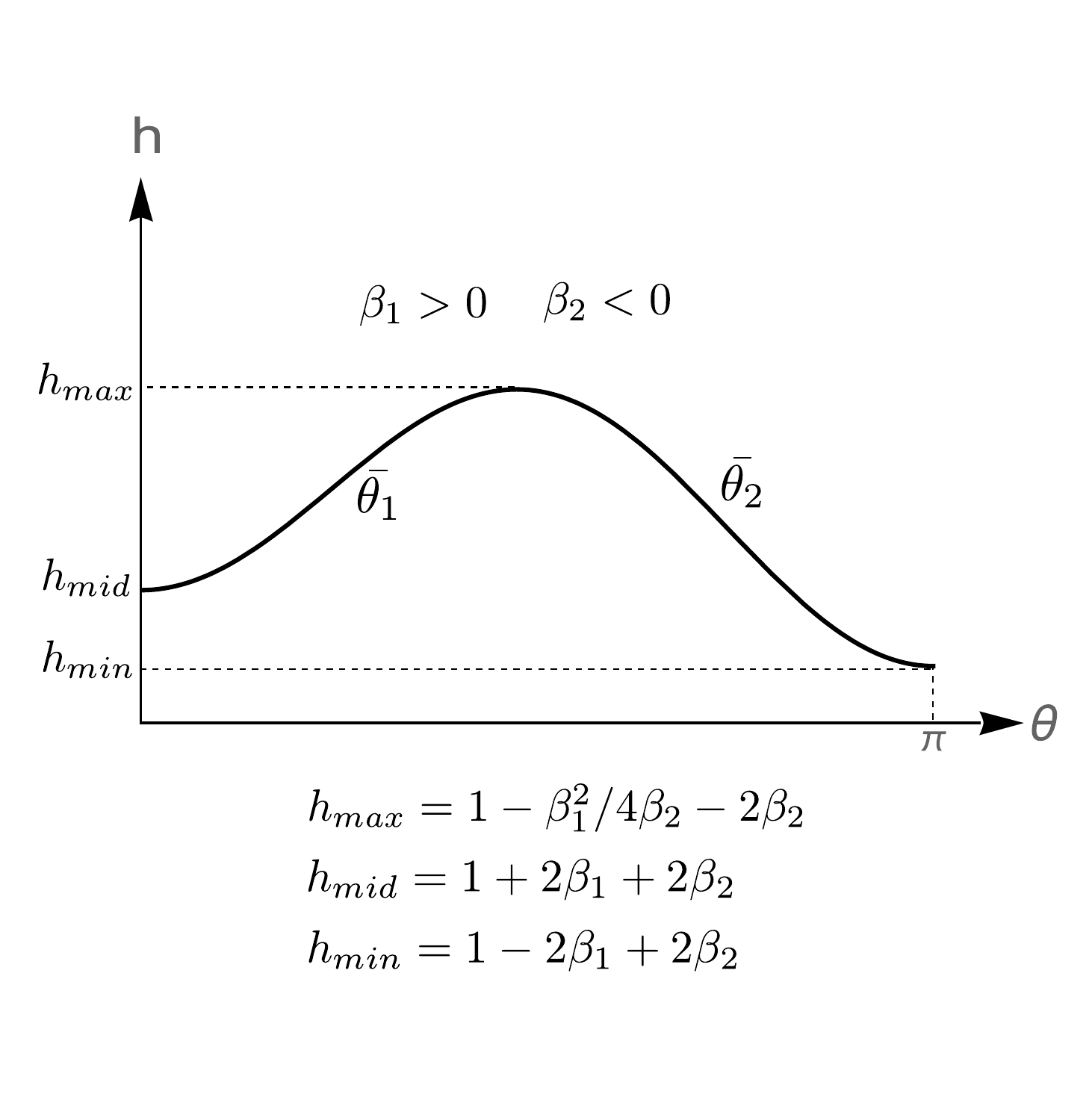}
\caption{$\b_2 <0$ }
\end{subfigure}
\caption{$h$ vs $\q$ for no-gap phase for $\b_1<4|\b_2|$. The black
  line correspond to $h_+(\q)$.}
\label{fig:hthetanogap}
\end{figure}

\begin{figure}[h]
\begin{subfigure}{.5\textwidth}
\centering
\includegraphics[width=6cm,height=5cm]{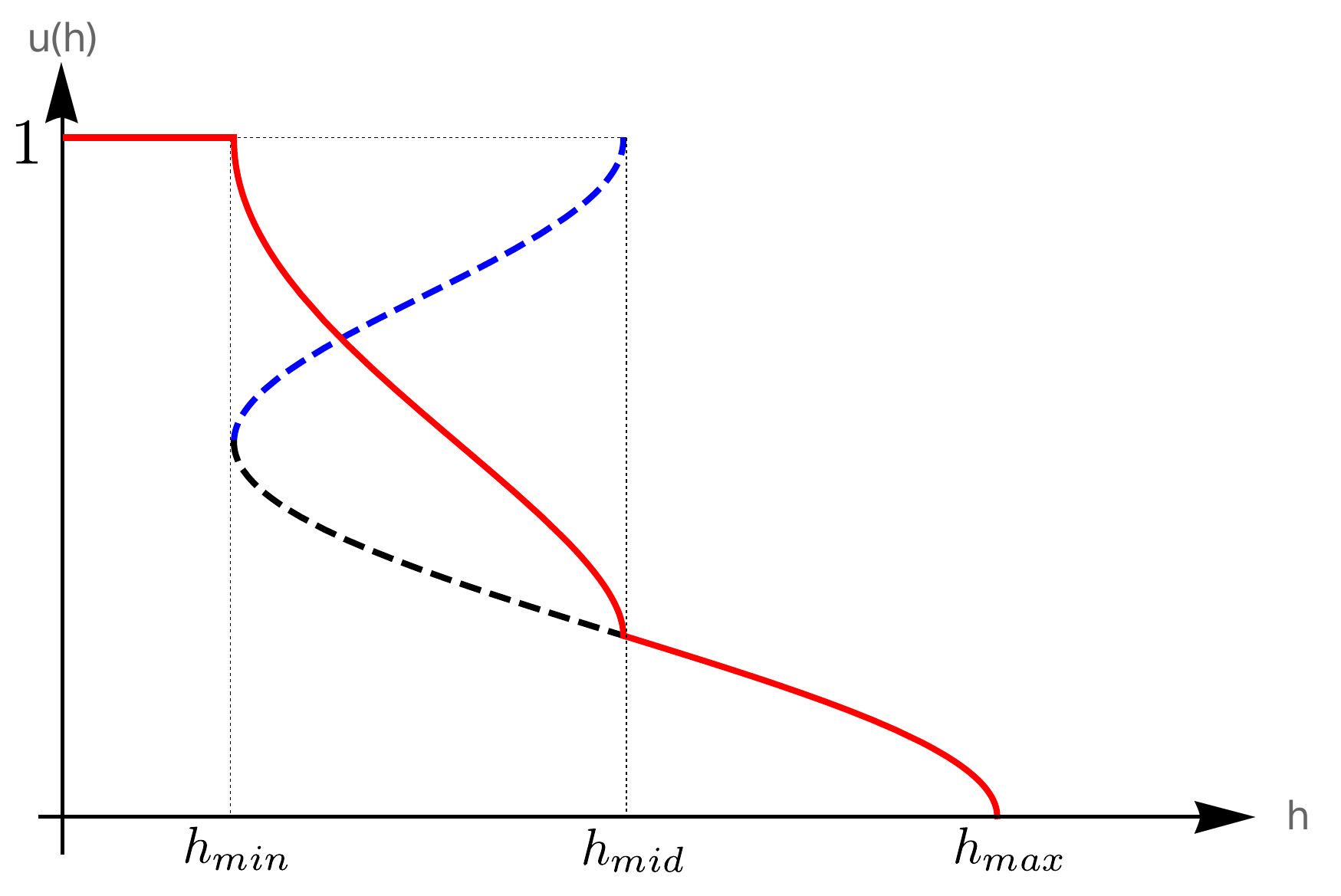}
\caption{$\b_2 >0$}
\end{subfigure}
\begin{subfigure}{.5\textwidth}
\centering
\includegraphics[width=6cm,height=5cm]{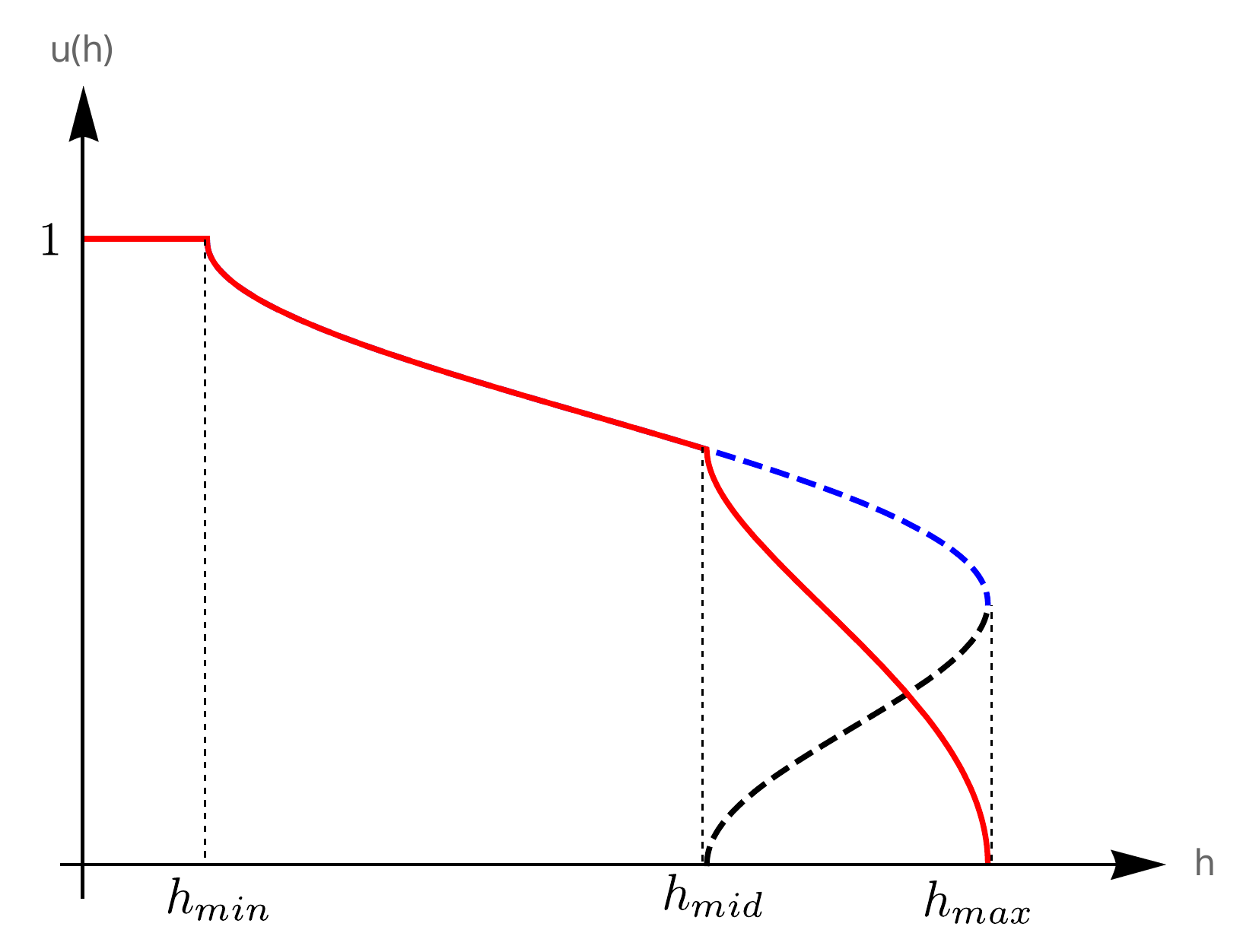}
\caption{$\b_2 <0$}
\end{subfigure}%
\caption{Young tableaux distribution for no-gap phase of $\b_1-\b_2$
  model. Blue and black dotted lines are $\bar \q_2$ and $\bar\q_1$
  respectively. Red line depicts $u(h)$.}
\label{fig:uhnog}
\end{figure}

For $\b_1 < |4\b_2|$, $\q$ has multiple solutions depending on $h$.
In figure \ref{fig:hthetanogap} we plot $h$ vs $\q$ for no-gap phase
and for $\b_1 < |4\b_2|$. The left and right plots correspond to
positive and negative $\b_2$ respectively.
Following definition (\ref{eq:udef}),
Young distribution for positive $\beta_2$ is given by
\ben\label{eq:uhnogap1}
\begin{split}
  u(h) & = {\bar \q_1(h)\over \pi} \hspace{1.5cm} \for \quad h_{mid}
  \leq  h \leq h_{max} \\
  &= 1 - \frac{\bar \q_2- \bar \q_1}\pi \hspace{.55cm} \for \quad
  h_{min}\leq h \leq h_{mid} .
\end{split}
\een
For negative $\b_2$ the distribution is given by,
\ben\label{eq:uhnogap2}
\begin{split}
  u(h) & = {\bar \q_2(h)-\bar \q_1(h)\over \pi} \hspace{.5cm} \for \quad h_{mid}
  \leq  h \leq h_{max} \\
  &= \frac{\bar \q_2(h)}\pi \hspace{1.8cm} \for \quad
  h_{min}\leq h \leq h_{mid} .
\end{split}
\een
These definitions geometrically imply that $u(h)$ is given by the
length of the white arc of radius $h$ in figure \ref{fig:thetaintng}
as discussed in section (\ref{sec:Youngfromphase}).  Young
distribution $u(h)$ as a function of $h$ is shown in figure
\ref{fig:uhnog} for both positive and negative $\b_2$ and
$\b_1<4|\b_2|$.
We see that, there is a kink in the distribution. This correspond to
sudden change in distribution function for boxes in a diagram. As in
\cite{duttagopakumar}, Young tableaux for no-gap phase has finite
number of rows empty. $n(x)=0$ for $1 < x < 1 - h_{min}$. After that
number of boxes starts increasing and reaches a maximum value
$n(0) = h_{max}-1$. The distribution of boxes from $x=0$ to
$x=1-h_{min}$ is governed by $u(h)$ given by equation
(\ref{eq:uhnogap1} or \ref{eq:uhnogap2}). Here we see a kink at
$h=h_{mid}$. For $0< h < h_{min}$, $u(h)=1$.
\begin{figure}[h]
\centering
\includegraphics[width=5cm]{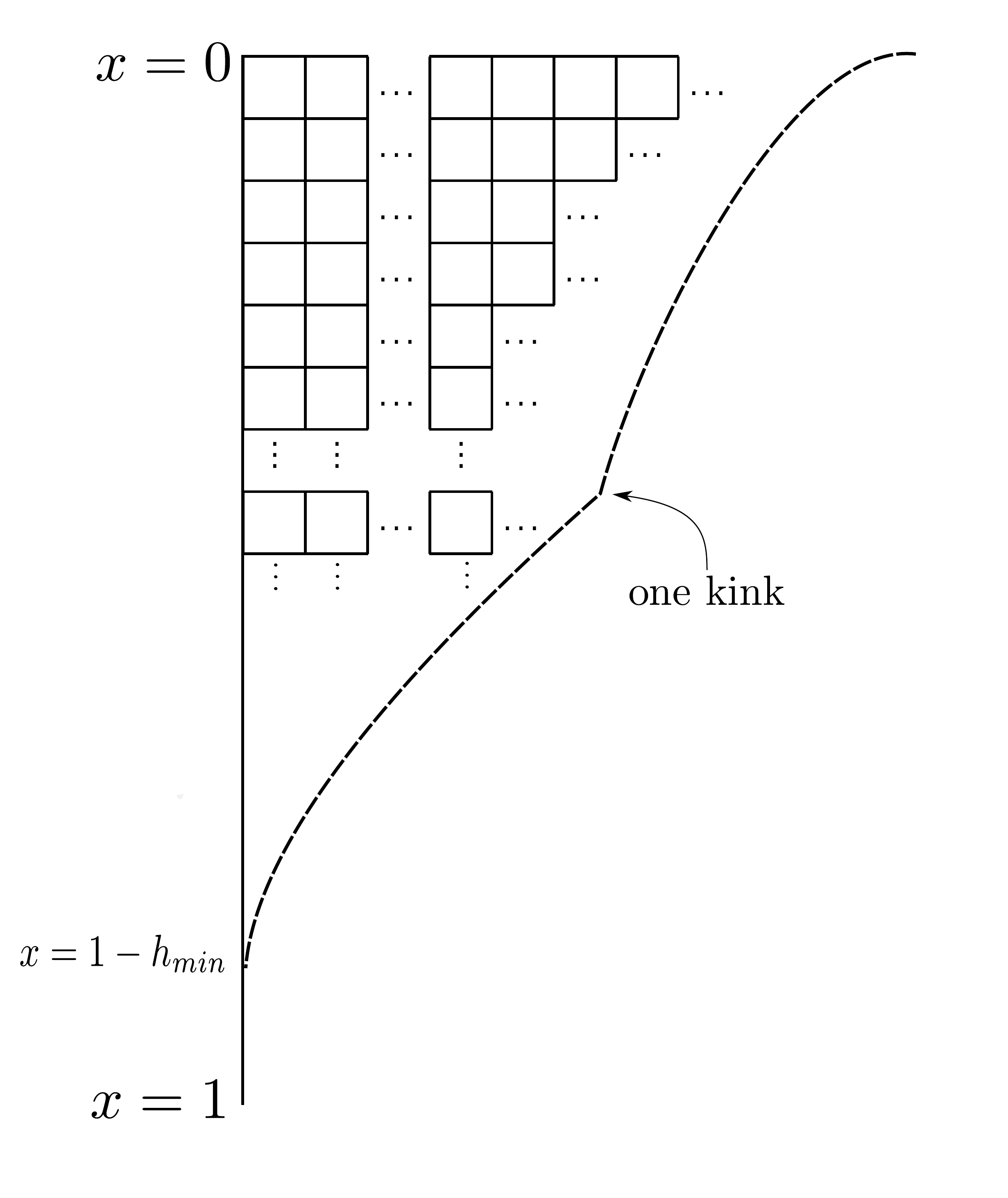}
\caption{A typical Young tableaux for no-gap phase. Dotted line
  correspond to edge of the distribution of boxes in large $N$
  limit. A kink corresponds to $u'(h)$ is discontinuous.}
\label{fig:YTnogap}
\end{figure}
A typical Young diagram for no-gap phase has been shown in figure
\ref{fig:YTnogap}.  For negative values of $\b_1$ the story remains
the same.

\subsubsection*{One-gap phase}

\begin{figure}[h]
\begin{subfigure}{.5\textwidth}
\centering
\includegraphics[width=7cm,height=5cm]{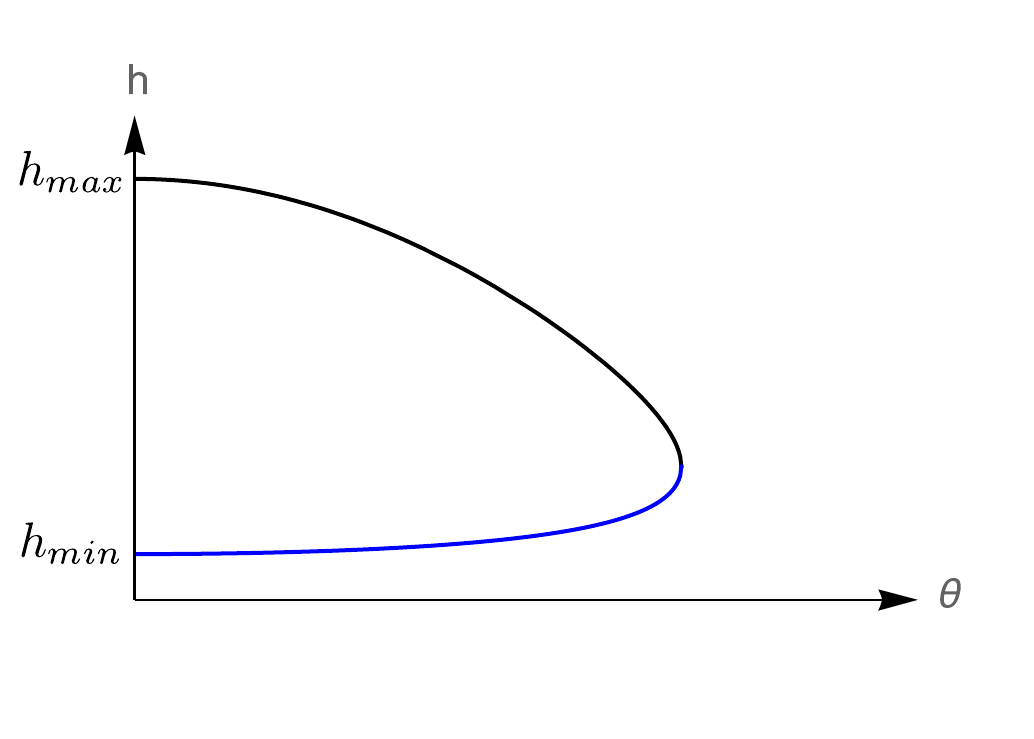}
\caption{$\b_1, \b_2 >0$ or $\b_2<0$, $\b_1>4|\b_2|$.}
\label{fig:hthetaA1a}
\end{subfigure}%
\begin{subfigure}{.5\textwidth}
\centering
\includegraphics[width=7cm,height=5cm]{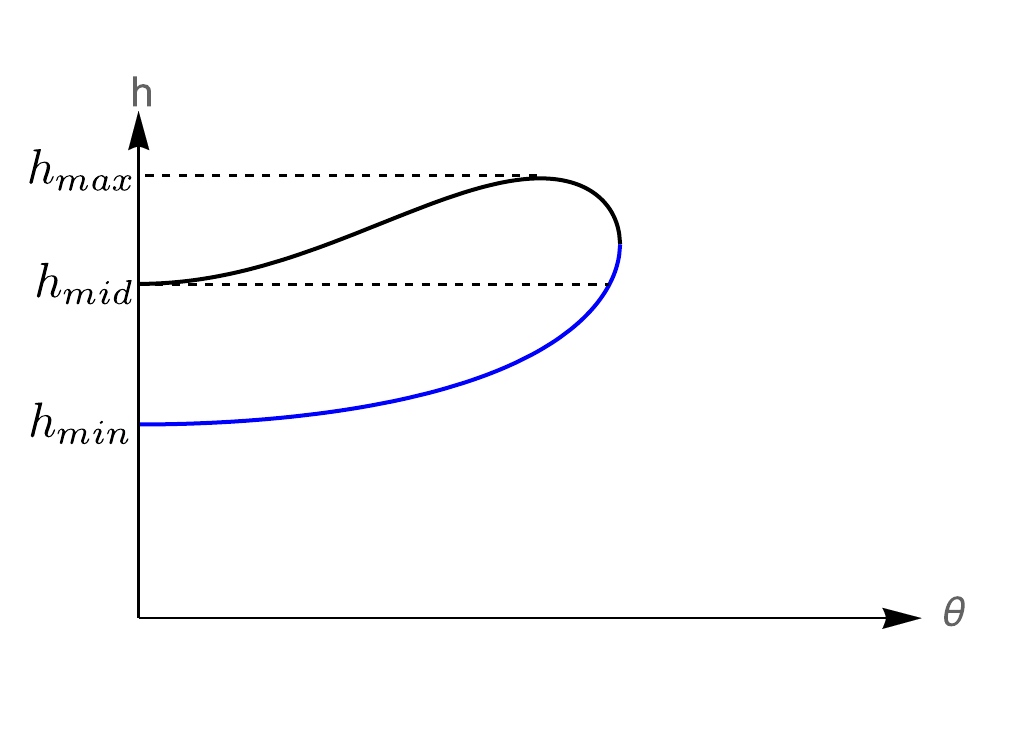}
\caption{$\b_2 <0 $, $\b_1<4|\b_2|$.}
\label{fig:hthetaA1b}
\end{subfigure}
\caption{Young tableaux distribution for one-gap phase of $\b_1-\b_2$
  model. Blue and black lines are $h_-(\q)$ and $h_+(\q)$ lines
  respectively.}
\label{fig:hthetaA1}
\end{figure}

\begin{figure}[h]
\begin{subfigure}{.5\textwidth}
\centering
\includegraphics[width=6cm,height=4cm]{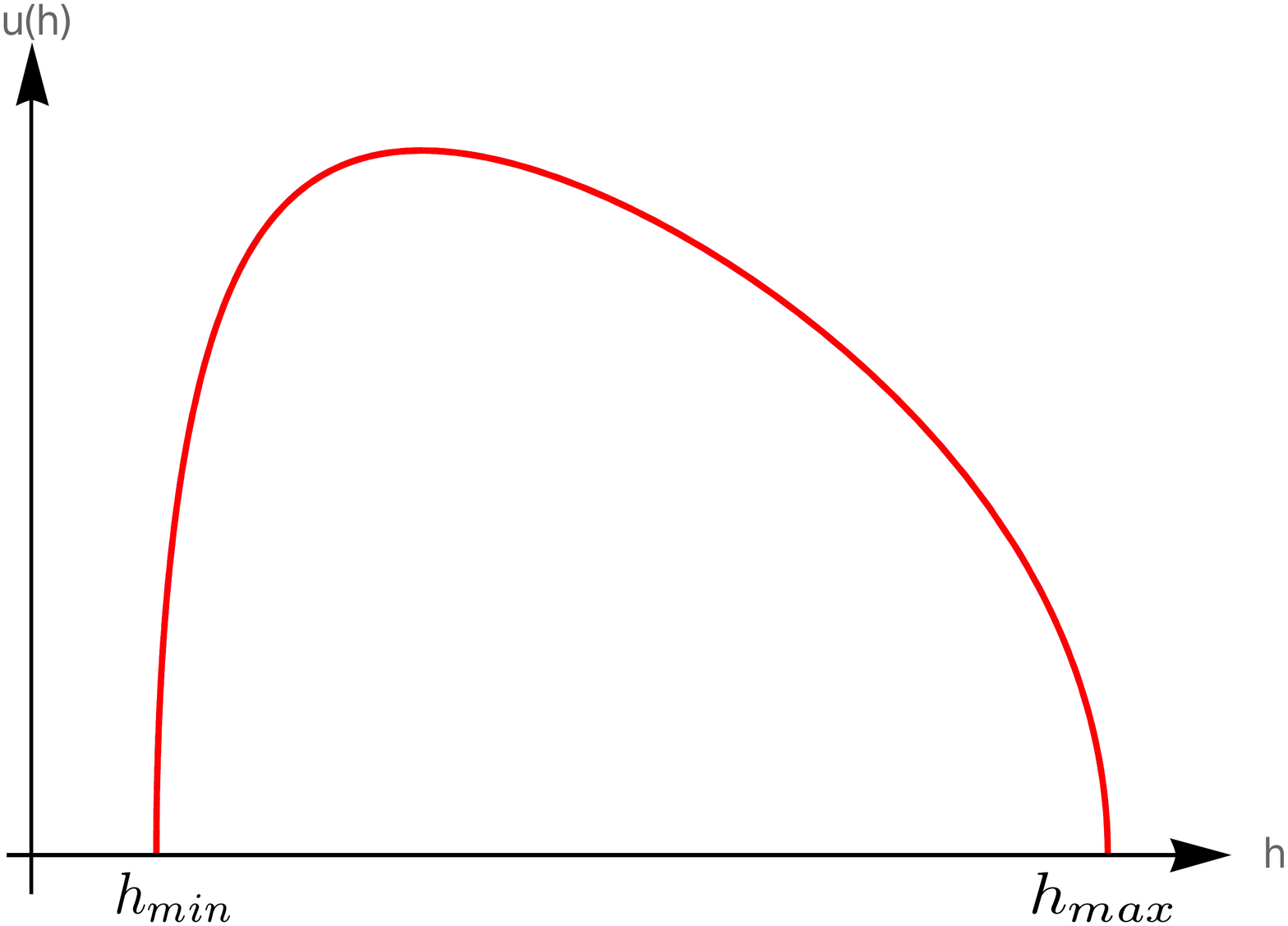}
\caption{$\b_1, \b_2 >0$ or $\b_2<0$, $\b_1>4|\b_2|$.}
\label{fig:uhoga}
\end{subfigure}%
\begin{subfigure}{.5\textwidth}
\centering
\includegraphics[width=6cm,height=4cm]{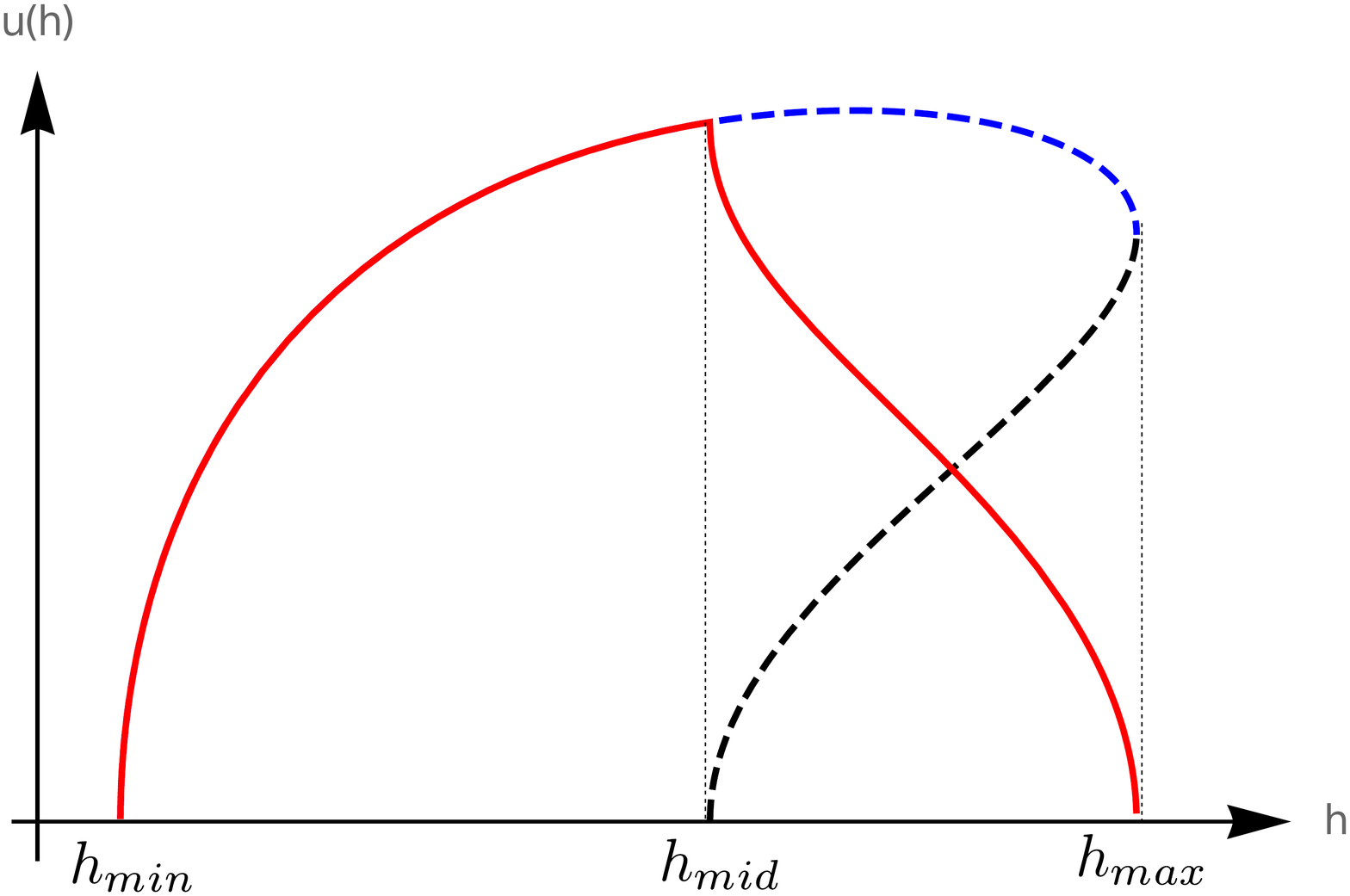}
\caption{$\b_2 <0 $, $\b_1<4|\b_2|$.}
\label{fig:uhogb}
\end{subfigure}
\caption{Young tableaux distribution for one-gap phase of $\b_1-\b_2$
  model. Blue and black dotted lines are $\bar \q_2$ and $\bar\q_1$
  respectively. Red line depicts $u(h)$.}
\label{fig:uhog}
\end{figure}

\begin{figure}[h]
\begin{subfigure}{.5\textwidth}
\centering
\includegraphics[width=5cm,height=6cm]{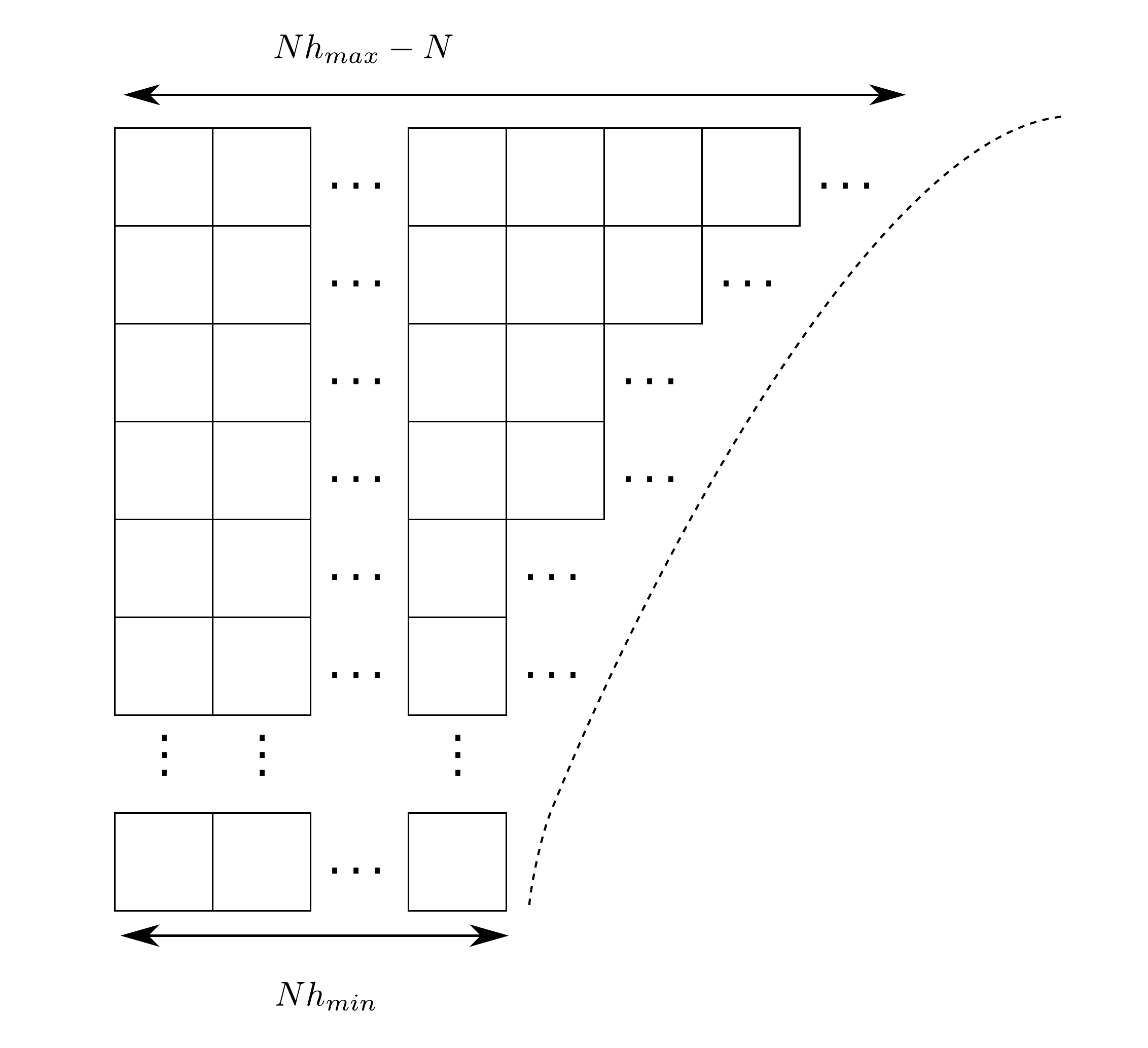}
\caption{$\b_1, \b_2 >0$ or $\b_1>4|\b_2|$ for $\b_2<0$. The dotted
  black line denotes a smooth distribution of boxes. $u'(h)$ is
  continuous everywhere.}
\label{fig: YTonegap a}
\end{subfigure}%
\hspace{.6cm}
\begin{subfigure}{.5\textwidth}
\centering
\includegraphics[width=5cm,height=6cm]{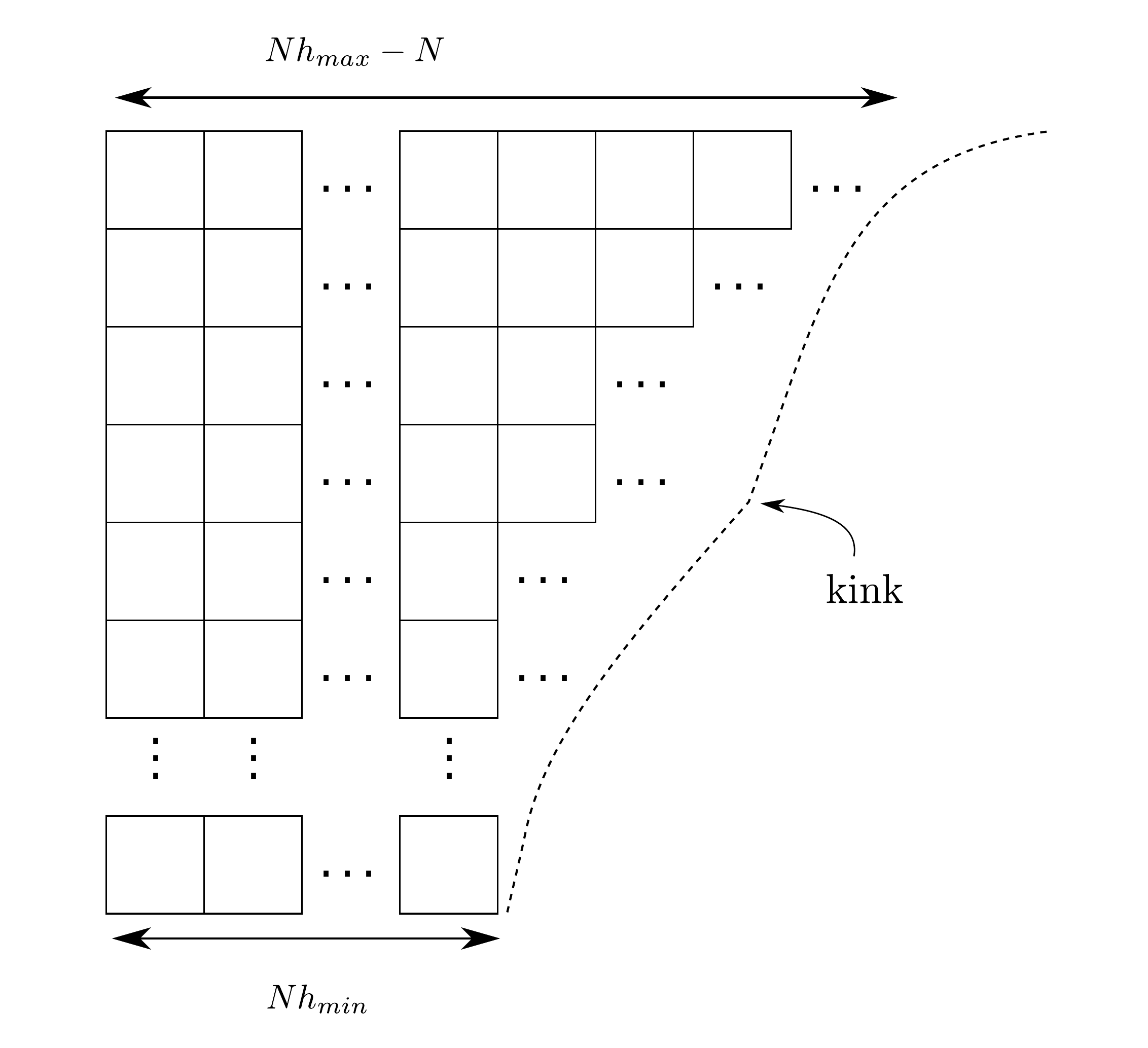}
\caption{$\b_2 <0 $, $\b_1<4|\b_2|$. Here we see a kink in the
  distribution of boxes. $u'(h)$ is discontinuous.}
\label{fig: YTonegap b}
\end{subfigure}
\caption{Typical Young diagram for one-gap phase of $\b_1-\b_2$
  model.}
\label{fig:YTonegap}
\end{figure}

Eigenvalue distribution for one-gap phase is given by equation
(\ref{eq:rhoa1phase}). Boundary of phase space droplet is determined
by equation (\ref{eq:b1b2boundrel}) with $\rho$ replaced by
$\r_{A_1}$. In figure \ref{fig:hthetaA1} we plot $h$ vs. $\q$ for
one-gap phase. 
Here we see that for $\b_1, \b_2 >0$ or $\b_1>4|\b_2|$ for $\b_2<0$
there is only one solution to $\q$ for a given $h$ (figure
\ref{fig:hthetaA1a}). For $\b_2<0$ two solutions of $\q$ develop for
$\b_1<4|\b_2|$ (figure \ref{fig:hthetaA1b}). Corresponding Young
distributions are plotted in figure \ref{fig:uhog}.
Young diagrams have all the rows filled with $\cO(N)$ number of
boxes. The last row and the first row have $N h_{min}$ and
$N(h_{max}-1)$ number of boxes respectively. In the first case
(figure \ref{fig:uhoga}), number of boxes smoothly increases from the
last row to first row. In second case (figure \ref{fig:uhogb}), we see
that there is a kink in the Young diagram, figure \ref{fig:YTonegap}.
A characteristic difference between Young diagrams for no-gap phase
and one-gap phase is that for no-gap phase a finite number of rows are
empty where as for one-gap phase all the rows are filled. At the phase
transition point ($A_0-A_1$ line) one can check that $h_{min}$ goes to
zero in either side.

\subsubsection*{Two-gap phase}

There are two possible two-gap phases : $B_2$ and $A_2$. \\

\begin{figure}[h]
\begin{subfigure}{.5\textwidth}
\centering
\includegraphics[width=6cm,height=4cm]{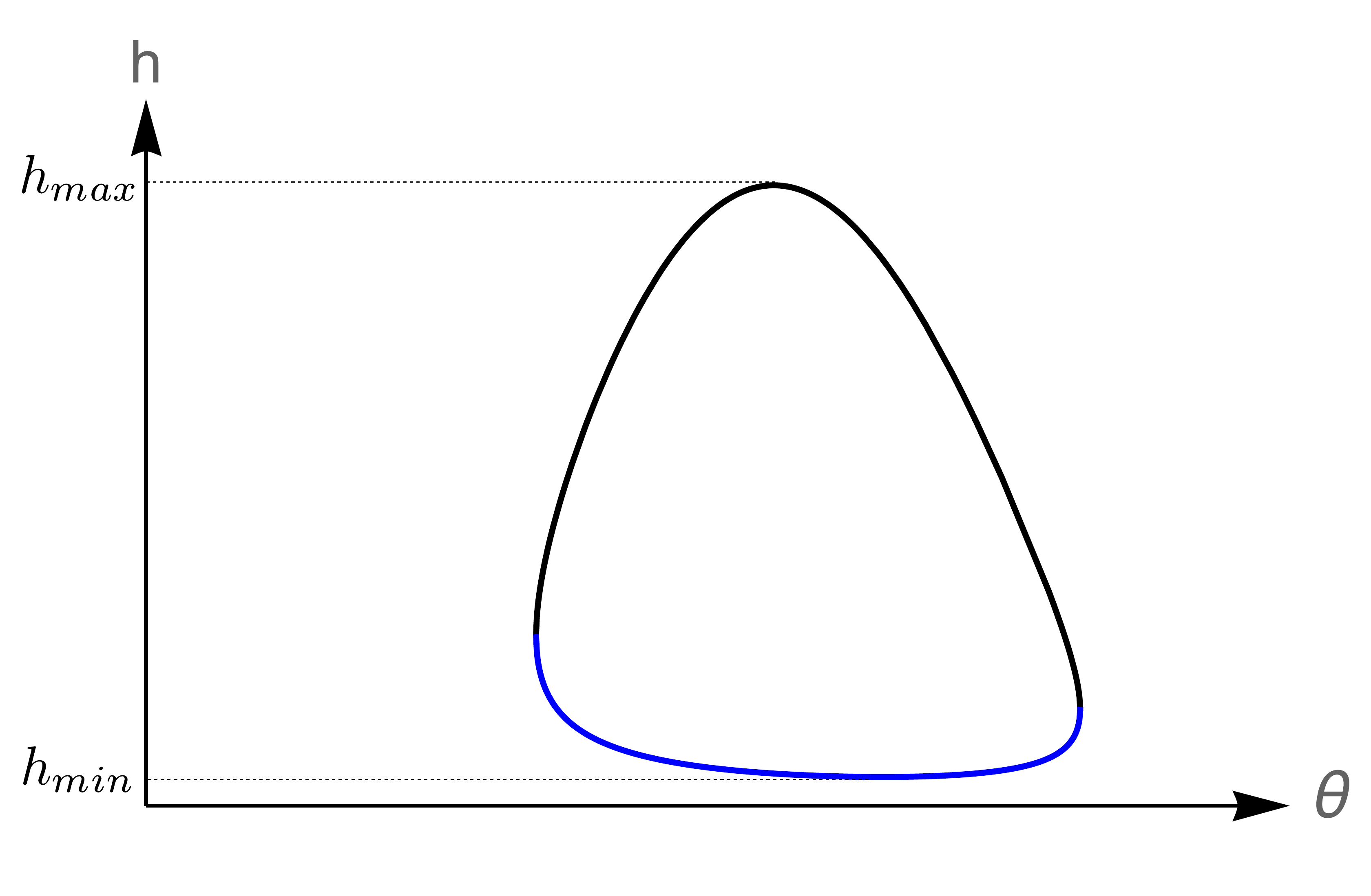}
\caption{$h$ vs. $\q$ for $B_2$ phase. Blue and black lines are
  $h_-(\q)$ and $h_+(\q)$ lines respectively.}
\end{subfigure}%
\hspace{.5cm}
\begin{subfigure}{.5\textwidth}
\centering
\includegraphics[width=6cm,height=4cm]{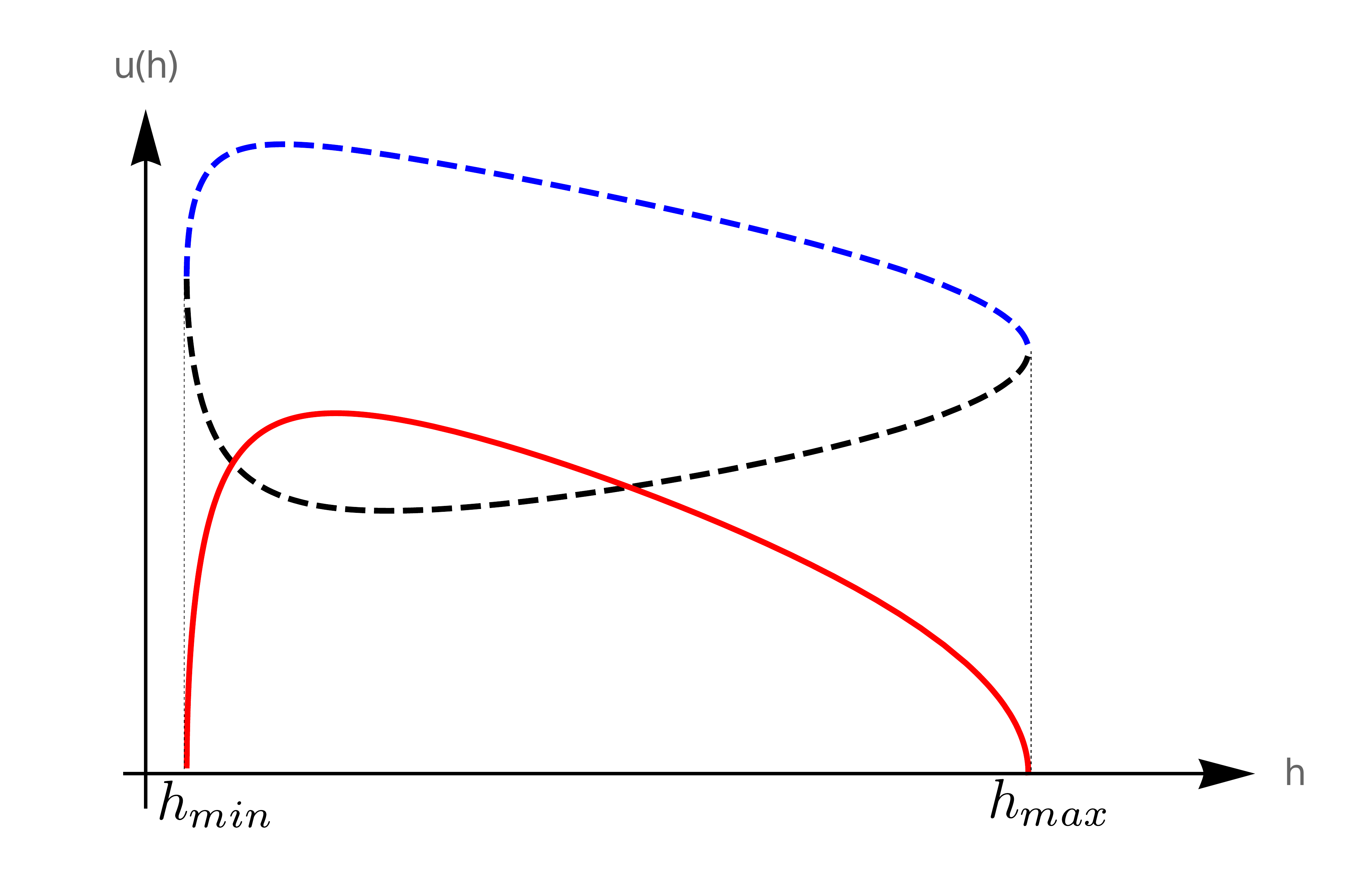}
\caption{$u(h)$ vs. $h$ for $B_2$ phase. Dotted blue and black lines
  are two possible solutions of $\q$ for a given $h$ and the red line
  denotes corresponding $u(h)$.}
\end{subfigure}
\caption{$h$ vs. $\q$ and corresponding $u(h)$ vs. $h$ for $B_2$
  phase.}
\label{fig:B_2phase}
\end{figure}

\noindent {\underline{\bf $B_2$ phase :}}\ \ For $B_2$ phase $h$ vs $\q$ and
corresponding $u(h)$ vs. $h$ are given in figure \ref{fig:B_2phase}.
A typical Young tableaux for $B_2$ phase is similar to that of $A_1$
phase without any kink.\\

\begin{figure}[h]
\begin{subfigure}{.5\textwidth}
\centering
\includegraphics[width=7cm,height=5cm]{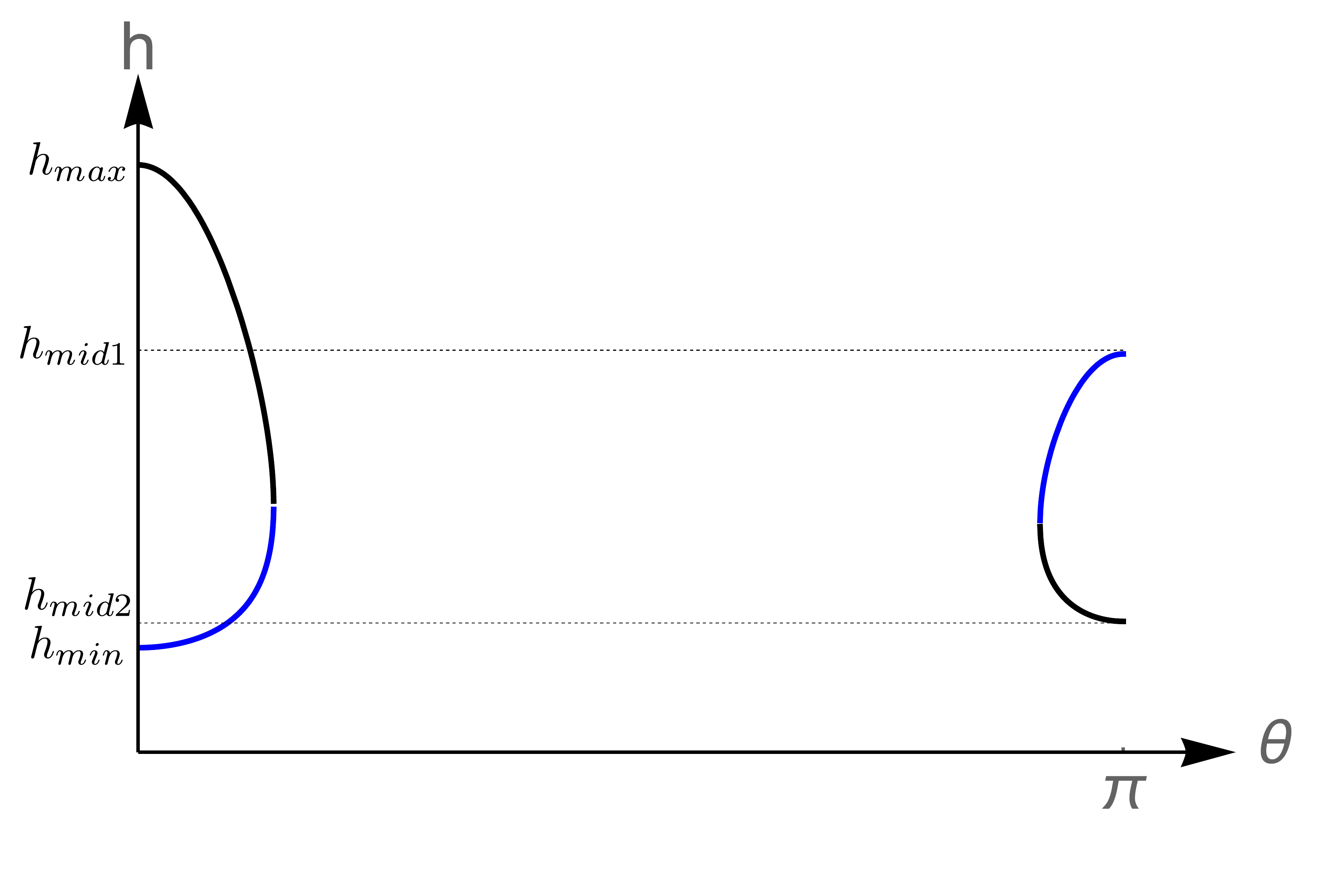}
\caption{$h$ vs. $\q$ for $A_2$ phase. Blue and black lines are
  $h_-(\q)$ and $h_+(\q)$ lines respectively.}
\end{subfigure}%
\hspace{.5cm}
\begin{subfigure}{.5\textwidth}
\centering
\includegraphics[width=6cm,height=4.3cm]{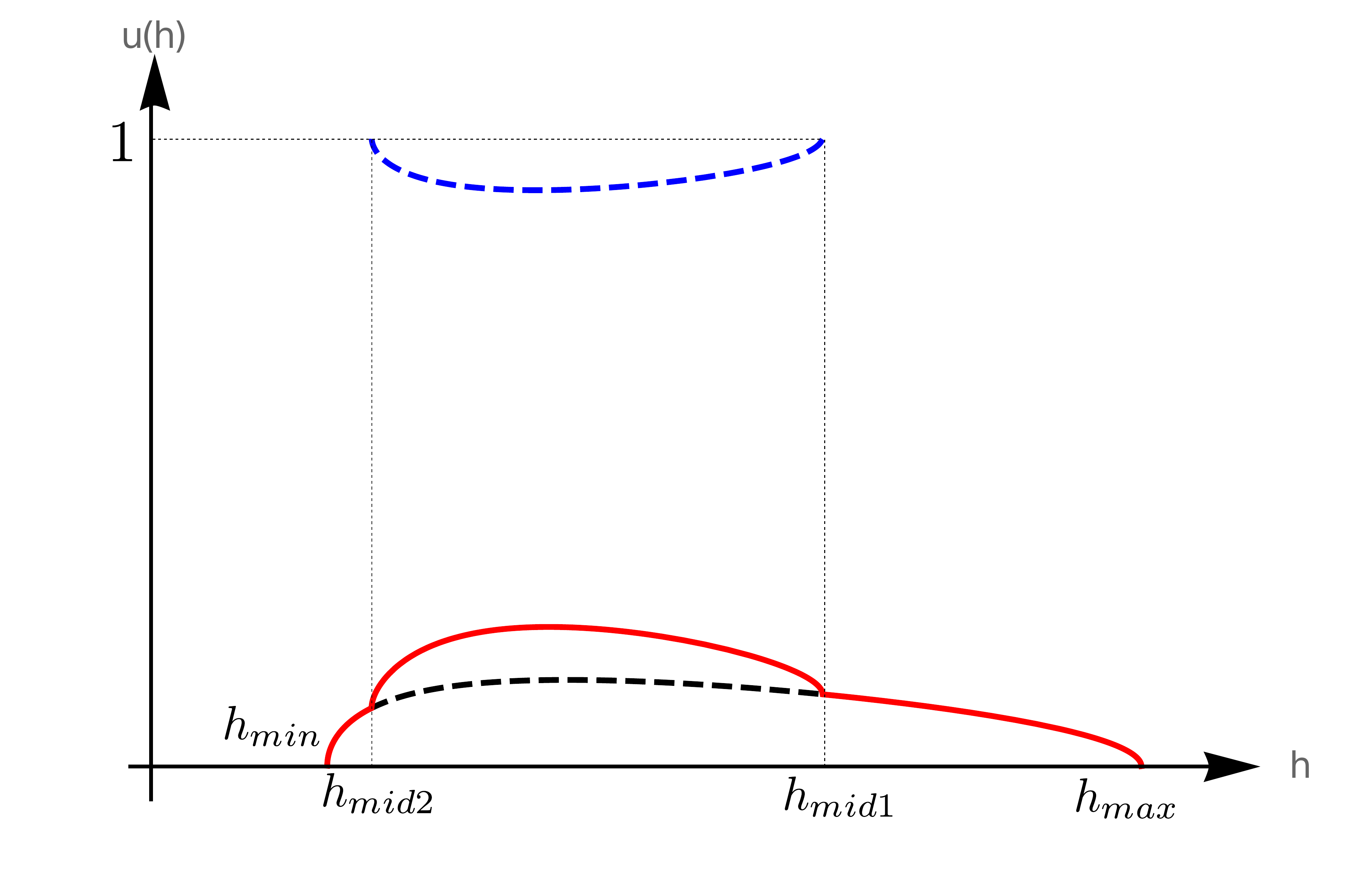}
\caption{$u(h)$ vs. $h$ for $A_2$ phase. Dotted blue and black lines
  are two possible solutions of $\q$ for a given $h$ and the red line
  denotes corresponding $u(h)$.}
\end{subfigure}
\caption{$h$ vs. $\q$ and corresponding $u(h)$ vs. $h$ for $A_2$
  phase.}
\label{fig:A_2phase}
\end{figure}

\begin{figure}[h]
\centering
\includegraphics[width=6cm,height=6cm]{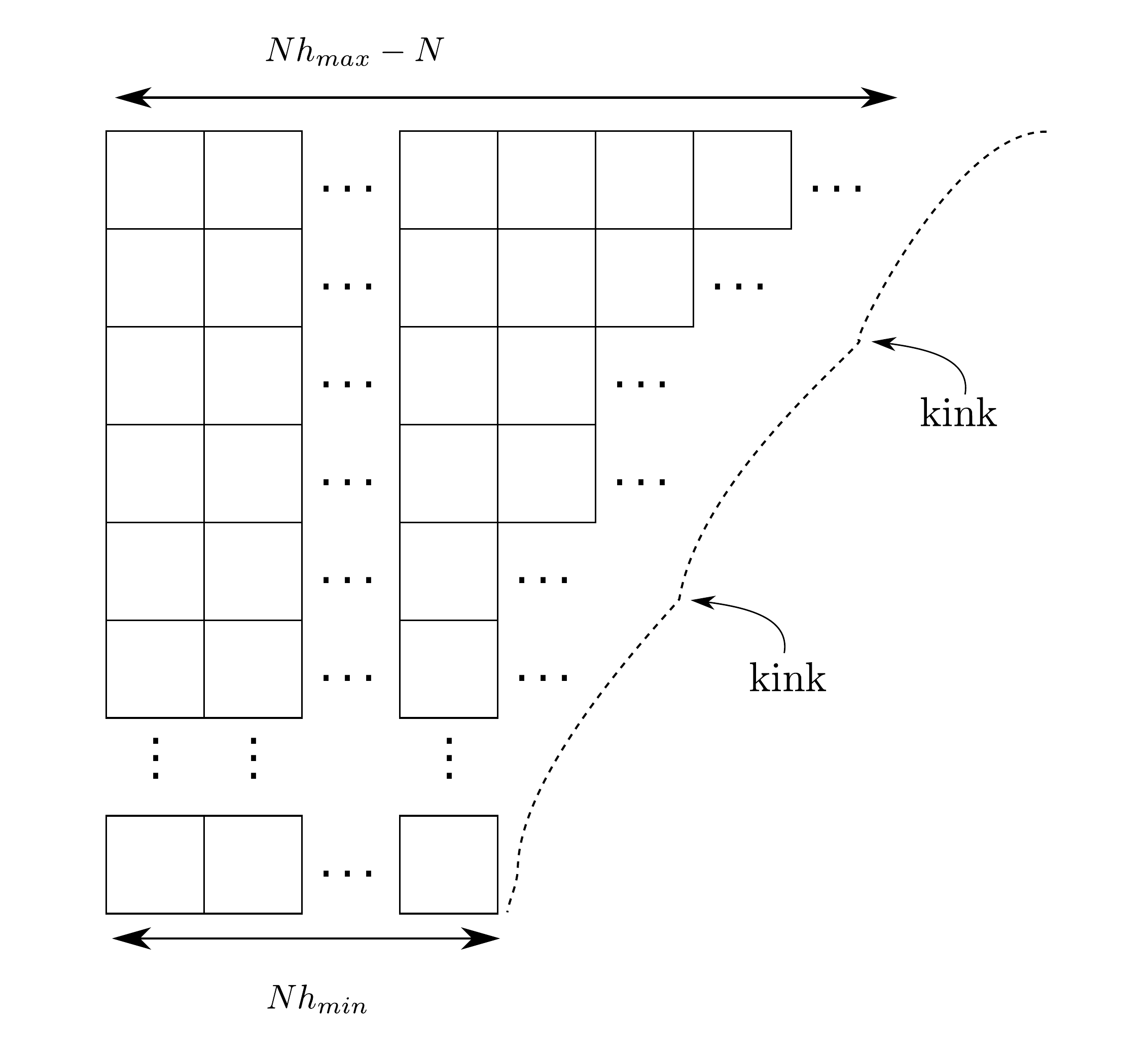}
\caption{A typical Young tableaux for $A_2$ phase.}
\label{fig:YT2g}
\end{figure}

\noindent {\underline{\bf $A_2$ phase :}}\ \ For $A_2$ phase also we
plot $h$ vs $\q$ and corresponding $u(h)$ vs $h$ in figure
\ref{fig:A_2phase}.
Corresponding Young tableaux for $A_2$ phase has two kinks. See figure
\ref{fig:YT2g}.
On $\b_1=0$ axis we see that there is no kink in Young
distribution. As we move towards left (or right, depending on sign of
$\b_1$) from $\b_1=0$ line, two kinks are developed at the top and
bottom of a tableaux. These two kinks approach towards each other as we
move towards $A_1$ (or $B_1$ phase). Finally on $A_2-A_1$ (or
$A_2$-$B_1$) transition line they meet and disappear.

On the other hand when we approach towards $A_0$ phase from any
arbitrary point in $A_2$ region, we find that the lower kink starts
moving towards $h_{min}$ and $h_{min}$ also starts decreasing. On
$A_2-A_0$ transition line, the lower kink merges with
$h_{min}=0$. This is exactly the $A_0$ phase on $A_0-A_2$ transition
line.


\section{Chern-Simons on $S^3$}\label{sec:CSonS3}
Chern-Simens (CS) theory plays an important role in different branches
of physics and mathematics. CS theory is also very extensively studied
in string theory. Partition functions of open and closed string theories can be related to partition functions of CS theory on compact manifolds \cite{Witten:1992fb,Gopakumar:1998ki,Aganagic:2002wv}.

CS theory on $S^3$ and on other three manifolds can be written in terms
of an ``exotic" matrix model \cite{Marino:2002fk, Aganagic:2002wv}. The
matrix model can be solved at large $N$ limit to obtain free energy of
open or closed topological string theory. Matrix model
analysis of CS theory is very useful to obtain informations about
topological string theory. Partition function for CS on $S^3$ can be
written as an integral over hermitian matrix ensemble with
logarithimic potential \cite{Marino:2002fk}. At large $N$, phase of CS
theory on $S^3$ is given by distribution of eigenvalues of hermitian
matrices on a straight line. It turns out that topological phase of
the theory corresponds to a ``gapped'' distribution of eigenvalues. In
this section we study large $N$ phase of CS on $S^3$ in terms of
distribution of eigenvalues of unitary matrices. This is required to
obtain phase space distribution of this theory. The phase space
picture will be helpful to understand underlying quantum mechanical
description of string degrees of freedom using dualities.

We show that a droplet picture exists for CS theory on $S^3$ for any
value of 't Hooft $\lambda = N g_s$. Since, CS on $S^3$ is topological
there exists no phase transition in the theory as we change the
coupling constant. We shall see that this property is reflected in
phase space distribution function, {\it i.e.}  as we change 't Hooft
coupling $\l$ the droplet changes its shape keeping the topology
intact.

Chern-Simons action with gauge group $G$ on a generic three-manifold
$M$ with level $k$ is given by
 \begin{equation}
  S=\frac{k}{4\pi}\int_M \Tr\left(A\wedge dA+\frac{2}{3}A\wedge
    A\wedge A\right). 
 \end{equation}
 Following \cite{Wittenjones} we can write the partition function for
 CS theory on the three-sphere $S^3$ with string coupling constant
 \be\label{eq:gs}
\displaystyle{g_s=\frac{2 \pi i}{k+N}={2\pi i \over \tilde k}},\quad
\where \quad \tilde k = k+N
\ee
as
\begin{equation}
 \zcs = \frac{1}{(k+N)^{N/2}} \prod_{a>0} 2\sin \left(\frac{\alpha
     \cdot \rho}{2}g_s\right). 
\end{equation} 
Where $\rho$ is the Weyl vector of $SU(N)$ and $\alpha$ is the
positive roots of $SU(N)\subset U(N)$ defined as
$\alpha_{ij}=\mu_i-\mu_j$ corresponding to the cartan subalgebra.

Using Weyl formula for character and properties of Jacobi's theta
function one can write the partition function for CS theory on $S^3$
as a UMM \cite{okuda}
\begin{equation}\label{eq:CS-S3-model1}
 \zcs(N,g_s)= {\cal A}(q,N) \int [DU]\, \text{exp}\left[
   N\sum_{n=1}^\infty 
 {\bar \b_n\over n} \lb \Tr U^n +\Tr U^{\dagger n}  \rb \right], \quad
\where \quad \bar\b_n =\frac{1}{N\ [n]},
\end{equation}
and
\begin{eqnarray}
 {\cal A}(q,N)=\left( \frac{g_s}{2\pi}\right)^{N/2}
  && e^{-\frac{\pi i}{4}N^2} q^{-\frac{N(N-1)}{12}}\left(\prod_{j=1}^\infty(1-q^j)\right)^N,\\
 q=e^{-g_s},\quad&& [n]=e^{g_{s}\frac{n}{2}}-e^{-g_{s}\frac{n}{2}}.
\end{eqnarray}
Thus we see that the partition function is exactly same (upto an
overall normalization) as that of a single plaquette model (equation
\ref{eq:singleplaq}) with the coefficients $1/n[n]$.

In large $N$ limit $|g_s|$ becomes small, one can expand the
exponential and find $[n] = g_s n$. Hence, we have 
\be
\bar \b_n = \frac{1}{N g_s n} \quad \text{in large $N$ limit.}
\ee
We shall study this model in large $N$ limit.

\subsection{Potential and eigenvalue distribution} \label{sec:cspot}

Since the partition function (\ref{eq:CS-S3-model1}) (action as well
as measure) is invariant under unitary transformation, one can go to a
diagonal basis where $U=\{e^{i\q_i}\},\ \ i = 1,\cdots N$, $\q_i$'s
are eigenvalues. In this basis CS action is given by
\be \label{eq:cspot} 
S_{CS}= N\sum_{n>0} {\bar \b_n\over n} \lb \Tr
U^n +\Tr U^{\dagger n} \rb = \frac N{N
  g_s}\sum_{i=1}^N\sum_{n=1}^{\infty} {2\cos n\q_i \over n^2} = \frac
N{N g_s}\sum_{i=1}^{N}\lB
\text{Li}_2(e^{i\q_i})+\text{Li}_2(e^{-i\q_i })\rB.  
\ee
Hence the partition function is 
\ben\label{eq:cspf-ev}
\cZ={\cal A} \int \lB\prod d\q_i\rB \exp\lB - \lb -
\frac N{N g_s}\sum_{i=1}^{N}\lB
\text{Li}_2(e^{i\q_i})+\text{Li}_2(e^{-i\q_i })\rB -
\sum_{{i=1}\atop{j < i}}^{N} \ln\lB \sin^2
\left(\frac{\theta_{i}-\theta_{j}}{2}\right)\rB\rb \rB. \nn \\
\een

Two important things to note here.
\begin{itemize}

\item 

  The second term in exponential in equation (\ref{eq:cspf-ev}) is
  respponsible for strong repulsion among the eigenvalues. This terms
  comes from Haar measure. The model will have a stable non-trivial
  eigenvalue distribution if the first term
  ($\text{Li}_2(e^{i\q_i})+\text{Li}_2(e^{-i\q_i })$) has minima
  between $-\pi$ and $\pi$. This term, in fact, gives a harmonic
  oscillator potential about $\q_i=\pi$. However, the coefficient of
  this potential term is $g_s=2\pi i/\tilde k$. If we study saddle
  point equation with this potential then we land up with a complex
  eigenvalue distribution.  Therefore to get a stable real and
  positive eigenvalue distribution we replace $g_s\ra i g_s$. In that
  case the potential becomes a harmonic oscillator potential about
  $\pi$ with real positive coefficient.

  The problem with complex potential is related to analyticity of
  resolvent. Resolvent computed for \ref{eq:CS-S3-model1} turns out to be non-analytic inside and outside the unit circle. Thus to make resolvent analytic in complex $z$ plane we need to replace $g_s\rightarrow ig_s$. Since free energy is an analytic function of $g_s$ an imaginary rotation in $g_s$ does not affect the calculation.

\item

  From equation (\ref{eq:cspot}) we see that the potential has a
  minima about $\q_i=\pi$. However, we are following a convension
  where everything is symmetric about $\q=0$. Thus to make the
  potential symmetric about $\q=0$ we give a $\pi$ shift to all
  eigenvalues. This is equivalent to $U\ra -U$. Since Haar measure is
  invariant under this shift ($[\cD(-U)]=[\cD U]$), the partition
  function (equation \ref{eq:CS-S3-model1}) becomes (considering both
  the points above),
\begin{equation}\label{eq:CS-S3-model}
 \zcs(N,g_s)= {\cal A}(q,N) \int [DU]\, \text{exp}\left[ N\sum_{n>0}
 {\b_n\over n} \lb \Tr U^n +\Tr U^{\dagger n}  \rb \right], \
\where \ \b_n =-\frac1\l\frac{(-1)^n}{ \ n},
\end{equation}
where 't Hooft coupling $\l$ is given by
\be
\l = \frac{2\pi N}{\tilde k} .
\ee
We shall work with this partition function.

\end{itemize}

In large $N$ limit, using saddle point analysis one can find the
eigenvalue distrubution which dominates the partition function. The
eigenvalue distribution function satisfies an integral equation, which
is in general difficult to solve to find a solution. Here we follow the discussion in section \ref{sec:resolvent} to find the eigen value distribution.

\subsection{Resolvent and Eigenvalue distribution}

Resolvent for partition function \ref{eq:CS-S3-model}
is given by
\begin{eqnarray}\label{eq:Rz-complex}
\begin{split}
&&R(z)=\frac{1}{2}\bigg[1-\frac1 \l\sum_{n=1}^{\infty}
(-1)^n\frac{z^{n}-z^{-n}}{n}+\sqrt{F(z)}\bigg]
\quad\quad|z|<1\\
&&R(z)=\frac{1}{2}\bigg[1-\frac1 \l\sum_{n=1}^{\infty}(-1)^n
\frac{z^{n}-z^{-n}}{n}-\sqrt{F(z)}\bigg]\quad\quad|z|>1
\end{split}
\end{eqnarray}
Note that the second term in $R(z)$ is antisymmetric under $z\ra
1/z$. Hence, performing resummation over appropriate regions we can
write $R(z)$ as
 \begin{eqnarray}\label{eq:CSResolv}
\begin{split}
{R}(z)&=&\frac{1}{2}\bigg[1+\frac{1}{\l}\ln[1+z]
-\frac{1}{\l}\ln[1+\frac{1}{z}]+\sqrt{{F}(z)}\bigg]\quad\quad|z|<1\\
{R}(z)&=&\frac{1}{2}\bigg[1+\frac{1}{\l}\ln[1+z]
-\frac{1}{\l}\ln[1+\frac{1}{z}]-\sqrt{{F}(z)}\bigg]\quad\quad|z|>1,
\end{split}
\end{eqnarray}

Since resolvent is analytic at $z=0$ by construction, $\sqrt{F(z)}$
must also have a logarithmic singularity at origin with equal and
opposite strength. This may happen when $F(z)$ is a perfect square,
i.e
$$F(z)=\lB1-\frac1 \l\sum_{i=1}^{\infty} (-1)^n \frac{z^{n}+z^{-n}}{n}
\rB^2.$$ In this case all the singular terms in equation
(\ref{eq:CSResolv}) are cancelled. Using equation (\ref{eq:evdef}) we
find spectral density is given by
\begin{equation}
  \r(\theta) = \frac1{2\pi}\left[
    1+\frac{2}{\l}\ln[2|\cos\frac{\theta}{2}|]\right],\quad -\pi \leq
  \theta\leq \pi. 
\end{equation}
This is {\bf no-gap solution (phase)} of CS on $S^3$. However,
$\r(\theta)$ becomes negative for some values of $\theta$, hence this
is not a physical solution or phase. We need to look for gapped phase
of the theory.

To find a more general class of solution we take the following anstaz
for $F(z)$
\begin{eqnarray}
\sqrt{F(z)}=\beta
\ln\bigg[\frac{g(z)-\sqrt{g^{2}(z)-f^{2}(z)}}{g(z)+
\sqrt{g^{2}(z)-f^{2}(z)}}\bigg]+\Gamma,
\end{eqnarray}
where $f(z)$ and $g(z)$ are polynomials of positive powers and
$\Gamma$ is some arbitrary constant\footnote{One can also take the
  ansatz for $F(z)$ to be
\begin{equation}
\sqrt{F(z)}=\beta\ln \left[ f(z) - \sqrt{g(z)}\right]+\Gamma.
\end{equation}
In this case it turns out that the functions $f(z)$ and $g(z)$ are not
positive polynomials.}.  Unknown functions $g(z)$ and $f(z)$ can be
obtained form analytic property and normalization condition of
$R(z)$. We also need to use the fact that $F(z)$ is symmetric under
$z\ra 1/z$ which implies that the function $\sqrt{g^{2}(z)-f^{2}(z)}$
has to be an even order polynomial. All the zeros of
$\sqrt{g^{2}(z)-f^{2}(z)}$ are located on $|z|=1$ line. Which means
branch cuts are exactly on unit circle in $z$ plane. Depending on
degrees of the polynomials $g(z)$ and $f(z)$ we have one gap or multi
gap solutions.

\subsubsection*{One Gap Solution}

One gap phase has two branch points on unit circle (therefore one
branch cut), hence $g(z)$ should be a order one polynomial.  We take
$g(z)=1+z$. From invariance of $F(z)$ under $z\ra1/z$ we have,
\begin{eqnarray}
  \sqrt{F(z)}=\beta \ln\bigg[\frac{z+1
  -i\sqrt{\alpha z-(z+1)^{2}}}{z+1+i\sqrt{\alpha z-(z+1)^{2}}}
  \bigg]+\Gamma, \ \text{where}\ \a, \ \b \ \text{are constants.}
\end{eqnarray}
Analyticity of $R(z)$ at $z=0$ implies that $\beta = 1/\l$. Finally,
from the normalization of $R(z)$ (i.e. $R_<(0)=1$) we have
$\alpha = 4 e^{-\l}$ and $\G=0$.  Therefore,
\be 
R(z)=\frac12\lB 1+\frac1\l \ln z +\frac1\l \ln \bigg[\frac{z+1
  -i\sqrt{4e^{-\l} z-(z+1)^{2}}}{z+1+i\sqrt{4e^{-\l} z-(z+1)^{2}}}
\bigg] \rB,\quad |z|<1.
\ee
Hence, eigenvalue distribution for one-gap phase is given by,
\begin{eqnarray}\label{eq:csev1gap}
  \r(\theta)&=& \frac1{2\pi} \lb 2 \Re\lB
                R(e^{i\q})\rB-1\rb={\sqrt{F(e^{i\theta})} \over 2\pi} =
                \frac{1}{\pi \l}\tanh^{-1}\left[ 
                \sqrt{1- \frac{e^{-\l}}{\cos^{2}\frac{\theta}{2}}}\right].
\end{eqnarray}
Since $\r(\q)\geq 0$, this implies eigenvalues are distributed over
the range
\begin{equation}
-2\cos^{-1}e^{-\l/2}<\theta< 2\cos^{-1}e^{-\l/2}.
\end{equation}
This analysis is similar to analysis of \cite{Marino:2004eq}. However,
\cite{Marino:2004eq} studied distribution of eigenvalues of hermitian
matrices. The gap depends on value of $\l$, it increases (spreading of
eigenvalue distribution decreases) as we decrease $\l$. At very small
value of $\l$ all the eigenvalues are concentrated about $\q=0$ point.
This can be understood from the fact that at small values of $\l$
attractive part of the potential in equation (\ref{eq:CS-S3-model})
becomes infinitely strong and hence all the eigenvalues are localised
at $\q=0$.

\subsubsection*{Two Gap Solution}

Continuing with the same sprit, one can construct the function $F(z)$
for two-gap solution. In this case $g(z)$ is a second order polynomial.
\begin{equation}\label{eq:twogapans}
\sqrt{F(z)}=\beta\log \left[\frac{z^2+\gamma  z+1-i \sqrt{\alpha 
   z^2-\left(z^2+\gamma  z+1\right)^2}}{z^2+\gamma  z+1 + i \sqrt{\alpha 
   z^2-\left(z^2+\gamma  z+1\right)^2}}\right]+\G.
\end{equation}
From the properties of $R(z)$ we find,
\begin{equation}
\beta = \frac{1}{2\l}, \qquad \alpha = 4 e^{-2\l} \quad \text{and}
\quad \G=0.
\end{equation}
The parameter $\gamma$ can not be fixed from the analyticity or
normalization conditions. Hence it is an one parameter family of
solution at this point.  The eigenvalue distribution for this branch
is given by,
\begin{eqnarray}
  \r(\theta) &=& \frac1{2\pi \l} \tanh^{-1} \left[\frac{\sqrt{
                     (2\cos\theta +\gamma)^2-4 e^{-2\l} }}{2\cos\theta
                     +\gamma} 
                     \right]\nonumber\\
                 &=& \frac1{2\pi \l} \tanh^{-1}
                     \left[\frac{4\sqrt{\left(\cos^2{\theta/2}
                     -\cos^2{\theta_1/2}\right)
                     \left(\cos^2{\theta/2}-\cos^2{\theta_2/2}\right)}} 
                     {2\cos\theta +\gamma}\right],
\end{eqnarray}
where
\begin{equation}
\theta_1 = 2 \cos^{-1}
\left[\frac{\sqrt{2(1+e^{-\l})-\gamma}}{2}\right],\quad\theta_2 = 2
\cos^{-1} 
\left[\frac{\sqrt{2(1-e^{-\l})-\gamma}}{2}\right].
\end{equation}
Thus we see that one can also have a two-gap solution for any value of
$\l$. However, computing free energy of the system one can show that
the two gap solution is not thermodynamically favoured i.e. free
energy is always greater than that of one gap solution for all values
of $\l$ \cite{Morita:2017oev}. This can also be understood from the
potential of the matrix model. Since the attractive part of potential
has only one minima about $\q=0$ therefore only one-gap solution is a
thermodynamically stable solution. However, this kind of solution
appears when we consider CS theory on lens-space \cite{Halmagyi} or
ABJM theory \cite{Marino:2009jd}.


\subsection{Phase space distribution}

Having understood eigenvalue distribution for CS theory on $S^3$ we
now proceed to find the most dominant representation for the most
dominant phase of CS theory. The distribution of boxes in the most
dominant Young diagram can be obtained from the boundary relation.

The boundary relation for a generalised plaquette model is given by
equation (\ref{eq:boundrelplaq})
\begin{equation}
h_{\pm}(\q)={1\over 2}+\sum_n \beta_n \cosa{n\theta} \pm
\pi\rho(\theta). 
\end{equation}
For CS on $S^3$, $\b_n = -\frac1\l {(-1)^n \over n}$, hence the second
term on right hand side can be resummed over $n$
\begin{eqnarray}
  \sum_n \beta_n \cosa{n\theta}&=&\frac1\l \lb \ln 2 +\ln
                                   |\cos\frac\q2|\rb.  \nonumber\\
                               &=& {1\over \l}\tanh^{-1}\left[\frac{4
                                   \cos^2\left(\theta/2\right)-1}{4 
                                   \cos^2\left(\theta/2\right)+1}\right] 
\end{eqnarray}
Eigenvalue distribution is given by equation
(\ref{eq:csev1gap}). Hence boundary relation (Fermi surface) is given
by
\begin{eqnarray}\label{eq:fermisurCS}
  h_{\pm}(\q)={1\over 2}+{1\over \l}\tanh^{-1}\left[\frac{4
  \cos^2\left(\theta/2\right)-1}{4 
  \cos^2\left(\theta/2\right)+1}\right]
  \pm \frac{1}{ \l}\tanh^{-1}\left[ 
  \sqrt{1- \frac{e^{-\l}}{\cos^{2}\frac{\theta}{2}}}\right]. 
\end{eqnarray}
\begin{figure}[h]
\begin{subfigure}{.4\textwidth}
\centering
\includegraphics[width=4cm,height=4cm]{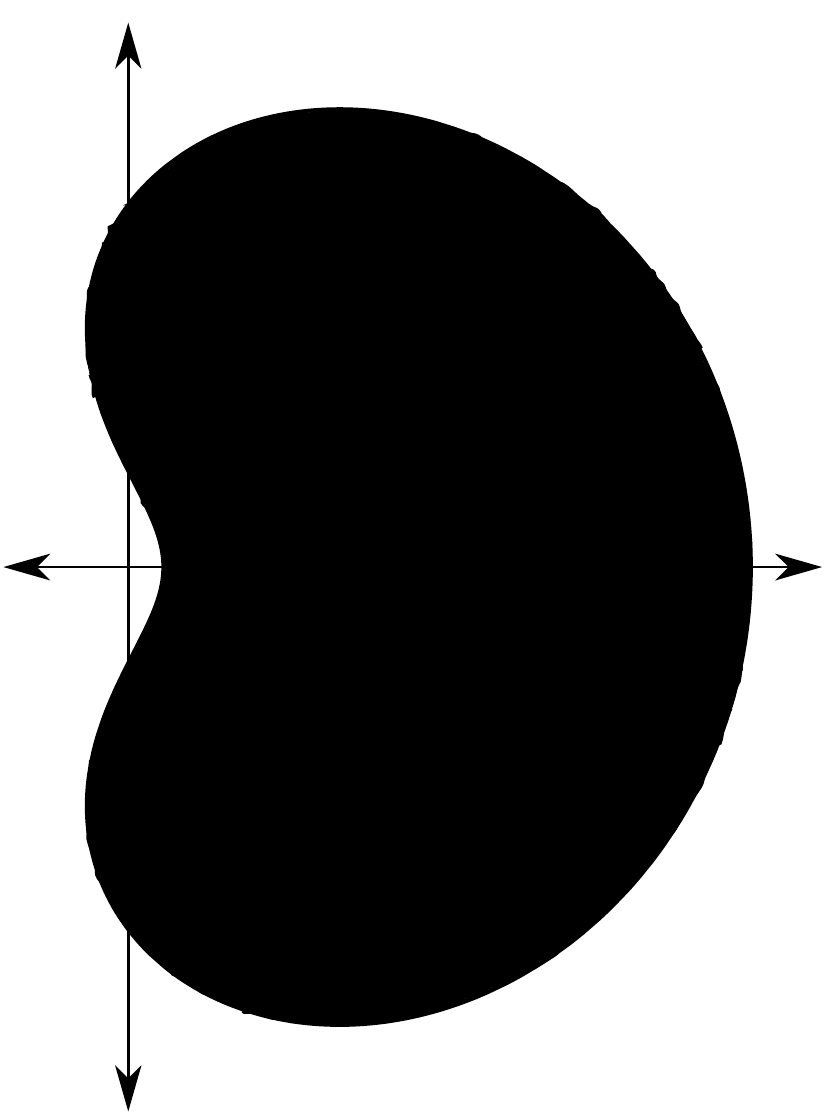}
\caption{$\l=.9$}
\end{subfigure}
\begin{subfigure}{.6\textwidth}
\centering
\includegraphics[width=8cm,height=4cm]{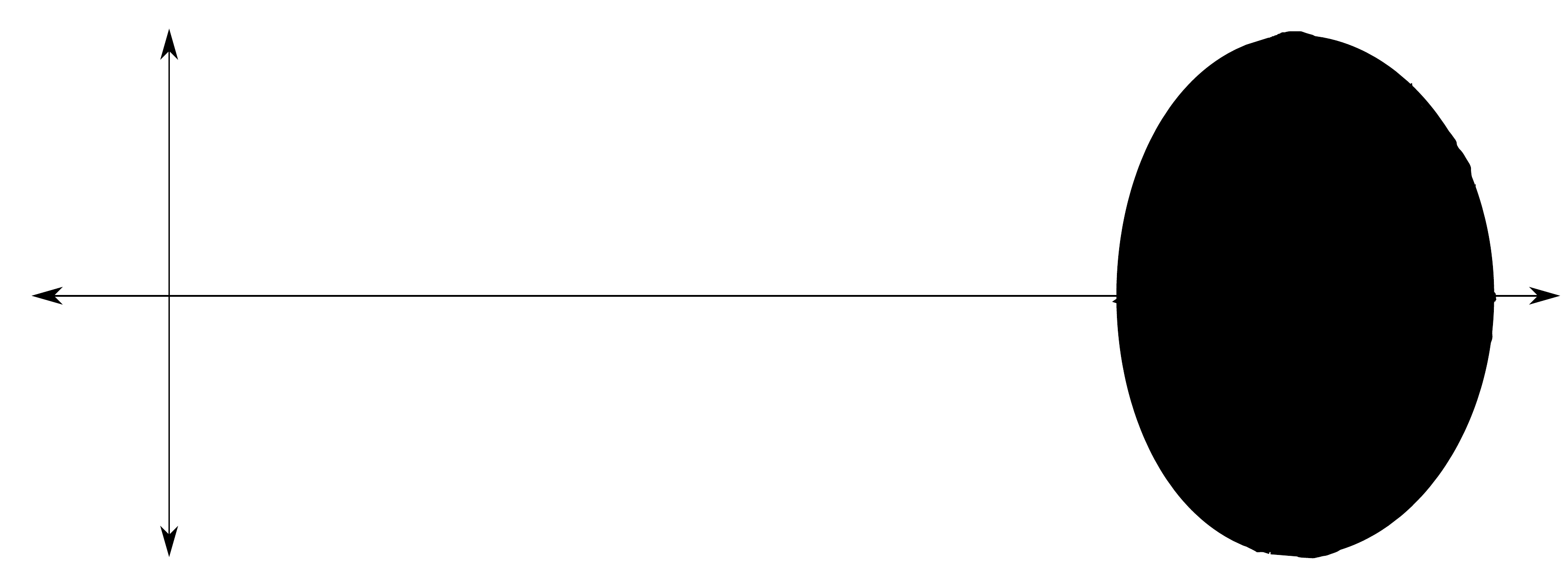}
\caption{$\l=.01$}
\end{subfigure}
\caption{Phase space distribution for Chern-Simons theory of $S^3$ for
  different values of $\l$.}
\label{fig:phasespaceCS}
\end{figure}
Phase space distribution for CS on $S^3$ has been plotted in figure
\ref{fig:phasespaceCS}. The droplets have the shape of ``kidney'' with
a cleavage on left. The origin is always outside the distribution
(character of gapped solution). As we increase values of $\l$, the
droplet moves towards the origin also the cleavage becomes sharper.
We also see that in valid ranges of $\l$ phase space distribution changes
its shape without crossing the origin. This implies that CS theory on
$S^3$ is topological. There is no phase transition in the theory.

\subsection{Young tableaux distribution}

\begin{figure}[h]
\centering
\includegraphics[width=13cm,height=9cm]{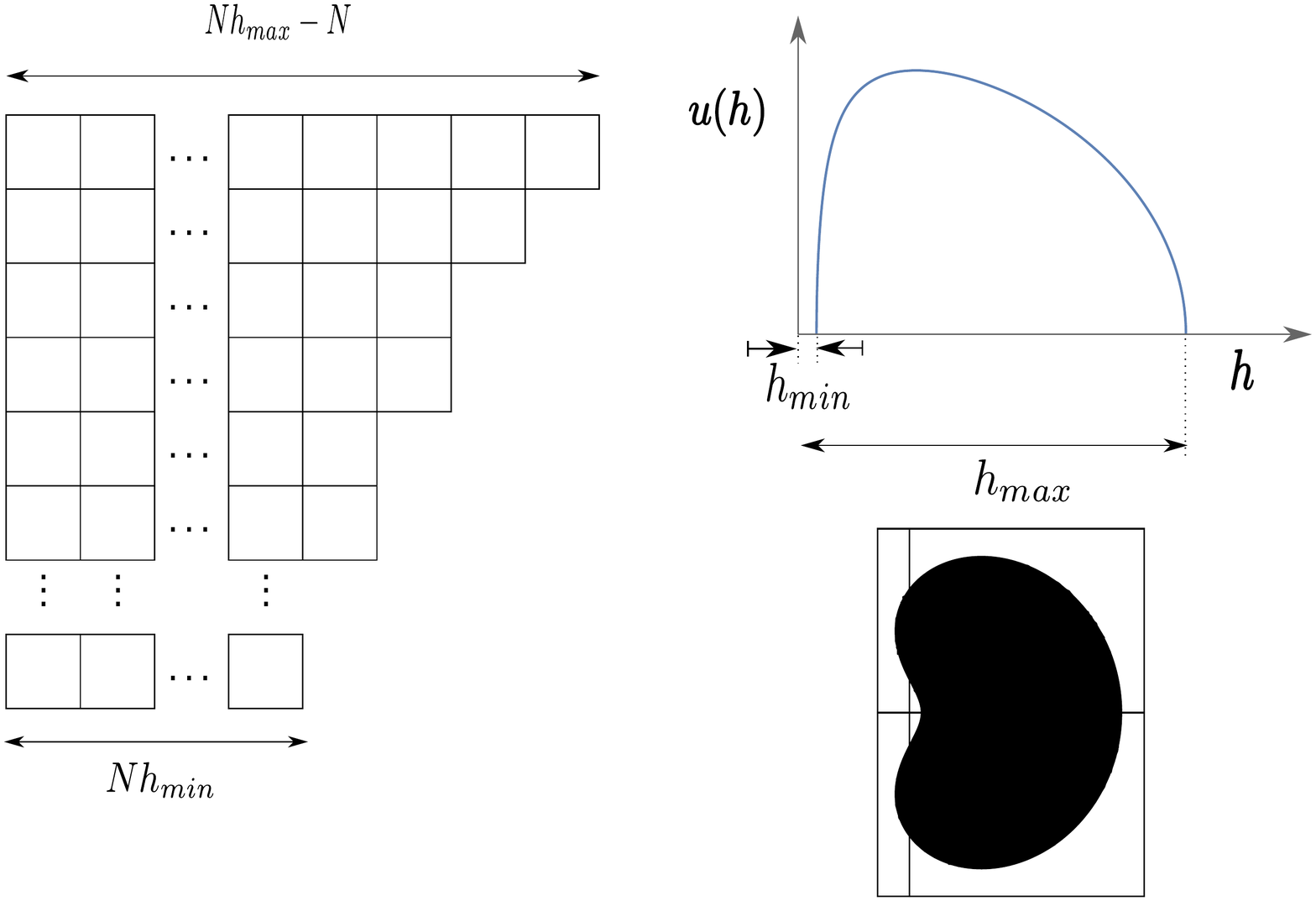}
\caption{Young tableaux distribution and density for a generic $\l$. This
  representation corresponds to a phase space distribution as shown in
  right bottom figure.}
\label{fig:uhcs}
\end{figure}

Since eigenvalues are identified to density of Young tableaux :
$\pi u(h) = \q$, one can invert the boundary relation
(\ref{eq:fermisurCS}) to obtaing distribution of boxes in the most
dominant phase of CS theory for a given $\l$. A typical relation
$u(h)$ and $h$ and corresponding Young diagram have been plotted in
figure \ref{fig:uhcs}. 
Minimum ($\hmin$) and maximum ($\hmax$) values
of $h$ correspond to number of boxes in the last row and first row of
Young diagram respectively and are given by
\ben\label{eq:hmaxmin}
\begin{split}
\hmax &= h_+(0)=\frac12 +\frac{\ln2}\l+\frac1{2\l} \ln \lB
{1+\sqrt{1-e^{-\l}}\over 1-\sqrt{1-e^{-\l}}} \rB, \\
\hmin &= h_-(0)=\frac12 +\frac{\ln2}\l-\frac1{2\l} \ln \lB
{1+\sqrt{1-e^{-\l}}\over 1-\sqrt{1-e^{-\l}}} \rB.
\end{split}
\een

For small $\lambda$, $\hmax$ and $\hmin$ go as
\be 
\hmax \sim \frac{\ln2}\l +\frac1{\sqrt\l}, \quad \hmin \sim
\frac{\ln2}\l - \frac1{\sqrt\l}.
\ee
In small $\l$ limit we see that both $\hmax$ and $\hmin$ are very
large and difference between them goes as $2/\sqrt\l$.

\section{Discussion and outlook}
\label{sec:dis}

In this paper we show that large $N$ generic phases of a generic
unitary matrix model can be described in terms of geometry of two
dimensional droplets in $(h,\q)$ plane, where $h$ is related to number
of boxes in Young diagram and $\q$ is eigenvalue of unitary
matrices. We also see that these droplets are similar to phase space
distribution of unitary matrix quantum mechanics on a constant time
slice. Hence, we argue that for a given unitary matrix model it is
possible to find an underlying quantum mechanics such
that different large $N$ phases of unitary matrix models (and hence
interacting gauge theories) are captured on different time slices of the quantum system.

This observation allows us to identify momenta of fermionic system
with number of boxes in a Young diagram. Distribution of boxes in
Young diagram captures information about the distribution of momenta of $N$
fermions. We show that, Young distribution for different large $N$ phases can be 
computed from the geometry of two
dimensional droplets. We explicitly work out two examples :
$\b_1$-$\b_2$ model and Chern-Simons theory on $S^3$.

Phase space description of gauge theories can be used to provide a
dual quantum mechanical description of string degrees of freedom,
since different large $N$ phases of gauge theory are dual to classical
solutions of string theory in AdS space. Therefore, one can extend this
idea for a supersymmetric QFT. One can consider
ABJM theory \cite{Aharony:2008ug}, supersymmetric CS theory in $2 + 1$
dimensions. Partition function of ABJM can be calculated exactly for
all values of coupling constant. There exists an interpolating
function between weak and strong coupling \cite{Marino:2009jd,
  Drukker:2010nc}. Hence, it would be interesting to write ABJM
partition function in terms of a unitary matrix model and find two 
dimensional droplets associated with large $N$ phases of the theory. One can
interpolate this droplet picture to strong coupling side to explore
properties/geometries of dual string/M-theories in the spirit of LLM
\cite{Lin:2004nb}. This will also serve as a precision test of
$AdS_4/CFT_3$. Understanding underlying matrix quantum mechanics for
CS theory coupled with matter fields \cite{Jain:2013py, Jain:2013gza,
  Jain:2012qi, Marino:2012az, Codesido:2014oua} would also be an
interesting problem to look at from the point of \cite{Dandekar:2014era}.

One should note that the inviscid Burgers' equation \ref{eq:fermisurfeom} 
is a generic equation for random matrix theories. 
In the continuum limit of QCD \cite{Durhusolesen}, 
first observed that spectral density of the Wilson operator 
follows this equation as a consequence of Makeenko-Migdal 
equation and proposed a order(gapped)-disorder(gapless) 
transition for $2D$ YM at large $N$. \cite{Rossi} Also came 
across this same quasi-linear equation while dealing with 
the continuum version of the loop equation for $QCD_2$. 
Starting with the Burgers' equation \cite{Nowak,NeubergerBurger} 
supplemented the proposal of  \cite{Durhusolesen}, 
and demonstrated the emergence of a spectral shock wave 
of Wilson loop eigenvalues at the large $N_c$ limit. 
Recently \cite{Neuberger} extended this study numerically 
for $d=4$. As matrix integrals can be thought of as an 
alternate way of writing the combinatorics of maps, this 
kind of connections between random matrix theory and loop 
equations can also be extended for combinatorial problems. 
To further explore this combinatorial avenue to understand 
the identification of loop equations with Tutte’s equations 
(recursively deleting or contracting edges) and furthermore 
to grasp the current status of ``random matrix" approach to 
2D quantum gravity one useful resource would be \cite{eynard_book}.

On a different ground, UMM has 
plenty of applications in mathematics
specially in number theory. Hilbert and P\' oyla speculated that there
is a spectral origin for non-trivial zeros of Riemann zeta
function. It was observed in \cite{berry-keating} that if the non-trivial
zeros are analogous to EV of some Hamiltonian, then the corresponding
conjugate variables, time period of some primitive orbits, are related to
prime numbers (proportional to $\ln p$ where $p$ is prime
number). Hence, if EV distribution of a UMM captures information about
non-trivial zeros, then we expect that corresponding Young distribution
function captures information about prime number distribution
\cite{duttadutta}. It would be nice to understand if there is any
quantum mechanical system which captures information about these two
distributions.

\noindent
{\bf Acknowledgement }

We would like to thank Rajesh Gopakumar for many helpful
discussion. We are grateful to Shiraz Minwala and Gautam Mandal for
their useful comments on our work. We are grateful to Debashis Ghoshal
for innovative discussion. AC would like to acknowledge the hospitality of
IISER Pune where part of this work is done. SD acknowledges the Simons
Associateship, ICTP. SD also thanks the hospitality of ICTP, Trieste
and JNU where part of this work has been done.  Work of SD is
supported by DST under a project with number {\it EMR/2016/006294}. PD is partially supported by research grant number 5304-2, "{\it Symmetries and Dynamics: Worldsheet and Spacetime}" from CEFIPRA/IFCPAR. Finally, we are grateful to people of India for their unconditional support towards researches in basic sciences.

\appendix

\section{Calculation of boundary relation for three trace action
}\label{app:details}

Exponentiating the effective action (\ref{eq:seffthreetrace}) we find,
\begin{eqnarray}
\begin{split}
e^{S_{eff}}&=\prod_{\vec n\ge 0}
\expb{{a_{n_1n_2}\over N^3}\Tr{U^{n_1}}\Tr{U^{n_2}}\Tr{U^{\dagger
      (n_1+n_2)}}} 
\expb{{a_{n_1n_2}\over N^3}\Tr{U^{\dagger n_1}}\Tr{U^{\dagger
      n_2}}\Tr{U^{(n_1+n_2)}}}\\ 
&=\prod_{\vec n\ge 0}
\sum_{[k]=0}^\infty {(a_{n_1n_2})^{k_{n_1n_2}}\over
  N^{3k_{n_1n_2}}k_{n_1n_2!}} 
(\Tr{U^{n_1}})^{k_{n_1n_2}}(\Tr{U^{n_2}})^{k_{n_1n_2}}
(\Tr{U^{\dagger(n_1+n_2)}})^{k_{n_1n_2}}\\
& \qquad \quad \sum_{[l]=0}^\infty {(a_{n_1n_2})^{l_{n_1n_2}}\over
  N^{3l_{n_1n_2}} 
l_{n_1n_2!}}(\Tr{U^{\dagger n_1}})^{l_{n_1n_2}}
(\Tr{U^{\dagger n_2}})^{l_{n_1n_2}}(\Tr{U^{(n_1+n_2)}})^{l_{n_1n_2}}
\end{split}
\end{eqnarray}
where, $\vec n$ is a two dimensional vector with entries $n_1$ and $n_2$,
$[k]$ and $[l]$ are matrices with entries $k_{n_1n_2}$ and $l_{n_1n_2}$.
Interchanging the sum and product we write
\ben
\begin{split}
e^{S_{eff}}&= \sum_{[k],[l]} \prod_{\vec n\ge 0}
{(a_{n_1n_2})^{k_{n_1n_2}+l_{n_1n_2}} 
\over k_{n_1n_2}! l_{n_1n_2}!} \\ 
& \qquad \prod_{\vec n\ge 0} \lb \tu{n_1}\rb^{k_{n_1n_2}} \lb
\tu{n_2}\rb^{k_{n_1n_2}} 
\lb \tud{n_1+n_2}\rb^{k_{n_1n_2}}\\
&\qquad \prod_{\vec n\ge 0} \lb \tud{n_1}\rb^{l_{n_1n_2}} \lb
\tud{n_2}\rb^{l_{n_1n_2}} 
\lb \tu{n_1+n_2}\rb^{l_{n_1n_2}}.
\end{split}
\een
This expression can be further simplified to,
\ben
\begin{split}
 e^{S_{eff}}&= \sum_{[k],[l]} \cT_0 \ N^{p_0+q_0+r_0+s_0}
 \Upsilon_{\vec p+ \vec r}(U) \Upsilon_{\vec q+\vec s}(\udag{})
\end{split}
\een
where,
\ben
\begin{split}
 p_i &= \sum_{j=0}^{\infty}k_{ij}+k_{ji}, \quad 
 q_i = \sum_{m,m=0}^{\infty} k_{mn}\delta(m+n-i) \quad \forall\ i
 =0,1, 2,\cdots\\ 
 s_i & = \sum_{j=0}^\infty l_{ij}+l_{ji}, 
 \quad r_i=\sum_{m,n=0}l_{mn}\,\delta(m+n-i)\quad \forall\ i =0,1,
 2,\cdots ,
\end{split}
\een
$\vec p=\{p_1,p_2,\cdots\}$, $\vec q=\{q_1,q_2,\cdots\}$, $\vec
r=\{r_1,r_2,\cdots\}$, 
$\vec s=\{s_1,s_2,\cdots\}$ and
\be
\Upsilon_{\vec p +\vec r}(U) = \prod_{n=1}^\infty \lb \tu n\rb^{p_n+r_n}.
\ee
The partition function, therefore, is given by,
\begin{eqnarray} \label{eq:pf3trace}
\cZ&=&\int [DU]\sum_{[k]}\sum_{[l]}\, \cT_0\,
\,
       N^{q_0+p_0+r_0+s_0}\,\, \Upsilon_{\vec{p}+\vec{r}}(U)
       \Upsilon_{\vec{q}+\vec{s}}(U^\dagger) .
\end{eqnarray}
$\Upsilon$'s can be written, like before, in terms of the characters
of the permutation group using the following identity
\begin{eqnarray}
\Upsilon_{\vec{p}+\vec{r}}(U)&=&\sum_R \chi_R(C(\vec{p}+\vec{r}))\Tr_{R} U\\
\Upsilon_{\vec{q}+\vec{s}}(U)&=&\sum_{R^{'}} \chi_{R^{'}}(C(\vec{q}+\vec{s}))\Tr_{R^{'}} \udag{}
\end{eqnarray}
Substituting $\Upsilon$ in equation (\ref{eq:pf3trace}) and using 
orthogonality relation (\ref{eq:normalization_TrU}) we find,
\begin{eqnarray}
\cZ=\sum_R \sum_{[k]}\sum_{[l]}\,\cT_0\,\,
\chi_R(c(\vec{p}+\vec{r}))\chi_R(c(\vec{q}+\vec{s})).
\end{eqnarray}
One can write sum of representations of $U(N)$ as sum over different
Young diagrams.  Finally, partition function is given by,
\ben\label{eq:pf3trace2}
\cZ = \sum_{\{\lambda_\a\}} \sum_{[k][l]} \cT_0 \
\chi_R(C(\vec{p}+\vec{r}))\chi_R(C(\vec{q}+\vec{s})) 
\delta(\Sigma_i i (p_i+r_i)-\Sigma_\a \l_\a)
\delta(\Sigma_i i (q_i+s_i)-\Sigma_\a \l_\a).\nn\\
\een

\subsection{Extremization of partition function}

In large $N$ limit we define,
\begin{equation}
a_{n_1n_2}=N^2a_{n_1n_2}^{'};\quad k_{n_1n_2}=N^2k_{n_1n_2}^{'};\quad  
l_{n_1n_2}=N^2l_{n_1n_2}^{'};\quad K=N^2 K'
\end{equation}
and using the expression for character given by equation
(\ref{eq:character1}), partition function (\ref{eq:pf3trace2}) can be
written as, (see next subsection \ref{appsec:detail} for details)
\begin{eqnarray}\label{eq:Zfinalgeneric}
\cZ&=& {N^4\over (2\pi i)^{2N}}\int
d[k'] d[l']\int[Dh]{[Dz]\over z(x)}
{[Dw]\over w(x)}\int dt\,ds \exp\lB- N^2 \seff{h,z,w} \rB
\end{eqnarray}
where,
$$
\qquad d[k'] = \prod_{n_1,n_2=0} dk'_{n_1n_2}, \quad 
d[l'] = \prod_{n_1,n_2=0} dl'_{n_1n_2}
$$
and
\begin{eqnarray}
-\seff{h,z,w}&=&\sum_{n_1,n_2=0}^{\infty}\Bigg[ k_{n_1n_2}'\lb 1+\lna{a_{n_1n_2}'
Z_{n_1}Z_{n_2}W_{n_1+n_2}\over k_{n_1n_2}'}\rb \nn\\
&& + l_{n_1n_2}'\lb 1+\lna{a_{n_1n_2}'
W_{n_1}W_{n_2}Z_{n_1+n_2}\over l_{n_1n_2}'}\rb\Bigg]
 \nonumber\\
&&-
\int_0^1dx\,h(x)\,\lna{z(x)w(x)}
+\frac12 \int_0^1 dx \Xint-_0^1 dy\, \lna{|z(x)-z(y)||w(x)-w(y)|}\nn\\
&&+i(t+s)(\sum_{n_1,n_2=0}^\infty (n_1+n_2)(k_{n_1n_2}'+l_{n_1n_2}')-K')
\end{eqnarray}
where,
\be
Z_n = \int  z^n(x) dx \quad \tand \quad W_n = \int  w^n(x) dx.
\ee
In the large $N$ limit dominant contribution to partition function comes 
from extremum value of this effective action. Varying this 
effective action with respect to $h(x)$ we find,
\begin{eqnarray}\label{eq:heom}
z(x)w(x)=e^{-i(t+s)}=\text{constant}.
\end{eqnarray}
Again, varying the effective action with respect to $k_{n_1n_2}'$ and $l_{n_1n_2}'$ 
one can write
\begin{eqnarray}\label{eq:kleom}
\begin{split}
 {a_{n_1n_2}'Z_{n_1}Z_{n_2}W_{n_1+n_2}\over k_{n_1n_2}'}&=e^{-i(t+s)(n_1+n_2)}
 =\text{constant},\\
{a_{n_1n_2}'W_{n_1}W_{n_2}Z_{n_1+n_2}\over l_{n_1n_2}'}&=e^{-i(t+s)(n_1+n_2)}
=\text{constant}.
\end{split}
\end{eqnarray}
Finally, varying $\seff{h,z,w}$ with respect to $z(x)$ we find,
\begin{eqnarray}
\Xint-_0^1 dy\,{z(x)\over z(x)-z(y)}-h(x)
+\sum_{n_1n_2=0}^{\infty}\Bigg[
k_{n_1n_2}' \lb {n_1 z^{n_1}(x)
\over Z_{n_1}} +{n_2 z^{n_2}(x) \over Z_{n_2}}\rb
+ l_{n_1n_2}' {(n_1+n_2) z^{n_1+n_2}(x)\over Z_{n_1+n_2}}\Bigg]
=0 .\nonumber\\
\end{eqnarray}
A similar equation can be obtained when we vary the effective with
respective to $w(x)$ with $z(x)$ replaced by $w(x)$.

Since, in equation (\ref{eq:Zfinalgeneric}) $z(x)$ and $w(x)$ vary
over some contour around zero, we take these contours to be unit
circle about the origin and hence one consistent set of solutions to
equations (\ref{eq:heom}) and (\ref{eq:kleom}) are given by,
\ben
z(x) = e^{i\q(x)}, \quad w(x)=e^{-\q(x)}, \quad t=s=0.
\een
Therefore,
\ben
Z_n = W_n \quad (\text{for all}\ n) \quad \text{and} \quad
k'_{n_1n_2}=l'_{n_1n_2}= a'_{n_1n_2} Z_{n_1}Z_{n_2}Z_{n_1+n_2}.
\een
Using these solutions the saddle point equation for
$z(x)=\exp(i\q(x))$ is given by equation (\ref{eq:sadeqnthreetrace}).


\subsection{Details of partition function calculation}
\label{appsec:detail}

We have
\begin{eqnarray}
  \cZ&=&\prod_{n_1,n_2=0}{N^4\over (2\pi i)^{2N}}\int\begin{matrix}
    &dk_{n_1n_2}^{'}\\&dl_{n_1n_2}^{'}\end{matrix}\int[Dh]{[Dz] \over
                        z(x)}{[Dw] \over w(x)}\int dt\,ds\,\exp N^2
                        \Bigg(\sum_{n_1,n_2=0}\Big[(k_{n_1n_2}^{'}+l_{n_1n_2}^{'})
                        \lna{a_{n_1n_2}^{'}\over
                        N}\nonumber\\ 
     &&-2k_{n_1n_2}'\lna{N}-k_{n_1n_2}'\lna{k_{n_1n_2}'}+k_{n_1n_2}'
        \nonumber
        -2l_{n_1n_2}'\lna{N}-l_{n_1n_2}'\lna{l_{n_1n_2}'}+l_{n_1n_2}' \Big]\\
     &&+(k_{00}'+l_{00}'+p_0'+q_0')\lna{N} + it (\sum_j
        j(p_j'+r_j')-K') +is(\sum_j j(q_j'+s_j')-K')\nonumber\\
     &&+\sum_{n=1}^{\infty}(p_n'+r_n')\lna{NZ_n} + \sum_{n=1}^{\infty}
        (q_n'+s_n')\lna{NW_n}+\frac12 \int_0^1 dx \Xint-_0^1 dy\,
        \lna{|z(x)-z(y)||w(x)-w(y)|}\nonumber\\ 
     &&-\int_0^1dx\,h(x)\,\lna{z(x)w(x)}\Bigg).
\end{eqnarray}
Now let us look into some relations
\begin{eqnarray}
\sum_{n=1}^{\infty}p_n'\lna{NZ_n}&=&\sum_{n=0}^{\infty}p_n'
                                     \lna{NZ_n}-p_0'\lna{NZ_n}\nonumber\\ 
&=&\sum_{n_1=0}^{\infty}\sum_{n_2=0}^{\infty}(k_{n_1n_2}'+k_{n_2n_1}')
    \lna{NZ_{n_1}}-p_0'\lna{NZ_0}\nonumber\\ 
&=&\sum_{n_1,n_2=0}^{\infty}k_{n_1n_2}'\lna{N^2Z_{n_1}Z_{n_2}}-p_0'\lna{N}
\end{eqnarray}
\begin{eqnarray}
\sum_{n_1=0}^{\infty}\sum_{n_2=0}^{\infty}\sum_{m=0}^{\infty}l_{n_2m}'
  \delta(m+n_2-n_1)&=&l_{00}'+(l_{01}'+l_{10}')+(l_{02}'+l_{11}'+l_{20})
                       +\cdots\nonumber\\
&=&l_{00}'+(l_{01}'+l_{02}'+\cdots)+(l_{10}'+l_{11}'+l_{12}'+\cdots) +
    \cdots\nonumber\\
&=&\sum_{n_1,n_2=0}^{\infty}l_{n_1n_2}
\end{eqnarray}
\begin{eqnarray}
\sum_{n=1}^{\infty}r_n'\lna{NZ_n}&=&\sum_{n=0}^{\infty} r_n'\lna{NZ_n}
                                     -r_0'\lna{NZ_n}\nonumber\\ 
&=&\sum_{n_1=0}^{\infty}\sum_{n_2=0}^{\infty}\sum_{m=0}^{\infty}l_{n_2m}'
    \delta(m+n_2-n_1)\lna{NZ_{n_1}}-r_0'\lna{NZ_0}\nonumber\\
&=&\sum_{n_1,n_2=0}^{\infty}l_{n_1n_2}'\lna{NZ_{n_1+n_2}}-r_0'\lna{N}
\end{eqnarray}
So finally we have the partition function as
\begin{eqnarray}
\cZ&=&\prod_{n_1,n_2=0}{N^4\over (2\pi i)^{2N}}\int\begin{matrix}
&dk_{n_1n_2}^{'}\,dt\\&dl_{n_1n_2}^{'}\,ds\end{matrix}\int[Dh]\int
                        \,{[Dz]\over z(x)}{[Dw]\over w(x)}\,\exp
                        N^2\Bigg[\begin{matrix}
                          &it(\sum_{j=1}^{\infty}
                          j(p_j'+r_j')-K')\\&+is(\sum_{j=1}^{\infty}
                          j(q_j'+s_j')-K')\end{matrix}\nonumber\\ 
&&+\sum_{n_1,n_2=0}^{\infty}\Bigg(k_{n_1n_2}'(1+\lna{a_{n_1n_2}'Z_{n_1}Z_{n_2}
   W_{n_1+n_2}\over
   k_{n_1n_2}'})+l_{n_1n_2}'(1+\lna{a_{n_1n_2}'W_{n_1}W_{n_2}
   Z_{n_1+n_2} \over l_{n_1n_2}'})\Bigg)\nonumber\\
&&-\int_0^1dx\,h(x)\,\lna{z(x)w(x)}+\frac12 \int_0^1 dx  \Xint-_0^1
   dy\, \lna{|z(x)-z(y)||w(x)-w(y)|}\Bigg] 
\end{eqnarray}
Some more relations
\begin{eqnarray}
\sum_{j=1}^{\infty}jp_j'&=&\sum_{j=0}^{\infty}jp_j'=\sum_{n_1,n_2=0}n_1(k_{n_1n_2}'+k_{n_2n_1}')\nonumber\\
&=&\sum_{n_1,n_2=0}^\infty(n_1+n_2)k_{n_1n_2}',
\end{eqnarray}
\begin{eqnarray}
\sum_{j=1}^{\infty}jr_j'&=&\sum_{j=0}^{\infty}jr_j' =
                            \sum_{n_1,n_2,m=0}n_1\,l_{n_2m}'
                            \delta(m+n_2-n_1) \nonumber\\
&=&\sum_{n_1,n_2=0}^\infty(n_1+n_2)l_{n_1n_2}'.
\end{eqnarray}
Equivalently
\begin{eqnarray}
\sum_{n_1=0}^\infty n_1(p_{n_1}'+r_{n_1}')&= & \sum_{n_1,n_2=0}^\infty
                                               (n_1+n_2)(k_{n_1n_2}'+l_{n_1n_2}'),\\ 
\sum_{n_1=0}^\infty n_1(q_{n_1}'+s_{n_1}')&= & \sum_{n_1,n_2=0}^\infty
                                               (n_1+n_2)(k_{n_1n_2}'+l_{n_1n_2}'). 
\end{eqnarray}
One can finally write the effective action as
\begin{eqnarray}
-S_{eff}&=&\sum_{n_1,n_2=0}^{\infty}\Bigg(k_{n_1n_2}'(1+\lna{a_{n_1n_2}'
            Z_{n_1}Z_{n_2}W_{n_1+n_2}\over
            k_{n_1n_2}'})+l_{n_1n_2}'(1+
            \lna{a_{n_1n_2}'W_{n_1}W_{n_2} Z_{n_1+n_2}\over
            l_{n_1n_2}'}) \Bigg)\nonumber\\
&&+i(t+s)(\sum_{n_1,n_2=0}^\infty
   (n_1+n_2)(k_{n_1n_2}'+l_{n_1n_2}')-K') -\int_0^1dx\,h(x)\,
   \lna{z(x)w(x)}\nonumber\\ 
&&+\frac12 \int_0^1 dx \Xint-_0^1 dy\, \lna{|z(x)-z(y)||w(x)-w(y)|} 
\end{eqnarray}
From this point onwards one can do the saddle point analysis. Varying
this effective action with respect to $h(x)$
\begin{eqnarray}
z(x)w(x)=e^{-i(t+s)}
\end{eqnarray}
Now varying the effective action with respect to $k_{n_1n_2}'$ and
$l_{n_1n_2}'$ one can write
\begin{eqnarray}
{a_{n_1n_2}'Z_{n_1}Z_{n_2}W_{n_1+n_2}\over k_{n_1n_2}'}&=&e^{-i(t+s)(n_1+n_2)}\\
{a_{n_1n_2}'W_{n_1}W_{n_2}Z_{n_1+n_2}\over l_{n_1n_2}'}&=&e^{-i(t+s)(n_1+n_2)}
\end{eqnarray}
If we choose $t=s=0$ then the equation of motion is
\begin{eqnarray}
\sum_{n_1n_2=0}^{\infty}\begin{pmatrix}
&n_1a_{n_1n_2}'Z_{n_2}W_{n_1+n_2}z^{n_1}(x)+n_2a_{n_1n_2}'
Z_{n_1}W_{n_1+n_2}z^{n_2}(x)\\ 
&+(n_1+n_2)a_{n_1n_2}'W_{n_1}W_{n_2}z^{n_1+n_2}(x)-h(x) + \Xint-_0^1
dy\,{z(x)\over z(p)-z(y)} 
\end{pmatrix}=0
\end{eqnarray}

\bc
----------------------------
\ec

\bibliography{bibforphase}{}

\providecommand{\href}[2]{#2}\begingroup\raggedright\begin{thebibliography}{10}

\bibitem{BIPZ}
E.~Brezin, C.~Itzykson, G.~Parisi and J.~B. Zuber, \emph{{Planar Diagrams}},
  \href{http://dx.doi.org/10.1007/BF01614153}{\emph{Commun. Math. Phys.} {\bf
  59} (1978) 35}.

\bibitem{douglas2}
M.~R. Douglas, \emph{{Conformal field theory techniques in large $N$ Yang-Mills
  theory}},  in \emph{{NATO Advanced Research Workshop on New Developments in
  String Theory, Conformal Models and Topological Field Theory Cargese, France,
  May 12-21, 1993}}, 1993.
\newblock \href{http://arxiv.org/abs/hep-th/9311130}{{\tt hep-th/9311130}}.

\bibitem{polchinski}
J.~Polchinski, \emph{{Classical limit of $(1+1)$-dimensional string theory}},
  \href{http://dx.doi.org/10.1016/0550-3213(91)90559-G}{\emph{Nucl. Phys.} {\bf
  B362} (1991) 125--140}.

\bibitem{Sundborg:1999ue}
B.~Sundborg, \emph{{The Hagedorn transition, deconfinement and $\mathcal{N}=4$
  SYM theory}},
  \href{http://dx.doi.org/10.1016/S0550-3213(00)00044-4}{\emph{Nucl. Phys.}
  {\bf B573} (2000) 349--363}, [\href{http://arxiv.org/abs/hep-th/9908001}{{\tt
  hep-th/9908001}}].

\bibitem{AMMKR}
O.~Aharony, J.~Marsano, S.~Minwalla, K.~Papadodimas and M.~Van~Raamsdonk,
  \emph{{The Hagedorn - deconfinement phase transition in weakly coupled large
  N gauge theories}},
  \href{http://dx.doi.org/10.4310/ATMP.2004.v8.n4.a1}{\emph{Adv. Theor. Math.
  Phys.} {\bf 8} (2004) 603--696},
  [\href{http://arxiv.org/abs/hep-th/0310285}{{\tt hep-th/0310285}}].

\bibitem{Basu:2005pj}
P.~Basu and S.~R. Wadia, \emph{{R-charged $AdS_5$ black holes and large N
  unitary matrix models}},
  \href{http://dx.doi.org/10.1103/PhysRevD.73.045022}{\emph{Phys. Rev.} {\bf
  D73} (2006) 045022}, [\href{http://arxiv.org/abs/hep-th/0506203}{{\tt
  hep-th/0506203}}].

\bibitem{Yamada:2006rx}
D.~Yamada and L.~G. Yaffe, \emph{{Phase diagram of $\mathcal{N}=4$
  super-Yang-Mills theory with R-symmetry chemical potentials}},
  \href{http://dx.doi.org/10.1088/1126-6708/2006/09/027}{\emph{JHEP} {\bf 09}
  (2006) 027}, [\href{http://arxiv.org/abs/hep-th/0602074}{{\tt
  hep-th/0602074}}].

\bibitem{AlvarezGaume:2005fv}
L.~Alvarez-Gaume, C.~Gomez, H.~Liu and S.~Wadia, \emph{{Finite temperature
  effective action, $AdS_5$ black holes, and $1/N$ expansion}},
  \href{http://dx.doi.org/10.1103/PhysRevD.71.124023}{\emph{Phys. Rev.} {\bf
  D71} (2005) 124023}, [\href{http://arxiv.org/abs/hep-th/0502227}{{\tt
  hep-th/0502227}}].

\bibitem{Harmark:2006di}
T.~Harmark and M.~Orselli, \emph{{Quantum mechanical sectors in thermal
  $\mathcal{N}=4$ super Yang-Mills on $R\times S^3$}},
  \href{http://dx.doi.org/10.1016/j.nuclphysb.2006.08.022}{\emph{Nucl. Phys.}
  {\bf B757} (2006) 117--145}, [\href{http://arxiv.org/abs/hep-th/0605234}{{\tt
  hep-th/0605234}}].

\bibitem{duttagopakumar}
S.~Dutta and R.~Gopakumar, \emph{{Free fermions and thermal $AdS/CFT$}},
  \href{http://dx.doi.org/10.1088/1126-6708/2008/03/011}{\emph{JHEP} {\bf 03}
  (2008) 011}, [\href{http://arxiv.org/abs/0711.0133}{{\tt 0711.0133}}].

\bibitem{duttadutta}
P.~Dutta and S.~Dutta, \emph{{Phase Space Distribution for Two-Gap Solution in
  Unitary Matrix Model}},
  \href{http://dx.doi.org/10.1007/JHEP04(2016)104}{\emph{JHEP} {\bf 04} (2016)
  104}, [\href{http://arxiv.org/abs/1510.03444}{{\tt 1510.03444}}].

\bibitem{riemannzero}
P.~Dutta and S.~Dutta, \emph{{Phase Space Distribution of Riemann Zeros}},
  \href{http://dx.doi.org/10.1063/1.4982737}{\emph{J. Math. Phys.} {\bf 58}
  (2017) 053504}, [\href{http://arxiv.org/abs/1610.07743}{{\tt 1610.07743}}].

\bibitem{kerov_book}
S.~Kerov, \emph{{Asymptotic Representation Theory of the Symmetric Group and
  its Application in Analysis}}.
\newblock {Translations of mathematical monographs}. American Mathematical
  Society, 2003.

\bibitem{Biane}
P.~Biane, \emph{{Representations of Symmetric Groups and Free Probability}},
  \href{http://dx.doi.org/https://doi.org/10.1006/aima.1998.1745}{\emph{Advances
  in Mathematics} {\bf 138} (1998) 126 -- 181}.

\bibitem{pallab}
P.~Basu, B.~Ezhuthachan and S.~R. Wadia, \emph{{Plasma balls/kinks as solitons
  of large N confining gauge theories}},
  \href{http://dx.doi.org/10.1088/1126-6708/2007/01/003}{\emph{JHEP} {\bf 01}
  (2007) 003}, [\href{http://arxiv.org/abs/hep-th/0610257}{{\tt
  hep-th/0610257}}].

\bibitem{Matytsin}
A.~Matytsin, \emph{{On the large-$N$ limit of the Itzykson-Zuber integral}},
  \href{http://dx.doi.org/https://doi.org/10.1016/0550-3213(94)90471-5}{\emph{Nuclear
  Physics B} {\bf 411} (1994) 805 -- 820}.

\bibitem{Grossmatytsin}
D.~J. Gross and A.~Matytsin, \emph{{Some properties of large $N$
  two-dimensional Yang-Mills theory}},
  \href{http://dx.doi.org/10.1016/0550-3213(94)00570-5}{\emph{Nucl. Phys.} {\bf
  B437} (1995) 541--584}, [\href{http://arxiv.org/abs/hep-th/9410054}{{\tt
  hep-th/9410054}}].

\bibitem{das-jevicki}
S.~R. Das and A.~Jevicki, \emph{{String Field Theory and physical
  interpretation of $D=1$ strings}},
  \href{http://dx.doi.org/10.1142/S0217732390001888}{\emph{Modern Physics
  Letters A} {\bf 05} (1990) 1639--1650}.

\bibitem{sakita}
A.~Jevicki and B.~Sakita, \emph{{The quantum collective field method and its
  application to the planar limit}},
  \href{http://dx.doi.org/http://dx.doi.org/10.1016/0550-3213(80)90046-2}{\emph{Nuclear
  Physics B} {\bf 165} (1980) 511 -- 527}.

\bibitem{zalewski}
J.~Jurkiewicz and K.~Zalewski, \emph{{Phase structure of $U(N\ra \infty)$ gauge
  theory on a two-dimensional lattice for a broad class of variant actions}},
  \href{http://dx.doi.org/http://dx.doi.org/10.1016/0550-3213(83)90221-3}{\emph{Nuclear
  Physics B} {\bf 220} (1983) 167 -- 184}.

\bibitem{Mandal:1989ry}
G.~Mandal, \emph{{Phase Structure of Unitary Matrix Models}},
  \href{http://dx.doi.org/10.1142/S0217732390001281}{\emph{Mod. Phys. Lett.}
  {\bf A5} (1990) 1147--1158}.

\bibitem{lasalle}
M.~Lassalle, \emph{{Explicitation of characters of the symmetric group}},
  \href{http://dx.doi.org/http://dx.doi.org/10.1016/j.crma.2005.09.016}{\emph{Comptes
  Rendus Mathematique} {\bf 341} (2005) 529 -- 534}.

\bibitem{hamermesh}
M.~Hamermesh, \emph{Group Theory and its Application to Physical Problems}.
\newblock Dover Publication, 1989.

\bibitem{fulton-harris}
W.~Fulton and J.~Harris, \emph{Representation Theory: A First Course (Graduate
  Texts in Mathematics)}.
\newblock Springer, 1999.

\bibitem{Douglas:1993iia}
M.~R. Douglas and V.~A. Kazakov, \emph{{Large N phase transition in continuum
  QCD in two-dimensions}},
  \href{http://dx.doi.org/10.1016/0370-2693(93)90806-S}{\emph{Phys. Lett.} {\bf
  B319} (1993) 219--230}, [\href{http://arxiv.org/abs/hep-th/9305047}{{\tt
  hep-th/9305047}}].

\bibitem{Kazakov:1995ae}
V.~A. Kazakov, M.~Staudacher and T.~Wynter, \emph{{Character expansion methods
  for matrix models of dually weighted graphs}},
  \href{http://dx.doi.org/10.1007/BF02101902}{\emph{Commun. Math. Phys.} {\bf
  177} (1996) 451--468}, [\href{http://arxiv.org/abs/hep-th/9502132}{{\tt
  hep-th/9502132}}].

\bibitem{thomasfermi1}
L.~H. Thomas, \emph{{The calculation of atomic fields}},
  \href{http://dx.doi.org/10.1017/S0305004100011683}{\emph{Mathematical
  Proceedings of the Cambridge Philosophical Society} {\bf 23} (1927)
  542–548}.

\bibitem{thomasfermi2}
E.~Fermi, \emph{{Un metodo statistico per la determinazione di alcune priorieta
  dell’atome}}, {\emph{Rend. Accad. Naz. Lincei} {\bf 6} (1927) 32}.

\bibitem{jevicki}
A.~Jevicki and B.~Sakita, \emph{{Collective field approach to the large-$N$
  limit: Euclidean field theories}},
  \href{http://dx.doi.org/http://dx.doi.org/10.1016/0550-3213(81)90365-5}{\emph{Nuclear
  Physics B} {\bf 185} (1981) 89 -- 100}.

\bibitem{pallabnew}
T.~Azuma, P.~Basu and P.~Samantray, \emph{{Phase Transitions of a (Super)
  Quantum Mechanical Matrix Model with a Chemical Potential}},
  \href{http://arxiv.org/abs/1707.02898}{{\tt 1707.02898}}.

\bibitem{koch}
R.~de~Mello~Koch, \emph{{Geometries from Young Diagrams}},
  \href{http://dx.doi.org/10.1088/1126-6708/2008/11/061}{\emph{JHEP} {\bf 11}
  (2008) 061}, [\href{http://arxiv.org/abs/0806.0685}{{\tt 0806.0685}}].

\bibitem{Friedan:1980tu}
D.~Friedan, \emph{{Some Nonabelian Toy Models in the Large $N$ Limit}},
  \href{http://dx.doi.org/10.1007/BF01942328}{\emph{Commun. Math. Phys.} {\bf
  78} (1981) 353}.

\bibitem{gross-witten}
D.~J. Gross and E.~Witten, \emph{{Possible Third Order Phase Transition in the
  Large N Lattice Gauge Theory}},
  \href{http://dx.doi.org/10.1103/PhysRevD.21.446}{\emph{Phys. Rev.} {\bf D21}
  (1980) 446--453}.

\bibitem{wadia}
S.~R. Wadia, \emph{{$N = \infty$ Phase Transition in a Class of Exactly Soluble
  Model Lattice Gauge Theories}},
  \href{http://dx.doi.org/10.1016/0370-2693(80)90353-6}{\emph{Phys. Lett.} {\bf
  93B} (1980) 403--410}.

\bibitem{Witten:1992fb}
E.~Witten, \emph{{Chern-Simons gauge theory as a string theory}}, {\emph{Prog.
  Math.} {\bf 133} (1995) 637--678},
  [\href{http://arxiv.org/abs/hep-th/9207094}{{\tt hep-th/9207094}}].

\bibitem{Gopakumar:1998ki}
R.~Gopakumar and C.~Vafa, \emph{{On the gauge theory / geometry
  correspondence}}, {\emph{Adv. Theor. Math. Phys.} {\bf 3} (1999) 1415--1443},
  [\href{http://arxiv.org/abs/hep-th/9811131}{{\tt hep-th/9811131}}].

\bibitem{Aganagic:2002wv}
M.~Aganagic, A.~Klemm, M.~Marino and C.~Vafa, \emph{{Matrix model as a mirror
  of Chern-Simons theory}},
  \href{http://dx.doi.org/10.1088/1126-6708/2004/02/010}{\emph{JHEP} {\bf 02}
  (2004) 010}, [\href{http://arxiv.org/abs/hep-th/0211098}{{\tt
  hep-th/0211098}}].

\bibitem{Marino:2002fk}
M.~Marino, \emph{{Chern-Simons theory, matrix integrals, and perturbative three
  manifold invariants}},
  \href{http://dx.doi.org/10.1007/s00220-004-1194-4}{\emph{Commun. Math. Phys.}
  {\bf 253} (2004) 25--49}, [\href{http://arxiv.org/abs/hep-th/0207096}{{\tt
  hep-th/0207096}}].

\bibitem{Wittenjones}
E.~Witten, \emph{{Quantum Field Theory and the Jones Polynomial}},
  \href{http://dx.doi.org/10.1007/BF01217730}{\emph{Commun. Math. Phys.} {\bf
  121} (1989) 351--399}.

\bibitem{okuda}
T.~Okuda, \emph{{Derivation of Calabi-Yau crystals from Chern-Simons gauge
  theory}}, \href{http://dx.doi.org/10.1088/1126-6708/2005/03/047}{\emph{JHEP}
  {\bf 03} (2005) 047}, [\href{http://arxiv.org/abs/hep-th/0409270}{{\tt
  hep-th/0409270}}].

\bibitem{Marino:2004eq}
M.~Marino, \emph{{Les Houches lectures on matrix models and topological
  strings}},  2004.
\newblock \href{http://arxiv.org/abs/hep-th/0410165}{{\tt hep-th/0410165}}.

\bibitem{Morita:2017oev}
T.~Morita and K.~Sugiyama, \emph{{Multi-cut Solutions in Chern-Simons Matrix
  Models}},  \href{http://arxiv.org/abs/1704.08675}{{\tt 1704.08675}}.

\bibitem{Halmagyi}
N.~Halmagyi and V.~Yasnov, \emph{{The Spectral curve of the lens space matrix
  model}}, \href{http://dx.doi.org/10.1088/1126-6708/2009/11/104}{\emph{JHEP}
  {\bf 11} (2009) 104}, [\href{http://arxiv.org/abs/hep-th/0311117}{{\tt
  hep-th/0311117}}].

\bibitem{Marino:2009jd}
M.~Marino and P.~Putrov, \emph{{Exact Results in ABJM Theory from Topological
  Strings}}, \href{http://dx.doi.org/10.1007/JHEP06(2010)011}{\emph{JHEP} {\bf
  06} (2010) 011}, [\href{http://arxiv.org/abs/0912.3074}{{\tt 0912.3074}}].

\bibitem{Aharony:2008ug}
O.~Aharony, O.~Bergman, D.~L. Jafferis and J.~Maldacena, \emph{{$\mathcal{N}=6$
  superconformal Chern-Simons-matter theories, $M2$-branes and their gravity
  duals}}, \href{http://dx.doi.org/10.1088/1126-6708/2008/10/091}{\emph{JHEP}
  {\bf 10} (2008) 091}, [\href{http://arxiv.org/abs/0806.1218}{{\tt
  0806.1218}}].

\bibitem{Drukker:2010nc}
N.~Drukker, M.~Marino and P.~Putrov, \emph{{From weak to strong coupling in
  ABJM theory}},
  \href{http://dx.doi.org/10.1007/s00220-011-1253-6}{\emph{Commun. Math. Phys.}
  {\bf 306} (2011) 511--563}, [\href{http://arxiv.org/abs/1007.3837}{{\tt
  1007.3837}}].

\bibitem{Lin:2004nb}
H.~Lin, O.~Lunin and J.~M. Maldacena, \emph{{Bubbling AdS space and $1/2$ BPS
  geometries}},
  \href{http://dx.doi.org/10.1088/1126-6708/2004/10/025}{\emph{JHEP} {\bf 10}
  (2004) 025}, [\href{http://arxiv.org/abs/hep-th/0409174}{{\tt
  hep-th/0409174}}].

\bibitem{Jain:2013py}
S.~Jain, S.~Minwalla, T.~Sharma, T.~Takimi, S.~R. Wadia and S.~Yokoyama,
  \emph{{Phases of large $N$ vector Chern-Simons theories on $S^2 \times
  S^1$}}, \href{http://dx.doi.org/10.1007/JHEP09(2013)009}{\emph{JHEP} {\bf 09}
  (2013) 009}, [\href{http://arxiv.org/abs/1301.6169}{{\tt 1301.6169}}].

\bibitem{Jain:2013gza}
S.~Jain, S.~Minwalla and S.~Yokoyama, \emph{{Chern Simons duality with a
  fundamental boson and fermion}},
  \href{http://dx.doi.org/10.1007/JHEP11(2013)037}{\emph{JHEP} {\bf 11} (2013)
  037}, [\href{http://arxiv.org/abs/1305.7235}{{\tt 1305.7235}}].

\bibitem{Jain:2012qi}
S.~Jain, S.~P. Trivedi, S.~R. Wadia and S.~Yokoyama, \emph{{Supersymmetric
  Chern-Simons Theories with Vector Matter}},
  \href{http://dx.doi.org/10.1007/JHEP10(2012)194}{\emph{JHEP} {\bf 10} (2012)
  194}, [\href{http://arxiv.org/abs/1207.4750}{{\tt 1207.4750}}].

\bibitem{Marino:2012az}
M.~Mariño and P.~Putrov, \emph{{Interacting fermions and $\mathcal{N}=2$
  Chern-Simons-matter theories}},
  \href{http://dx.doi.org/10.1007/JHEP11(2013)199}{\emph{JHEP} {\bf 11} (2013)
  199}, [\href{http://arxiv.org/abs/1206.6346}{{\tt 1206.6346}}].

\bibitem{Codesido:2014oua}
S.~Codesido, A.~Grassi and M.~Mariño, \emph{{Exact results in $ \mathcal{N}=8
  $ Chern-Simons-matter theories and quantum geometry}},
  \href{http://dx.doi.org/10.1007/JHEP07(2015)011}{\emph{JHEP} {\bf 07} (2015)
  011}, [\href{http://arxiv.org/abs/1409.1799}{{\tt 1409.1799}}].

\bibitem{Dandekar:2014era}
Y.~Dandekar, M.~Mandlik and S.~Minwalla, \emph{{Poles in the $S$-Matrix of
  Relativistic Chern-Simons Matter theories from Quantum Mechanics}},
  \href{http://dx.doi.org/10.1007/JHEP04(2015)102}{\emph{JHEP} {\bf 04} (2015)
  102}, [\href{http://arxiv.org/abs/1407.1322}{{\tt 1407.1322}}].

\bibitem{Durhusolesen}
B.~Durhuus and P.~Olesen, \emph{{The spectral density for two-dimensional
  continuum QCD}},
  \href{http://dx.doi.org/https://doi.org/10.1016/0550-3213(81)90230-3}{\emph{Nuclear
  Physics B} {\bf 184} (1981) 461 -- 475}.

\bibitem{Rossi}
P.~Rossi, \emph{{Continuum $QCD_2$ from a fixed-point lattice action}},
  \href{http://dx.doi.org/https://doi.org/10.1016/0003-4916(81)90075-0}{\emph{Annals
  of Physics} {\bf 132} (1981) 463 -- 481}.

\bibitem{Nowak}
J.-P. Blaizot and M.~A. Nowak, \emph{{Large-${N}_{c}$ Confinement and
  Turbulence}},
  \href{http://dx.doi.org/10.1103/PhysRevLett.101.102001}{\emph{Phys. Rev.
  Lett.} {\bf 101} (Sep, 2008) 102001}.

\bibitem{NeubergerBurger}
H.~Neuberger, \emph{{Burgers' equation in $2D$ $SU(N)$ YM}},
  \href{http://dx.doi.org/https://doi.org/10.1016/j.physletb.2008.06.064}{\emph{Physics
  Letters B} {\bf 666} (2008) 106 -- 109}.

\bibitem{Neuberger}
R.~Lohmayer and H.~Neuberger, \emph{{Nonanalyticity in Scale in the Planar
  Limit of QCD}},
  \href{http://dx.doi.org/10.1103/PhysRevLett.108.061602}{\emph{Phys. Rev.
  Lett.} {\bf 108} (Feb, 2012) 061602}.

\bibitem{eynard_book}
B.~Eynard, \emph{Counting Surfaces: CRM Aisenstadt Chair lectures}.
\newblock {Progress in Mathematical Physics 70}. Birkhauser Basel, 2016.

\bibitem{berry-keating}
M.~V. Berry and J.~P. Keating, \emph{{The Riemann Zeros and Eigenvalue
  Asymptotics}}, \href{http://dx.doi.org/10.1137/S0036144598347497}{\emph{SIAM
  Review} {\bf 41} (1999) 236--266},
  [\href{http://arxiv.org/abs/https://doi.org/10.1137/S0036144598347497}{{\tt
  https://doi.org/10.1137/S0036144598347497}}].

\end{thebibliography}\endgroup
\bibliographystyle{jhep}

\end{document}